\documentclass[manuscript, nonacm]{acmart}

\usepackage{adjustbox}

\usepackage[utf8]{inputenc} 
\usepackage[T1]{fontenc}    
\usepackage{refcount}
\usepackage{url}            
\usepackage{booktabs}       
\usepackage{amsfonts}       
\usepackage{nicefrac}       
\usepackage{microtype}      
\usepackage{xcolor}         
\usepackage{subcaption}
\usepackage{graphicx}
\usepackage[inkscapelatex=false]{svg}
\svgpath{{figures/all_datasets/}} 

\usepackage{listings}
\usepackage{wrapfig} 

\begin{document}

\title{Sound Check: Auditing Audio Datasets}

\author{William Agnew}
\email{wagnew@andrew.cmu.edu}
\affiliation{%
  \institution{Carnegie Mellon University}
  \country{USA}
}

\author{Julia Barnett}
\email{juliabarnett@u.northwestern.edu}
\affiliation{%
  \institution{Northwestern University}
  \country{USA}
}

\author{Annie Chu}
\email{anniechu@u.northwestern.edu}
\affiliation{%
  \institution{Northwestern University}
  \country{USA}
}

\author{Rachel Hong}
\email{hongrach@cs.washington.edu}
\affiliation{%
  \institution{University of Washington}
  \country{USA}
}

\author{Michael Feffer}
\email{mfeffer@andrew.cmu.edu}
\affiliation{%
  \institution{Carnegie Mellon University}
  \country{USA}
}

\author{Robin Netzorg}
\email{robert_netzorg@berkeley.edu}
\affiliation{%
  \institution{University of California Berkeley}
  \country{USA}
}

\author{Harry H. Jiang}
\email{hhj@andrew.cmu.edu}
\affiliation{%
  \institution{Carnegie Mellon University}
  \country{USA}
}

\author{Ezra Awumey}
\email{eawumey@andrew.cmu.edu}
\affiliation{%
  \institution{Carnegie Mellon University}
  \country{USA}
}

\author{Sauvik Das}
\email{sauvik@cmu.edu}
\affiliation{%
  \institution{Carnegie Mellon University}
  \country{USA}
}

\renewcommand{\shortauthors}{Agnew et al.}

\begin{abstract}
Generative audio models are rapidly advancing in both capabilities and public utilization---several powerful generative audio models have readily available open weights, and some tech companies have released high quality generative audio products.
Yet, while prior work has enumerated many ethical issues stemming from the data on which generative visual and textual models have been trained, we have little understanding of similar issues with generative audio datasets, including those related to bias, toxicity, and intellectual property.
To bridge this gap,
we conducted a literature review of hundreds of audio datasets and selected seven of the most prominent to audit in more detail.
We found that these datasets are biased against women, contain toxic stereotypes about marginalized communities, and contain significant amounts of copyrighted work. To enable artists to see if they are in popular audio datasets and facilitate exploration of the contents of these datasets, we developed a web tool audio datasets exploration tool at \url{https://audio-audit.vercel.app/}.
\end{abstract}







\maketitle

\section{Introduction}


Deep learning and ML-based techniques achieve state-of-the-art performance for a broad range of audio processing tasks ranging from speech transcription \cite{radford2023robust}, pitch estimation \cite{kim2018crepe, liu2023fast} to acoustic event classification \cite{wang2022deep}. Beyond solving foundational problems in areas like signal processing, speech processing, and music information retrieval (MIR), these technologies support an increasing number of higher-level human-AI interactions. For example, virtual avatars, assistive technologies for individuals with visual impairments, as well as novel user interface navigation paradigms utilize advances in audio AI techniques to improve the quality and fidelity of user experiences where audio is a primary interaction modality \cite{DanielescuNB23,UpadhyayOlder24,YuInterface23}. More recently, generative AI has led to the development of audio based technologies that can perform several different tasks including making reservations, and text-to-song prompting systems \cite{RichardCA19,WuTTM24}. 


While the harms of discriminative audio AI have received some treatment by prior work in HCI \cite{Wenzel2024Values},the harms of generative audio technologies are less understood. Recent events suggest generative AI audio will have unique social and legal implications spanning individuals' right of publicity, misinformation and copyright law, especially when these systems are trained on stolen data. For example, the recently deployed generative AI platform SUNO AI, has been accused of unlawfully using artists' copyrighted music to provide its users with a steady stream of royalty-free music \cite{Newton-Rex_2024_Suno}. Another AI-music generation platform, Udio, has been sued by huge players in the music industry such as UMG Recordings, Capitol Records, and Sony who alleged they unlawfully used their recordings to train their model, which Udio essentially conceded to in pre-litigation documents \cite{udio_litigation}. Other companies such as OpenAI have been embroiled in legal disputes over the unauthorized use of individuals' personal identities, like when the company released a voice eerily similar to Scarlet Johansson's to narrate its GPT-4 Chatbot after the actor rejected OpenAI's request to use her actual voice \cite{Johansson2024}. Generative audio is also poised to open new and exacerbate existing abuse vectors from misinformation peddling (e.g., deepfake audio imitating President Biden \cite{Murphy_2024}) to voice cloning scams (e.g., where attackers scam unsuspecting grandparents with phone calls impersonating the voice of loved ones \cite{fox59_ai2024}). Since the harms and uses of generative AI technologies are largely driven by the data on which they are trained, it is imperative that we have a better understanding of the datasets that are being used to train these models.

Our paper lays the groundwork for the comprehensive analysis of the audio data currently used to train AI models spanning the domains of music, speech, and sound. This work is motivated by the fact that while the downstream ethical risks and harms of generative text and vision models have been the subject of significant prior work \cite{birhane2024dark, birhane2021multimodal, bianchi2023easily, hong2024s}, there has been comparatively little focus and understanding of these issues in the context of generative audio, leading to a ``documentation debt'' \cite{bender2021dangers} where widely used audio datasets are often poorly documented and understood. Inspired by audits of vision and text datasets that helped crystallize discussion of the ethical harms present in those modalities of generative AI, in this paper we ask the following questions:

\begin{itemize}
    \item[(RQ1)] What audio datasets are being used currently? What is the dsitribution of their use?
    \item[(RQ2)] From where are these audio datasets sourced? What licenses are these data under?
    \item[(RQ3)] Do these datasets contain toxic content?
    \item[(RQ4)] Who is represented and not represented in audio datasets?
\end{itemize}


To answer these questions, we conducted a comprehensive literature review of audio datasets, studying their size, sourcing, contents, usage, and other aspects. From this literature review, we identify seven audio datasets that are representative of the field due to their size, popularity, and overlaps with other popular datasets. We analyze the transcripts, metadata, and genres of these datasets to understand what their contents are, who is represented in these datasets, and where their data were sourced from. We choose to conduct a broad audit of audio dataset practices given the paucity of existing studies of audio datasets. While audio is a deep and rich modality, we believe an assessment of the modality as a whole is needed to scope more specific audits and help uncover relations between audio and other modalities.


Our literature review identified 175 unique audio datasets that were used between May 2023 and May 2024. These datasets vary significantly in size and popularity, with a few datasets and data sources constituting the majority of publicly available audio (RQ1). We find that 36\% of datasets were scraped from the web, 49\% were created, 13\% were augmented (modifying an existing dataset to the extent such that it constitutes a new dataset), and 2\% were purchased from online marketplaces. 35\% of datasets are potentially copyright infringing, meaning there is at least some portion of the data for which access beyond private listening requires purchasing licenses (RQ2). We find audio datasets have a wide range of contents, with representative audio datasets containing narrations of public domain (and therefore typically decades old) books, readings of sentences from The Glasglow Herald, and samples of audio from Youtube and different royalty-free music databases (RQ3). We find who is in these datasets is frequently not present in public documentation, complicating efforts to asses bias and representation. These datasets are most frequently in English, with several notable exceptions of datasets that have made explicit efforts to collect non-English data. Finally, we find that mentions of marginalized groups are less likely to appear in the transcripts of these datasets (RQ4). 

To summarize, our paper has four primary contributions. We (1) \textbf{survey audio datasets} and models to understand \textbf{dataset practices, uses, and creation methods}. We (2) \textbf{assess} transcripts, genres, and artist demographics for \textbf{bias and toxicity}. We (3) extract and infer \textbf{information about copyright and artists to understand IP issues}. Finally, we (4) project how dataset composition may \textbf{affect downstream models} and \textbf{provide recommendations for mitigation}.

Before detailing the methods and results of our audio dataset and model audits, we first briefly describe prior work in audio AI, research regarding ethical considerations of such developments, and work in dataset audits more generally.

\subsection{Audio AI}


The main architectures used in current audio modeling are quite similar to those utilized in image and text, with slight adjustments made to handle the specifics of audio or speech data, such as the Audio Spectrogram Transformer \cite{gong2021ast}, which utilize frequency representations of audio like spectrograms to better model the different audio modalities. LALMs often will combine pre-trained LLMs with audio-specific encoders to extend LLM capabilities to audio modalities \cite{tang2024salmonn, gong_ltuas, chu2023qwenaudio}. Generative approaches to audio often borrow from LLMs \cite{borsos2023audiolm} or diffusion models that instead act on spectrograms or waveforms directly \cite{liu2023audioldm}. To derive models from audio datasets, a number of methods have been introduced mirroring developments in deep learning for text and images, including RNNs \cite{sturm2019machine}, CNNs \cite{google_bach,huang2017counterpoint}, and combinations of the two \cite{donahue2017dance}, have been proposed for the audio domain. Similarly, researchers have more recently turned to techniques leveraging transformer architectures~\cite{Agostinelli2023musiclm, Donahue2023singsong, garcia2023vampnet} and diffusion~\cite{forsgren2022riffusion, wang2023audit}, especially as generative AI (GenAI) has captured public interest.


While there are many semantically distinct audio modalities, they largely fall into three categories: music, speech, and (environmental) sound--largely referred to as audio by the community.  Historically, these modeling in each of these modalities has primarily fallen under separate fields, but recent advances in Large Audio Language Models (LALMs) have seen these distinct modalities being united under single frameworks \cite{ghosh2024gama, gong_ltuas}.These models aim to perform classical tasks, such as Automatic Speech Recognition, Audio Captioning, Note Identification, under a single modeling paradigm \cite{tang2024salmonn} and support applications spanning accessibility technologies to music generation.


At present a number of challenges exist in the the domain of AI Audio. Many of these problems stem from the limitations of AI-models regarding their ability to pick up on contextual cues traditionally only understood through knowledge and familiarity with the nuances of human communication, such as idiomatic or sarcastic language)\cite{TomarSarcasm23}. Similarly, accents and speech impediments may prove obstacles to model performance\cite{Wenzel2024Values}. Proposed solutions to these problems include the use of synthetic data to approximate data from populations not well represented in the AI training data like stutterers \cite{Hao2024SyntheticDI}.

For further information on AI for audio, \citet{civit2022systematic} provide a review of music generation, \citet{mehrish2023review} review AI for speech processing, and \citet{nogueira2022sound, kelley2020review}, and \citet{palaniappan2014artificial} review AI for sound processing in several major application areas.


\subsection{Ethics of Audio AI}

In light of the recent turn to GenAI and improvements of other ML-based audio technologies, some researchers have started to grapple with corresponding ethical concerns and implications. However, as detailed by the literature review conducted by \citet{barnett2023ethical} surveying generative audio research papers, few authors have contemplated the potential negative impacts of their work. Even further, \citet{morreale2023data} find that audio datasets are often created without permission of audio owners and creators. Some of these harms have started to be addressed, especially recently, such such as training data attribution of generative audio models \cite{barnett2024exploring, bralios2024generation}. 

\citet{shelby2023sociotechnical} highlight the sociotechnical nature of AI harms, and emphasize that harms cannot exist independent of societal norms and structures---they have to exist in a set of systems. Ruha Benjamin \cite{benjamin2019race} sheds light on how technological harms (not unlike most societal harms) have a disproportionate effect on people of color; technologies were built for the people in power and ``often adopt the default norms and power structures of society.'' Audio harms are no exception; they can even be magnified when we do not understand the contents of the data. White voice actors doing ``black'' accents is akin to blackface through the audio medium, and voice actors note this is something that can poison datasets. In a similar vein, modern voice cloning models boast of being able to capture diverse voices, but both being able and being unable to model voice archetypes like a ``gay voice'' comes with a host of both safety and representational harms  \cite{sigurgeirsson2024just}---being unable to could result in harms of not representing all types of voices, while being able to could lead to harmful stereotyping and or require data collection that could put participants at risk. 


Audio deepfakes present a whole new set of harms separate from those already realized by visual and even video deepfakes. In 2023 alone, music featuring deepfake voices of popular artists went viral on social media \cite{Coscarelli_2023, feffer2023deepdrake}, prompting the music industry to start grappling with intellectual property concerns entailed by generative audio models \cite{Hoover,spotify_ai_removal,Patel_2023,Sisario_2024} and even take down online communities where deepfake audio was proliferating \cite{Hook_2023}. Audio deepfakes are particularly dangerous in the case of phishing and fraud, where bad actors can impersonate voices with high believability and deceive people or even bypass voice security systems \cite{habib2019semi,sisman2020overview,kim2022guided}.Many audio papers, especially text-to-speech papers, note the potential for misuse in form of audio deepfake \cite{wang2020deepsonar, kim2020glow, kim2022guided} and some even noted they had no plans to release their models due to the strong potential for misuse via deepfake \cite{kim2022guided}.






Even within audio harms, there are potential risks specific to sub-fields in audio such as speech generation, music generation, and even the sub-sub-fields like text-to-speech (TTS). \citet{HutiriHarms24} detail the specific harms inherent to speech generators such as voice clones of voice actors, ``bringing back the dead'', and audio deepfakes of public figures. \citet{batlle2023transparency} focus on the specific aspect of transparency within generative music and highlight the link between transparency and creativity, originality, and ownership of AI-generated music, suggesting that we should move towards more transparent AI-based music generation. Within TTS there are specific questions with regards to liability of harmful speech \cite{henderson2023s} as well as harms of the reverse---a high potential for hallucination in speech-to-text \cite{koenecke2024careless} with an estimate of about 1\% of audio transcriptions being entirely hallucinated.

Other ethical quandaries remain regarding the contribution of these models to climate change \cite{douwes2021energy}, speaker privacy and security \cite{o2024maskmark, champion2023anonymizing}, creativity 
\cite{khosrowi2023diffusing}, GenAI's effect on music creators as a whole \cite{barnett2023ethical, lee2022ethics} and the ethics of using voice synthesis on deceased people 
\cite{feffer2023deepdrake, lee2022ethics}.
Beyond risks for misinformation and economic harms to artists, recent high-profile instances of fraud (e.g., the transfer of millions of dollars to scammers leveraging GenAI to deceive targets \cite{Lo_2024,Milmo_2024}), physiognomy (e.g., gender and sexual orientation classification \cite{lee2024deepfakes}), and surveillance (e.g., gunshot detection for predictive policing \cite{crocco2016audio}) illustrate the real-world privacy, security, and ethical risks of these technologies.

HCI research has shed light on the often overlooked human labor underlying much of what AI systems are currently capable of \cite{fox2023patchwork}. The human labor involved, which might include data cleaning, annotation and tagging, up to the creation of the data AI are trained on contribute significantly to model performance \cite{li2021cleanml}. Researchers in this space must address challenges stemming from opaque data collection and processing techniques on the part of AI developers, which are now increasingly being trained on individuals personal or copyrighted data \cite{lemley2024generative, samuelson2023generative}. 

Audio AI is a complex and rapidly growing field, and if harms are not addressed early then they will be snowballed as the field progresses. In this work we examine the current state of audio datasets and highlight potential ways to employ these datasets more ethically and how to curate more ethically sourced datasets in the future. 

\subsection{Dataset Audits}



Audits of datasets have proven vital for understanding the behavior and forecasting biases, toxicity, and other harms of downstream models. \citet{prabhu2021large} found that the 80 Million Tiny Images dataset contained racist and non-consensual intimate imagery (NCII), leading to the creators to take down this dataset \citep{johnson2020MIT}. \citet{Birhane2021laion} and \citet{thiel2023identifying} uncover evidence of child sexual abuse material (CSAM) in the LAION5B text-image dataset \cite{schuhmann2022laion}, leading to its removal \cite{404laion}. While dataset audits incorporate a variety of methods and aims---representation, toxicity, privacy, or copyright concerns---they all help determine how the targeted dataset's contents align with expectations in efforts to achieve accountability \citep{birhane2024ai}. \citet{paullada2021data} surveyed dataset audits and found they reveal representational harms and the presence of problematic content overlooked during data curation. Despite the impact of audits, \citet{bender2021dangers} argue that machine learning faces a dataset ``documentation debt'', with many popular datasets having little if any documentation. Audio suffers acutely from documentation debt, with very few analyses of audio datasets outside of their suitability for increasing technical measures of performance, with the notable exception of bias and representation audits of Mozilla Common Voice \cite{langdetect}. In this paper, we undertake
a broad, domain-wide audit of audio datasets, assessing changing data practices and uncovering a range of risks and harms caused by audio datasets.


\section{Literature Review of Current Audio Datasets and Models}


As a first step to understanding current audio dataset practices, we conducted a broad literature review of audio datasets created or used between May 2023 and May 2024. 

In order to understand how many audio datasets exist, the distribution of their usage in the research community, and how these datasets were sourced, we conduct a literature review on audio modelling papers submitted to arXiv, a preprint platform previous studies have found to be an effective source for current and important audio AI papers \cite{barnett2023ethical}. We search for papers uploaded between May 1 2023 and May 1 2024 to capture 1 year of data and annotate the datasets included in these papers. We chose this time frame to capture the most recent usage of datasets in a field that has undergone rapid changes in recent years with the introduction of transformers \cite{vaswani2017attention} and rapidly growing academic and commercial interest in generative AI. We analyzed audio modelling papers until we approached saturation of datasets (i.e., until we found most of the datasets used were either one-offs or common repeats). We first detail how we conducted this review and then discuss the contents and usage of the identified datasets.

\subsection{Literature Review Methodology}
\begin{figure}[ht]
  \begin{center}
    \includegraphics[width=0.5\textwidth]{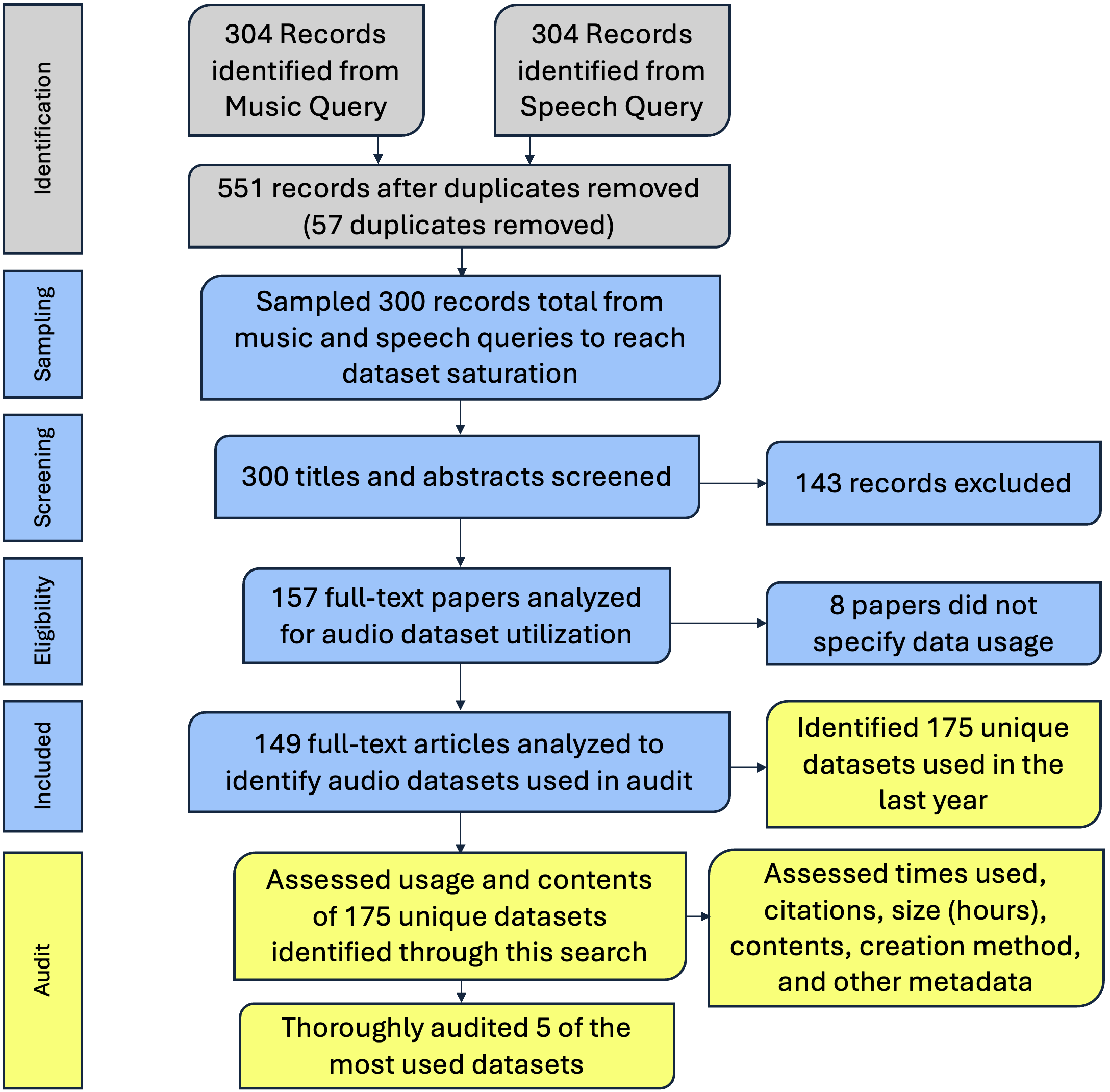}
  \end{center}
  \caption{Flow diagram inspired by PRISMA \cite{moher2015preferred} detailing the paper corpus used to produce the datasets audited in this paper. We started with 551 records from arXiv (grey) and analyzed 149 full-text articles (blue) to identify 175 unique datasets for analysis (yellow).} 
\label{fig:prisma}
\end{figure}



We were interested in identifying audio datasets currently used by researchers to both include in our audit and provide an overview of the data landscape. We chose to conduct a systematic literature review of one year of arXiv \cite{arxiv_page} computer science works about audio models. We chose one year since there has been a paradigmatic shift in audio generative models, driven in part by the recent advances in large language models and both their translation to text-to-audio models and adopted language-model-style generation as seen in AudioLM \cite{borsos2023audiolm}, MusicLM \cite{Agostinelli2023musiclm}, SoundStorm \cite{borsos2023soundstorm}, VampNet \cite{garcia2023vampnet}, and more---researchers and commercial developers alike are largely abandoning the early approaches we saw in 2020-2022 such as the music transformer \cite{huang2018music}. Since 2023, approaches towards building Large Audio Language Models (LALMs) largely involve combining many datasets from a broad range of tasks in speech, audio, and music processing \cite{tang2024salmonn, gong2023listen, chu2023qwenaudio}, often using proprietary language models like the GPT model family to supplement existing data with additional question-answer pairs \cite{ghosh2024gama, deshmukh2024audioentailment}. One such dataset, OpenASQA, combines a total of 13 publicly available audio, music, and speech datasets to train their LALM, LTU-AS, and uses GPT-3.5-Turbo to generate QA pairs \cite{gong_ltuas}. As researchers move towards curating giant meta-datasets of datasets, it becomes exceedingly vital to understand the origins, licensing, and limits of the many datasets that feed the creation of LALMs. 

We chose to focus on arXiv submissions because Barnett's \cite{barnett2023ethical} recent systematic literature review on the ethical implications of generative audio models resulted in a final corpus comprised of 91\% arXiv works even after starting from a 50/50 split of arXiv and ACM works. Though it is difficult to quantify the most influential papers in anything other than citations, which is certainly not a perfect metric, we verified that all the major audio generation papers such as the ones already mentioned in this paper were present in arXiv as well to justify this focus of our exploratory systematic literature review. 

Following Barnett's literature review \cite{barnett2023ethical}, we used the following query on arXiv (once for music, once for speech). This query searches all fields on arXiv including title, abstract, and keywords of papers in this database, but does not search the full text of articles.

\begin{lstlisting}
[
    [
        [All: "generative"] AND 
        [All: "$\langle$music/speech$\rangle$"] AND
        [All: "model"]
    ] 
    AND [date_range: from 2023-05-01 to 2024-05-01]
]; classification: Computer Science (cs)    
\end{lstlisting}

The music query produced 304 records and the speech query produced 1561. For context, the same date range from 2022-2023 included 114 records for music and 421 for speech, further confirming the rapid growth of this field. In order to have parity, we took the most recent 304 queries from speech. We then removed duplicates and sampled 300 records for further analysis.

Within these 300 records, we then screened the titles and abstracts to assess eligibility as follows. We restricted papers to full-length audio papers using audio data for analysis or training. Of the 143 records we excluded, most were entirely focused on another modality such as text or vision (72\%), conducted literature reviews or meta-analyses (10\%), researched dance or motion generation (10\%), were incomplete---e.g., stopped mid paragraph and was not a finished paper (3\%), or had another salient quality (5\%) such as not being in English or withdrawn from arXiv (which typically only occurs when a co-author did not grant permission for uploading to arXiv or something was substantially incorrect with the paper, withdrawals are not easily granted by request of authors). Of the remaining 157 papers in the corpus, we further excluded eight papers that did not specify their data usage even though they clearly used a dataset for their paper. This resulted in 149 full-text articles for inclusion.

Our final corpus included 66 papers about music, 59 papers about speech, 19 papers about general audio (either environmental non-music, non-speech sounds, or general purpose audio), and five papers about music and speech (typically singing voice synthesis)---these categories are mutually exclusive. The authors then went through these papers and identified any audio datasets used for training or evaluation. If the datasets were simply noted (e.g., in the related literature section), but not explicitly discussed as being part of the training or evaluation process described in the paper, they were not included in this analysis. This process yielded 175 unique audio datasets. See Figure \ref{fig:prisma} for a visualization of this process.

\subsection{Analysis of Current Audio Datasets}\label{sec:analy_current_audio_datasets}

\begin{table*}[t]
\addtolength{\tabcolsep}{-0.17em}
\begin{tabular}{c|c|c|ccc|ccc|c}\toprule
 \multicolumn{10}{c}{\textbf{Overview of Datasets}} \\
 \cmidrule(lr){1-10}
  & \textbf{Overall} &
  \multicolumn{1}{c|}{\textbf{Creation Method}} &
  \multicolumn{3}{c}{\textbf{Size (Hours)}} &
  \multicolumn{3}{|c|}{\textbf{Citations}} &
  \multicolumn{1}{c}{\textbf{Copyright}}\\
 \cmidrule(lr){1-10}
 \textbf{Category} &
 \textbf{Datasets} &
 \textbf{Scraped} &
 \textbf{Sum} & 
 \textbf{Median} &
 \textbf{Mean} &
 \textbf{Sum} & 
 \textbf{Median} &
 \textbf{Mean} &
 \textbf{Infringing}\\
 \cmidrule(lr){1-10}
 Music  & 61 & 36\% & 74,139 & 19 & 1,236  & 14,346 & 98  & 267 & 33\%\\
 Speech & 80 & 24\% & 573,522 & 59 & 7,546 & 39,511 & 202 & 590 & 20\%\\
 Sounds & 31 & 32\% & 37,178 & 64 & 1,377  & 11,998 & 159 & 461 & 35\%\\
 Music+Speech& 3 & 0\% & 44 & 15 & 1  & 30 & 15 & 15 & 0\%\\
 \midrule
\bottomrule
\end{tabular}
\caption{Descriptive statistics from 175 databases identified in review. Split by music, speech and sound, we list the count of datasets, percent that were scraped, size in hours, total number of citations (beyond our corpus), and conservative estimate of the percent likely copyright infringing. All of this information was determined by two authors independently evaluating each dataset by reading the original papers proposing the datasets (when present), investigating all possible information provided online about the datasets, and lacking both of those downloading the dataset and personally assessing this information.\vspace*{-5mm}
}
\label{tab:descriptive_results}
\end{table*}

\begin{figure}
  \begin{center}
    \includegraphics[width=0.48\textwidth]{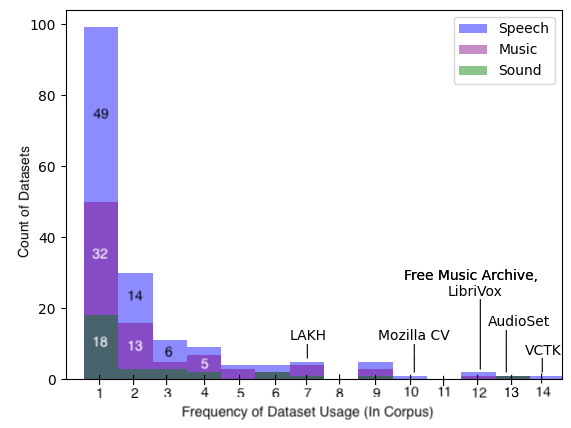}
  \end{center}
  \caption{Stacked bar plot displaying count of times datasets were used by papers in the corpus. Split by Speech, Music, and Non-music/Non-Speech sounds. The vast majority of datasets were only used once, while a select few were used multiple times.}
   \label{fig:lit_count_plot}
\end{figure}

We then analyzed the 175 audio datasets found through our literature review in order to understand practices, uses, and creation methods. For each dataset, we  noted the number of times papers in our corpus used it, the number of the times it had been cited overall (beyond our sample), its size in hours, the categorization of its contents (music/speech/general sound), how its corresponding data was collected, and concerning copyright infringement of the dataset (see Table \ref{tab:descriptive_results}). 

\subsubsection{Dataset Labeling and Categorization} \label{app:dataset_labeling_cats}
\hfill\\
\textit{\textbf{Calculating Dataset Content Duration (Hours)}} In order to calculate the duration of each dataset, we conducted a thorough investigation and made necessary assumptions. When explicitly stated, we listed the exact duration of the dataset. When we were provided with number of files and average file length, we performed the calculation. When we were provided with number of files and access to individual files, we got the average file length based on a small sample and multiplied by number of files in the dataset. If we did not have easy access to the data, then we made the following assumptions unless explicitly instructed to do otherwise. We assumed:
\begin{itemize}
    \item Average song length: 3 minutes
    \item Average childrens' song length: 1.5 minutes
    \item Single sentence utterance: average is 12 words, which using average speech length is 5 seconds in English
    \item Single word utterance: 1 second
    \item Phonetically dense sentences: 15 seconds
    \item Anything from YouTube: we examined the data and noted 10 or 30 seconds
\end{itemize}

\textit{\textbf{Original Content (Yes/No):}} A dataset is considered original content if audio samples were new recordings, synthesized data from an existing source, created new data using a model trained on an existing source (such as generating fake chorales with a neural network like JS Fake Chorales \cite{peracha2021js}), converted an existing source from one modality to another (for example, WSJ0 \cite{garofolo1993csr}, which includes speech recordings made from its sister WSJ text corpus), or added significant data derived from an existing source, such as crowdsourced annotations to tag songs (e.g., MagnaTagATune  \cite{law2009evaluation}, Free Music Archive \cite{defferrard2016fma}). They are not regarded as the original creator if the dataset solely consists of scraped or crawled videos or their links to another source.

\textit{\textbf{Potential for Copyright Infringement (Yes/No):}} We adopt a conservative and stringent approach to assessing whether a dataset has the potential for copyright infringement. For example, any datasets scraping from Youtube are classified as having potential for copyright infringement as YouTube hosts a wide variety of content, much of which is protected by copyright. Scraping and distributing this content without proper authorization violates YouTube's terms of service and copyright laws. For annotation-based datasets that provide links to audio recordings, we assess the original sources from which users can obtain the respective audio recordings. As such, we categorize MTG-Jamendo \cite{bogdanov2019mtg} (including its derivative, Song Describer Dataset \cite{manco2023thesong}, and Free Music Archive \cite{defferrard2016fma} as not infringing upon copyright. This is determined based on the fact that the sources, Jamendo.com and freemusicarchive.org, offer music within the public domain, allowing songs to be freely listened to and downloaded. Conversely, we categorize Million Song Dataset~\cite{Bertin-Mahieux2011} and MagnaTagATune \cite{law2009evaluation} as having potential for copyright infringement as their audio sources, 7digital.com and magnatune.com, necessitate purchasing licenses for access beyond private listening. So while music on these sites may be free to listen to, they are not necessarily free to download and use. 

\textit{\textbf{Method of Dataset Creation (Scraped, Created, or Augmented):}} We take a single-label over a multi-label approach. If any portion of the dataset undergoes scraping, we categorize it as `scraped'. For instance, if a dataset is partially created through scraping a website and partially created by augmented an existing dataset, we denote it as `scraped'. If the dataset is a direct subset of another, we label it as `augmented’. We note online marketplaces as such. For all other datasets, we denote as `created'.

We built a database of meta-data for each of these datasets, available at [redacted for blind review], which contains more granular information such as links to download, original purpose of the dataset, whether it was free to access, whether it was scraped or created, and if applicable, the language or genre of the contents.

\subsubsection{Distribution and Usage of Datasets}

Of the 175 datasets, the vast majority of them were only used in one ($n=99$; 57\%) or two papers ($n=45$; 26\%). The full distribution can be found in Figure \ref{fig:lit_count_plot}. Only a handful of datasets were used more than 5 times.
Speech datasets had the largest skew: most datasets were only used by one paper, while VCTK \cite{yamagishi2019cstr} was used by 14. Speech datasets were also the largest by number of hours (see Appendix \ref{app:dataset_labeling_cats}) We documented 573,522 hours of speech data (median = 59 hours), the vast majority of which came from VoxPopuli \cite{wang2021voxpopuli}, a 400,000 hour dataset consisting of European parliament event recordings, and the Spotify Podcast Dataset~\cite{clifton2020spotify}, 100,000 hours of Spotify Podcasts. Music datasets totaled 74,139 hours, with a similar skew (median of 19 hours) driven by The Million Song Dataset \cite{Bertin-Mahieux2011} (50,000 hours), Irish Massive ABC Notation Dataset \cite{wu2023tunesformer} (7,200 hours), and Free Music Archive \cite{defferrard2016fma} (5,920 hours). 
These findings stand in contrast to text and image modalities, where there exist a smaller number of very large, widely used datasets that have significant source overlaps \cite{schuhmann2022laion, gadre2024datacomp, raffel2020exploring, soldaini2024dolma}, and pose a challenge for auditors, policymakers, or practitioners attempting to assess or shape audio dataset practices.

\begin{figure*}
    \centering
    \includegraphics[width=0.99\textwidth]{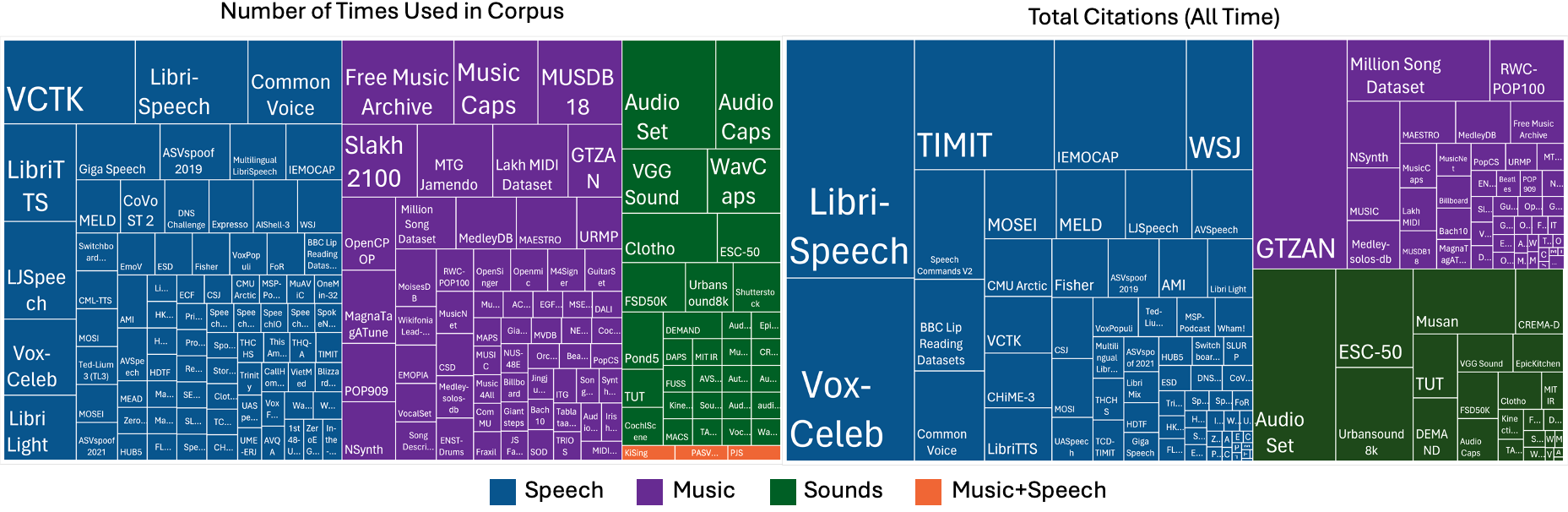}
    \caption{Area charts displaying the proportion of all 175 audited datasets by (1) number of times used in our paper corpus, (2) the cumulative total of citations received to date. Split into 4 categories: Speech, Music, Non-Music/Non-Speech sounds, Music and Speech combined.}
    \Description{Area charts displaying the proportion of all 175 audited datasets by (1) number of times used in our paper corpus, (2) the cumulative total of citations received to date. Split into 4 categories: Speech, Music, Non-Music/Non-Speech sounds, Music and Speech combined.}
    \includegraphics[width=\textwidth]{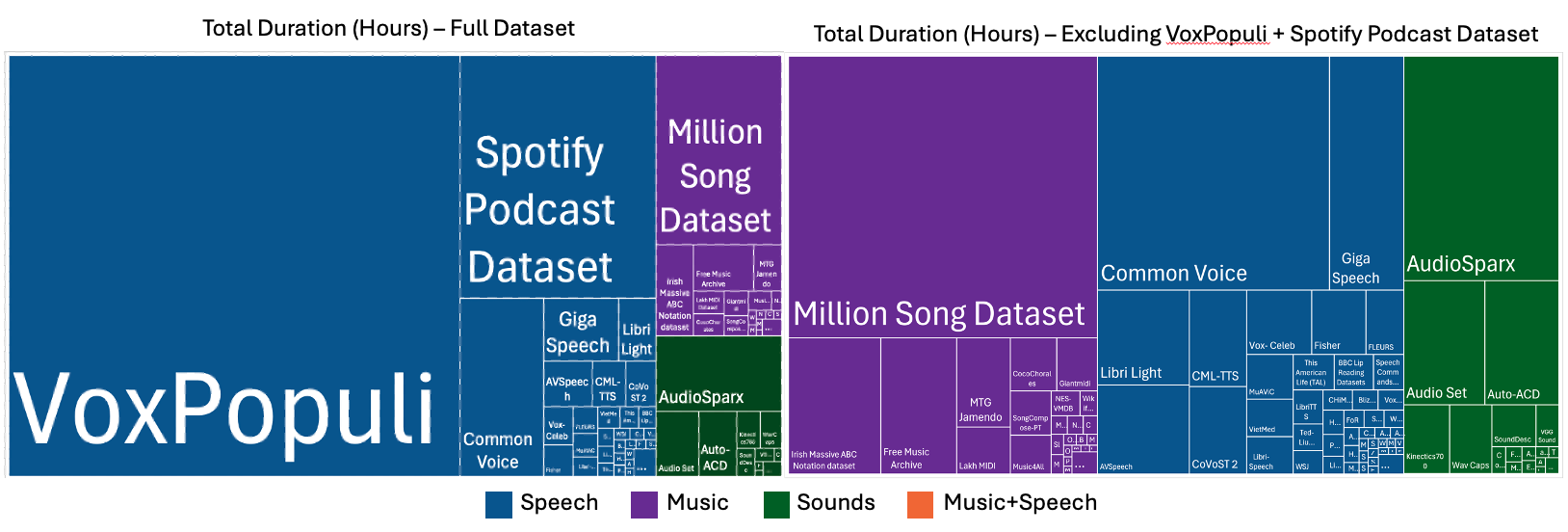}
    \caption{Area charts displaying the proportion of each dataset's estimated total duration relative to (1) all audited datasets and (2) all datasets excluding the two largest datasets, VoxPopuli and Spotify Podcast Dataset (right)}
    \label{fig:area_plots}
\end{figure*}

    

We also calculated the total citations these datasets had received\footnote{Citation data accessed from Google Scholar and may not be exhaustive.} to gauge popularity relative to usage---it is important to note that oftentimes when someone cites a dataset, they are doing so in acknowledgement of the field (e.g., in the related literature section) as opposed to actually using that data for training or evaluation. As seen in Figure \ref{fig:area_plots}, both usage and citations are quite fragmented, but actual citations have a heavier focus on a few important datasets. Similar to usage and hours, speech dominated the total citation count receiving 39,511 cumulative citations (median$=202$), with LibriSpeech \cite{panayotov2015librispeech} in the lead with 6,136 citations. Music was second with 14,346 cumulative citations (median$=98$); GTZAN was the most popularly cited music dataset with 4,345 citations. Datasets including sounds were least cited with 11,998 cumulative citations (median$=159$), with AudioSet being most popularly cited at 3,204 citations.

It was also noteworthy that many papers in our corpus did not release the data they used. Out of the 157 papers in our sample, 65 papers used at least one proprietary dataset, and there were 77 proprietary datasets in total. Of these 77 datasets, 79\% ($n=61$) were not released. Some likely reasons to not release internal, propietary datasets include protection of trade secrets or avoiding the risk of being outed for accidental or intentional copyright infringement, though these papers did not tend to specify why they did not release their data. We noted when we believed there was a high likelihood of the datasets in our corpus containing copyrighted content, such as when datasets scraped YouTube or TED Talks. We annotated 27\% of all datasets as having a high likelihood of comprising copyright infringing content---20\% of speech datasets, 33\% of music datasets, and 35\% of non-music non-speech sound datasets. We also annotated how datasets were created, including scraping (30\%), augmenting or taking a subset of an existing dataset (17\%), creating from scratch (50\%), or sourcing from an online marketplace of audio (2\%).


\subsubsection{Language Contents of Datasets}

When considering the linguistic diversity in speech datasets, the inclusion of underrepresented languages is frequently not prioritized, with many datasets predominantly featuring only one language. Out of the 77 speech datasets we examined, the majority---61 of them---were monolingual, with 50 solely in English, followed by 7 in Mandarin. In contrast, 16 datasets encompassed between 2-30 languages, while only two datasets included more than 50 languages. 

\subsubsection{Sources of Datasets}

The two most salient audio data sources were YouTube ($n=25$ datasets) and LibriVox ($n=13$). Other standouts were freesound.org ($n=6$), Spotify ($n=4$), and VCTK \cite{yamagishi2019cstr} ($n=3$). Other popular sources for audio content included podcasts ($n=6$), marketplace websites ($n=4$), TED talks ($n=4$), TV shows ($n=3$),  and parliament/public speeches ($n=3$). We found that LibriVox and its derivatives were referenced in 35 out of the 59 speech papers included in our literature review. LibriSpeech \cite{panayotov2015librispeech}, a dataset comprised of public domain audiobooks read by volunteers across various languages, serves as a foundation for 13 derivative datasets, including both direct derivatives like the LibriSpeech dataset\cite{panayotov2015librispeech} and Musan \cite{musan2015}, as well as derivatives of derivatives, such as LibriLight \cite{librilight} and LibriMix \cite{cosentino2020librimix}.

We find that 13 datasets stem from LibriVox, categorized into 5 direct derivatives (LibriSpeech \cite{panayotov2015librispeech}, LJSpeech \cite{ljspeech17}, Multilingual LibriSpeech \cite{Pratap2020MLSAL}, Musan \cite{musan2015}, CML-TTS \cite{Cmltts2023}) and 6 second-order derivatives, (LibriLight \cite{librilight}, DNS Challenge Dataset \cite{dubey2023icassp}, LibriTTS \cite{zen2019libritts}, HiFi TTS \cite{bakhturina2021hi}, LibriMix \cite{cosentino2020librimix}, RealEdit \cite{peng2024voicecraft}). 

The most cited derivative datasets include LibriSpeech ($n=12$) , LJSpeech ($n=9$), LibriTTS \cite{zen2019libritts} (2nd order) ($n=9$), and LibriLight with ($n=6$). This trend aligns with the overall popularity of these datasets, gauged by citation numbers, with the exception of Musan \cite{musan2015}, cited only once in our papers, yet potentially ranking as the 2nd most popular among the 13 derivatives. It's crucial to emphasize that LibriVox predominantly comprises century-old texts, encompassing outdated and potentially problematic language, cultural perspectives, and social norms. This finding corroborates our outcomes discussed later. Consequently, researchers utilizing LibriVox should be mindful of the potential introduction of bias and toxicity inherent in this dataset.

\subsubsection{Takeaways from Literature Review}

The composition of audio datasets used by the research and commercial audio community is vastly different to that of text and vision. It is extremely fragmented---beyond a few notable datasets (e.g., VCTK \cite{Veaux2017CSTRVC} and AudioSet \cite{audioset}), researchers are prone to using a few one-off datasets that suit their needs. These datasets also come with their own suite of problems; much of the contents in these datasets are likely stolen from creators without their knowledge, potentially infringing on someone's copyright, or using content at scales beyond original understanding of data providers. There is no standardized way to prepare, create, release, or even discuss audio datasets used in research. Much of the work done in this literature review was scrappy scaffolding to understand the composition of datasets---future datasets should prepared and released in a much more mindful and documented manner which we will suggest below.

\section{Audit of Audio Datasets}

In this section we audit seven representative datasets, assessing contents, biases, toxicity, sources, and copyright status to address RQ2, RQ3, and RQ4:
\begin{enumerate}
    \item \textit{AudioSet} \cite{audioset}, a dataset of 2 million 10-second YouTube clips, comprising a range of music, speech, and sounds;
    \item \textit{Mozilla Common Voice 17} \cite{ardila2019common}, a corpus of crowd-sourced sentences read by volunteers;
    \item \textit{VCTK} \cite{Veaux2017CSTRVC}, a dataset of sentences read from the Herald, a Scottish newspaper, and several shorter accent elicitation texts;
    \item \textit{LibriVox} \cite{librivox}, a dataset of volunteer recordings of public domain books; 
    \item \textit{Free Music Archive} \cite{defferrard2016fma}, a music-specific dataset scraped from an online repository;
    \item \textit{Jamendo} \cite{bogdanov2019mtg}, another scraped music-specific dataset, and lastly;
    \item the \textit{Lakh MIDI Dataset} \cite{raffel2016learning}, a subset of the Million Song Dataset \cite{Bertin-Mahieux2011}, which is a currently unavailable dataset of music taken by The Echo Nest, a now defunct music analytics company.\footnote{The Lakh dataset consists of MIDI files of songs from from the Million Song dataset. These MIDI files contain transcribed lyrics, but do not include audio of those lyrics themselves. We choose to include these lyrics in our audit as they are a proxy for the Million Song Dataset, which is no longer available, but still in use by recent papers.} 
\end{enumerate}
These datasets represent some of the largest and most popular datasets in our survey, and they cover music, sound, and speech subdomains. 

\paragraph{Transcription and Textual Analysis}

We first obtain high-quality transcripts. While Common Voice, VCTK, LibriVox, and the Lakh MIDI datasets contain transcripts, we found the transcripts included with AudioSet Youtube videos were of lower quality and not well-aligned. Therefore, we use Whisper-large \cite{radford2023robust} to transcribe over 50,000 randomly selected AudioSet YouTube clips. We use Whisper-large with prompting improvements for music transcription from \citet{zhuo2023lyricwhiz} to transcribe Jamendo and Free Music Archive. After converting to text, we use prior text dataset audit techniques and tools. For toxicity analysis, we use the \texttt{pysentimento} library used to detect hate content \cite{birhane2024into}, as well as Surge AI's profanity list \citep{profanity}. To detect language, we use the \texttt{langdetect} library \cite{langdetect} along with Whisper language predictions on AudioSet, Jamendo, and Free Music Archive clips. To investigate the dataset content in relation to sociodemographic identities, we search for a set of keywords encompassing race, gender, religion, and sexual orientation, using lists established from prior work \citep{dodge2021documenting, hong2024s, lu2020gender}. Finally, we track common words and calculate pointwise mutual information between common words and identity keywords in order to determine stereotypical associations \citep{rudinger2017social}.

We recognize the various limitations of these audit libraries and techniques, especially as hate speech models and profanity detection are more likely to wrongly flag data relating to LGBTQ+ or Black voices as toxic content \citep{dodge2021documenting, sap2019risk}. In addition, language predictions are unreliable and rely on text content \citep{lucy2024aboutme}. Despite these challenges, as our goal in this work was to present findings from a broad audit of these datasets to identify concerns that may warrant further, more in-depth investigation in future work, we argue that our findings are still useful indicators of broad trends of bias or toxicity.

\subsection{Overview of Audio Datasets}

\begin{figure}[ht]
  \begin{center}
\includegraphics[width=0.32\textwidth]{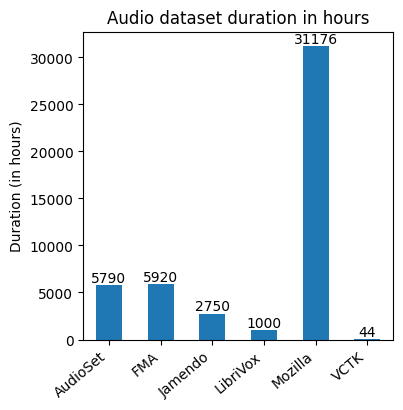}
  \end{center}
  \caption{Estimated total duration of audited datasets.}
    \label{fig:audit_duration}
\end{figure}

\paragraph{Size} AI in text, image, and video modalities has increasingly come to rely on extremely large datasets, often scrapped from the web, rather than smaller, purpose-made datasets. We assess the size of different audio datasets to understand which datasets and data sources may meet a need for larger audio datasets. In Figures \ref{fig:area_plots} and \ref{fig:audit_duration}, we present the total size of each dataset in hours of content. Despite the current popularity of each dataset, we find a wide variation in length: the Mozilla Common Voice dataset is nearly two orders of magnitude larger than VCTK dataset, despite both being speech datasets. In particular, we find that several of the largest audio datasets are no longer publicly available, including the Spotify Podcast Dataset and the Million Song Dataset. Of the remaining largest audio datasets, several contain highly specific contents, such as recordings of the proceedings of the European Parliment (VoxPopuli \cite{wang2021voxpopuli}), or traditional Irish music (Irish Massive ABC Notation Dataset \cite{wu2023tunesformer}).

\paragraph{Creation Dates} Text-based audio datasets have two relevant creation dates: audio creation date and text creation date. VCTK was recorded in 2013 and features newspaper articles up to 2013 \cite{Veaux2017CSTRVC}. LibriVox recordings were made between 2005 and present \cite{librivox}. However, the recordings leverage public domain texts, which are often over a century old because texts typically only enter the public domain 70 years after the death of their last living author \cite{public_domain}. The Lakh MIDI dataset is derived from the Million Song Dataset, composed of songs released before 2012 \cite{raffel2016learning}. Similarly, AudioSet is composed of YouTube videos released before 2016 \cite{audioset}. Of the audited datasets, only Mozilla Common Voice features both contemporary audio recordings and text, with creation dates between 2017 and present. Training cutoff date, or the date of the most recent training data, has emerged as a crucial concern of large language models, determining which events they may have knowledge of\cite{cheng2024dated}. The relative age of most audio datasets could lead to downstream models will perform worse on new, or newly popular, words, sounds, and music genres, reflecting a bias towards the past\cite{birhane2022values}.

     

\paragraph{Identity Representation}

\begin{table*}
\footnotesize
\begin{center}
    \centering
    \begin{tabular}{|l|l|l|l|l|l|l|l|l|}
    \hline
        \textbf{Dataset} & \textbf{Age} & \textbf{Gender} & \textbf{Sexual Orientation} & \textbf{Language} & \textbf{Locale/Country} & \textbf{Accent} & \textbf{Race/Ethnicity} & \textbf{Disability}\\ \hline
        Mozilla Common Voice & yes & yes & no & yes & yes & no & no & no\\ \hline
        VCTK & yes & yes (binary) & no & yes & yes & yes & no & no\\ \hline
        LibriVox & no & yes (binary) & no & yes & no & yes & no & no\\ \hline
        Lakh & no & no & no & no & no & no & no & no\\ \hline
        AudioSet & no & no & no & yes & yes & no & no & no\\ \hline
        Free Music Archive & no & no & no & yes & yes & no & no & no\\ \hline
        Jamendo & no & no & no & no & no & no & no & no\\ \hline
    \end{tabular}
\end{center}
\caption{Documentation of demographics in audited datasets.}
\label{fig:dem_doc}
\end{table*}

\begin{figure*}[t]
    \centering
    \includegraphics[width=\textwidth,trim={0 0 0 20}]{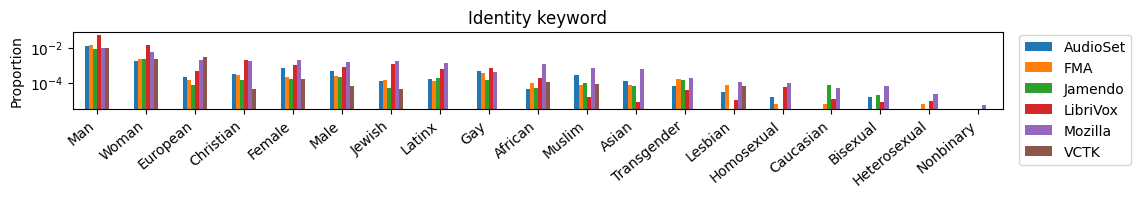}
    \caption{Proportion of identity keyword mentions for each dataset. Y-axis is in log-scale.}
    \label{fig:datasets_identity}
\end{figure*}

Who is represented in the datasets (RQ4) is a central question of dataset audits with significant downstream impacts on bias. In Table \ref{fig:dem_doc} we summarize available information about audio creator demographics. We find that datasets often have limited demographic information, and several important demographic categories, including sexual orientation, disability, race, and ethnicity, are not documented in any of the audited datasets. To assess representation and bias of different demographics, we instead study identity keywords in audio transcripts.
In Figure \ref{fig:datasets_identity}, we present counts of identity keywords, using keywords from \citet{dodge2021documenting}, for each dataset's transcripts, with plurals and common alternative spellings merged. With the exception of Mozilla Common Voice, we find ``man'' is used 2-100x more often than ``woman'' in these datasets. We also find that many identity keywords have a very low count in many datasets, with ``Muslim'' only appearing a few thousand times, 5-10x less than ``Christian'' and ``Nonbinary'' appearing only approximately ten times. Mozilla Common Voice is the only dataset that contains at least one instance of every identity keyword studied. While this list of identity words is not comprehensive, and some words have alternate non-identity meanings, our results evidence low representation of marginalized groups in audio datasets.

\begin{figure}
  \begin{center}
\includegraphics[width=0.45\textwidth]{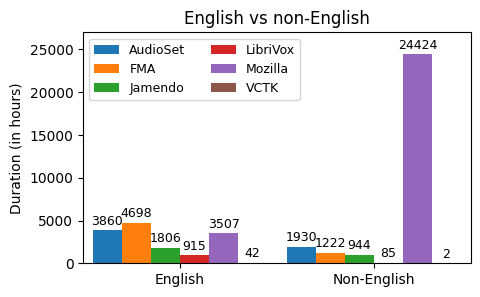}
  \end{center}
  \caption{Estimated duration of each dataset by hours in English and not English.}
    \label{fig:datasets_language_en}
\end{figure}

\begin{figure*}
  \begin{center}
\includegraphics[width=1.0\textwidth]{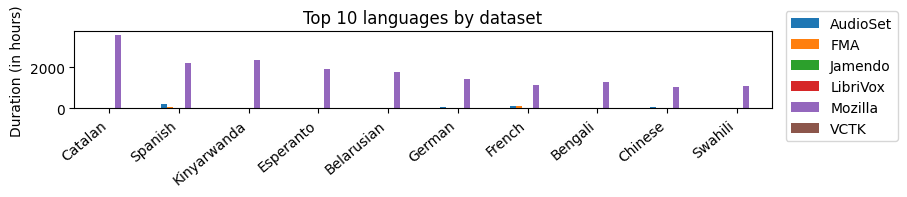}
  \end{center}
  \caption{Estimated amount of each dataset by language outside of English.}
    \label{fig:datasets_language}
    \vspace{-15pt}
\end{figure*}

\paragraph{Language Representation}
In Figure~\ref{fig:datasets_language_en}, we show the breakdown of English vs. non-English audio, and in Figure~\ref{fig:datasets_language}, we show the approximate number of hours of audio in each dataset broken down by language for non-English languages. We find that most audio datasets contain between 2x and 10x more data in English than non-English, with the exception of Mozilla Common Voice, which contains 24,424 hours of non-English data and 3507 hours of English data.

\begin{figure}
    \centering
    \includegraphics[width=0.5\textwidth]{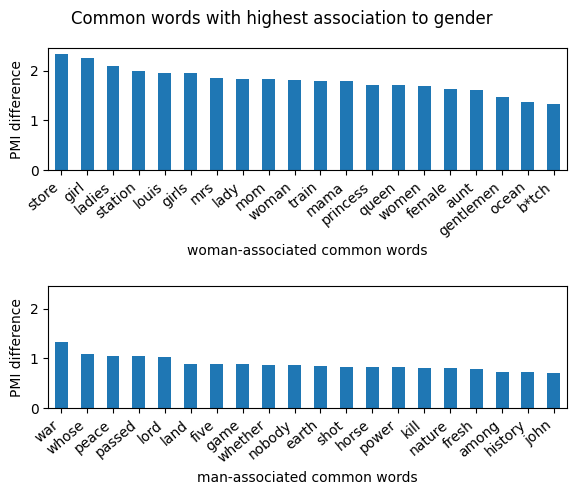}
    \caption{Common words with highest PMI to man-related versus woman-related keywords (from \citet{lu2020gender}) across all datasets (randomly sampling 40 thousand sentences each). We consider words that appear at least 100 times in man-related and woman-related samples.\label{fig:datasets_gender_pmi}}
    
\end{figure}

We assess binary gender bias by comparing words with the highest pointwise mutual information (PMI) difference between ``woman'' and ``man'' keywords across all datasets (Figure \ref{fig:datasets_gender_pmi}). PMI compares the probability of two words appearing independently with the probability of their appearing together in a sentence. A higher PMI indicates that words are more strongly associated with each other. We find ``woman''-related words are more associated with terms about families and childcare than ``man''-related words, while ``man''-related words are not correlated with typically gendered terms. In addition, we find that ``woman''-related words have stronger associations with ``store'' and ``b*tch,'' while ``man''-related words have stronger associations with ``war'' and ``kill.'' Overall, these associations provide evidence that women are commonly depicted in relation to families, childcare, and as subjects of the male gaze \cite{bloom2017reading, mulvey2013visual}. Nonbinary genders were not represented well enough in these datasets to assess bias. However, given the prevalence of biased and toxic content towards queer people online \cite{queerinai2023queer}, it is likely larger audio datasets will contain biased and toxic content towards these groups.

\begin{figure}
  \centering
  \begin{minipage}[b]{0.4\textwidth}
    \includegraphics[width=0.8\textwidth]{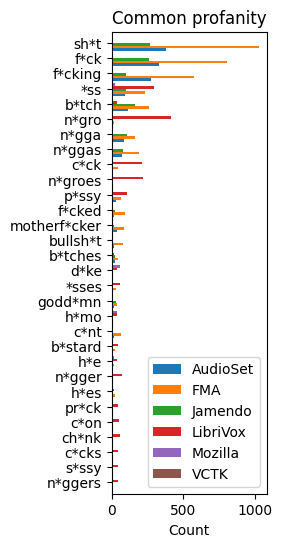}
   \caption{
   English profanity occurrences in all datasets.}
      \label{fig:datasets_profanity}
  \end{minipage}
  \hfill
  \begin{minipage}[b]{0.5\textwidth}

  \includegraphics[width=0.8\textwidth]{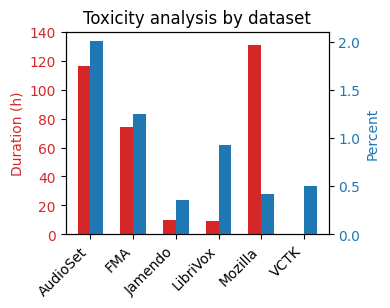}
%
   \caption{Estimated duration and percent of toxic-predicted English sentences split by dataset.}
     \label{fig:datasets_toxicity}
  \end{minipage}
\end{figure}


\subsection{Toxicity}

In this section we assess toxicity in selected audio datasets (RQ3). In Figure \ref{fig:datasets_profanity}, we show the most common profane words in the datasets. We note that profanity is not the same as toxicity, and in some contexts these words are neither profane or toxic. We find that, while all datasets contain at least some profane words, FMA and LibriVox contain by far the most, with many thousands of occurrences of racist and queerphobic terms. This finding may be due to LibriVox's sourcing from public domain texts, which are at a minimum 70 years old and generally much older, and represent times where where toxic dialogues about marginalized populations were more overt.

In Figure \ref{fig:datasets_toxicity}, we present the number of hours of content in each dataset classified as toxic by the \texttt{pysentimento} toxicity classifier. This classifier considers not just profanity but additional textual cues to assess whether a text is toxic. Examples of and further discussion of sentences classified as toxic are availible in Appendix \ref{app:audioset}. While we find that each dataset has only approximately 1\% of content flagged as toxic, this still amounts to hundreds of hours of toxic content. While levels of toxicity are low relative to the size of each dataset, large language models have displayed an ability to recall text encountered only a few times during training \cite{carlini2022quantifying}, raising the possibility that large audio models will exhibit similar behavior with even small amounts of profane or toxic content.

\subsection{Audio Datasets Licensing}

\begin{table}
\begin{center}
\begin{tabular}{|c | c |} 
 \hline
 \textbf{Dataset} & \textbf{License} \\ 
 \hline
 Mozilla Common Voice & CC0 \cite{cc0}\\
 \hline
 VCTK & Open Data Commons Attribution License \cite{odc} \\
 \hline
 LibriVox & CC BY 4.0 \cite{cc4} \\
 \hline
 Lakh & Echo Nest License \cite{echno_nest_license}\\
 \hline
 AudioSet & Youtube License \cite{yt_license} or Creative Common Licenses \cite{yt_cc}\\
 \hline
 Free Music Archive & Various Creative Commons\\
 \hline
 Jamendo & Various Creative Commons \\
 \hline
\end{tabular}
\end{center}
\caption{Licenses of audio datasets}
\label{fig:licenses}
\end{table}

\begin{figure}
    \centering
    \includegraphics[width=\textwidth]{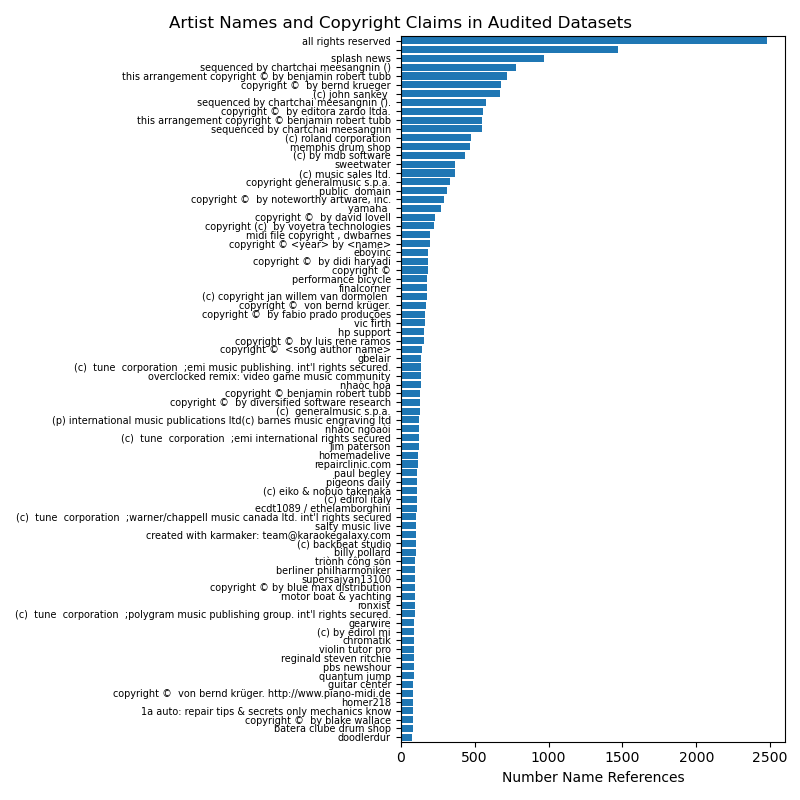}
    \caption{Count of appearances of artist names and copyright claims in the AudioSet and Lakh datasets.}
    \label{fig:artist_names}
\end{figure}

In Table \ref{fig:licenses}, we summarize licenses of audio datasets (RQ2). Mozilla Common Voice, VCTK, and LibriVox have permissive licenses that allow any use with minimal restrictions. The other four datasets all have licenses that potentially impact the ability of these datasets to be used for training or commercial applications. Lakh is derived from the Million Song Dataset, itself derived from Echo Nest, a music data service, which is subject to the Echo Nest License \cite{echno_nest_license}, which prohibits commercial use. Lakh also contains many copyright tags, and we present the most frequent tags in Figure \ref{fig:artist_names}. AudioSet is derived from YouTube videos, which are licensed either under Creative Common Licenses \cite{yt_cc} with different levels of permissiveness, or the YouTube License \cite{yt_license}, where the creator retains all ownership. However, YouTube can use or modify videos in connection with YouTube's business, and users of YouTube can use or modify videos only as a feature of YouTube. We present the most common YouTube channel names in Figure \ref{fig:artist_names} in our appendix. Free Music Archive and Jamendo data are covered under several Creative Commons and other licenses, including many that prohibit commercial use and derivatives, and that require artist attribution, derivative work to have the same license, and restrict use to personal use only. In short, our analysis uncovers extensive presence of copyrighted material in these datasets from a broad range of artists.

\section{Discussion}

In this paper we found 175 audio datasets that are in recent use. The distribution of their usage is long-tailed, with a small number used frequently, and a majority only used once or twice (RQ1). Many audio datasets were created in lab for research purposes, but approximately 1/3 were scrapped from online sources. These scrapped datasets are under a variety of licenses, with many audio data licensed under versions of creative commons licenses that prevent commercial use, remixing, or require attribution, complicating their usage for AI training (RQ2). The largest and most frequently used datasets have diverse contents, including public domain books, royalty-free music, Youtube videos, and readings from The Glasglow Herald. This reflect broader trends we found of audio datasets not reflecting the distribution of human speech or music, but rather focusing on specific genres or topics. Finally, we found these datasets contain many instances of profanity and toxicity (RQ3). Minimal documentation and metadata made assessing who is in audio datasets challenging, pointing to a need for better documentation practices. We found that marginalized communities were less likely to be explicitly mentioned in audio datasets, and we extracted creator and artists names from several audio datasets (RQ4).

\subsection{Impacts of Bias and Toxicity in Audio Datasets on Downstream Models}


Our analysis indicates that the datasets we analyze in depth are biased both statistically (i.e., skewed inclusion of certain types of data) and socially (i.e., certain keywords and terms correlated with identity), in addition to containing non-trivial amounts of profane and potentially toxic content. Given that models in other modalities have displayed an ability to recall data encountered only a few times during training\cite{carlini2022quantifying}, this raises the possibility that large audio models will exhibit similar behavior with even small amounts of profane or toxic content. These issues can also lead to disparate performance across socially salient categories. For instance, we observe heavy emphasis on the English language across datasets. 
Downstream of these datasets, \citet{radford2023robust} note that while their Whisper transcription model obtains state-of-the-art performance or close to it on several tasks, they are limited in that ``Whisper's speech recognition performance is still quite poor on many languages'' due to their ``pre-training dataset [being] currently very English-heavy due to biases of [their] data collection pipeline.'' Disparate performance of transcription via audio models across languages, and by extension cultures, is thus already documented as an impact of audio dataset biases\cite{fuckner2023uncovering, nacimiento2024gender}.

Beyond transcription, issues with other tasks like translation, voice recognition, and audio generation may also be on the horizon (if not currently extant but largely unknown or undocumented). For example, we discover through our PMI, identity keyword, and profanity analyses that there are gender associations with certain words, virtual omission of keywords pertaining to sexual orientation, and a high incidence of profanity related to sexual anatomy and racial slurs. If not properly addressed, these features of audio datasets may yield downstream audio generation models that perpetuate or even accelerate instances of social bias and toxicity that echo harmful stereotypes found in the wild (e.g., models that describe or portray women in a stereotyped way, models that are unable to create outputs relating to the LGBTQ experience, models that degenerate into profanity based on sexual anatomy or racial slurs without provocation, etc.).

This said, recent efforts have been made to improve the diversity of speech datasets, such as the Common Voice corpus \cite{ardila2019common}, the Corpus of Regional African American Language \cite{kendall2023corpus}, and the Mid-Atlantic Gender Expansive Speech Corpus \cite{hopenonbinary}. However, ethical dataset creation and use goes far beyond mere inclusion, and inclusion itself, especially in the context of AI and big tech, can often be predatory, harming included communities by exposing them to data- and AI-intensified surveillance, poorly performing AI, and loss of control over data and culture \cite{papareo}. Auditing datasets to uncover biases and a lack of representation is the start of the conversation, but effective solutions must be lead by the communities who own the data \cite{papareo}.


\subsection{Scale of Audio Datasets}
Audio datasets require considerably more storage per sentence than text datasets. AudioSet totals nearly 2.5TB in size and contains approximately 52 million spoken words. In contrast, C4~\cite{raffel2020exploring}, a text dataset approximately 30\% of the (file) size of AudioSet, contains 153 billion words~\cite{soldaini2024dolma}. Including the same breadth and depth of content in audio datasets as text datasets will require significantly more storage and compute which can both exacerbate known harms for large generative models, and create new ones.

For example, the requirement for more storage and compute limits which actors can train on such datasets. While the datasets of corporations training the largest audio models remain closed to external auditors, we know OpenAI Whisper \cite{radford2023robust} was trained on 680,000 hours of audio, and the BigSSL speech recognition model \cite{zhang2022bigssl} from Google was trained on one million hours of YouTube audio. Both of these datasets are over an order of magnitude larger than the largest freely accessible dataset in our survey. Thus, generative audio models may impose significantly higher barriers for participatory approaches to AI development than generative text models.

Moreover, while the high carbon and water impacts \cite{ftwater} of LLM training \cite{luccioni2023estimating} and inference \cite{everman2023evaluating,luccioni2023power} are well known, the scale of audio datasets raises concerns that audio and multimodal models will have even higher pollution and resource costs compared to purely text-based models~\cite{douwes2021energy}.

\subsection{Source of Audio Datasets}

The existence of proprietary datasets vastly larger than open datasets raises questions about the sources of these datasets. Google has explicitly sourced massive audio datasets from YouTube~\cite{zhang2022bigssl}, and Spotify released (and then removed) a 100,000 hour transcribed dataset from hosted podcasts~\cite{clifton2020spotify}. While OpenAI has not indicated how the Whisper dataset was sourced, nor are they publicly known to host massive audio datasets, the few indications of the source of massive audio datasets we have point to existing commercial hosting and streaming corporations\cite{verge_openai_yt}.

\subsection{Corporations Now Perceive Audio Data as Valuable}

Corporations once freely released massive audio datasets. Google released AudioSet, over 5,000 hours of YouTube audio, in 2017. However, AudioSet was released as links to YouTube videos that could be independently downloaded to obtain audio. When we tried to download this dataset in May 2024, we found YouTube had rate limiters and crawler IP blocking that would make downloading this dataset take several months. Similarly, Spotify released 100,000 hours of transcribed podcast audio in 2020, but then removed this dataset in December 2023, citing ``shifting priorities''~\cite{spotify_ai_removal}. Most explicitly, Universal removed its music from Tiktok in January 2024, partially citing concerns over AI generated music and AI covers of its songs~\cite{umgtiktok}. In our analysis of current audio datasets (Section \ref{sec:analy_current_audio_datasets}), there were 77 proprietary datasets utilized, of which 79\% ($n=61$) were not released. We argue these events constitute a pattern of increasing restrictiveness to proprietary audio datasets, mirroring recent trends in text data sources\cite{longpre2024consent}. These restrictions seem in part motivated by the new value of massive audio datasets to training GenAI to create or modify audio. As \citet{ojewale2024towards} have noted, restrictions on dataset access is a major barrier auditors face, making the trend towards closed-source and proprietary datasets especially concerning.

\subsection{Impacts of IP Theft on Audio Workers}

In recent years, musicians have turned to platforms like YouTube and Spotify to distribute their work, reach wider audiences, and generate revenue. These platforms provide exposure and monetization opportunities, allowing musicians to build their careers and fan bases. For example, artists may rely on music streaming  data to identify where to play shows and which songs to perform \cite{muhlbach2020behind}. However, this dependence now comes with the potential risk of these platforms using the artists' work to train proprietary GenAI models to produce content that mimics an artist’s unique style, which could potentially threaten their competitive edge. Copyright laws provide some protection against the unauthorized use of artists' works, though the rapid evolution of AI technologies is outpacing existing legal frameworks. However, audio workers are not without recourse. For example, The International Confederation of Music Publishers (ICMP), which represents 90 percent of the world's published music, launched a resource called RightsAndAI.com 
which allows creatives to reserve their rights against exploitation by generative AI \cite{MBW2023}. This is one of a growing number of initiatives to protect workers' creative outputs against generative AI copyright infringement. Recently proposed US Senate legislation seeks to protect the public from generative AI. If passed, the bipartisan 2024 No AI Fakes Act would serve to protect individuals' voices and likenesses from AI deepfakes \cite{no_fakes}.


\subsection{Multimodal Models}


Audio datasets are almost always multimodal, containing sounds in addition to the text created by those sounds. Even datasets primarily focused on general sounds have text-based descriptors and keywords that can be leveraged by a multimodal model. Unfortunately, this linkage between text and audio can doubly harm audio AI tuned to both modalities---each modality will share and amplify the risks and harms of the other such as those already noted by \citet{weidinger2021ethical}.

There is already precedent for this concern. Prior work with multimodal models (within text and images) has illustrated that the usage of several components to handle separate modalities increases models' brittleness and attack surfaces relative to those of unimodal models (e.g., research into attacks via CLIP embedding analysis \cite{tong2024mass}, counterfactual questions paired with images \cite{tu2023many}, images with harmful text instructions \cite{gong2023figstep}, and adversarial images with imperceptible perturbations \cite{qi2024visual}). Similar component coupling has already been employed to build text-to-audio models like MusicLM \cite{Agostinelli2023musiclm} and Riffusion \cite{forsgren2022riffusion}, and similar corresponding issues 
have been discovered in such models 
\cite{Newton-Rex_2024_Suno,Newton-Rex_2024_Udio}. From a modeling perspective, we echo previous recommendations of researchers with regard to multimodal training and development (e.g., better training to solidify correspondences between modalities \cite{tu2023many} and alleviation of \emph{algorithmic monoculture}, the lack of diversity of models and data \cite{bommasani2022picking, kleinberg2021algorithmic, tong2024mass}), but from a dataset perspective, we suggest that data in each modality should be thoroughly checked for harmful material and filtered if necessary (e.g., text descriptors alluding to stereotypes, speech with slurs, or music with offensive lyrics). 




\section{Recommendations}

In this paper, we have surveyed current audio datasets, documenting their sourcing, size, and usage statistics. We conducted deeper audits of seven representative music, voice, and sound datasets, studying who and what is in these datasets, including language, bias, and toxicity. Our audit found that non-English languages are underrepresented in most datasets, exhibit similar levels of bias and toxicity to text datasets, and contain little documentation about who is represented in the data. In addition, many audio datasets are scraped from sources with licenses prohibiting remixing, commercial use, or requiring attribution, raising questions of consent and downstream model compliance. To ensure harms that have plagued other modalities, including bias, toxicity, and intellectual property concerns, are not also present in future audio AI, we recommend that 1) audio dataset developers adopt improved documentation to enable better assessment of bias and representation and 2) audio dataset developers only use data that permits remixing and commercial use at a minimum, but ideally seek active and informed consent for usage in AI. Both of these aims may be achieved by continuing existing practices of creating datasets in-lab specifically for AI.

\subsection{Datasheets for Audio Datasets}

To enable effective audio AI dataset documentation, we adapt and extend Gebru et al.'s ``datasheets for datasets'' \citet{gebru2021datasheets} to the context of audio to guide ethical audio dataset development, documentation, and use. In particular, we add several questions that specifically speak to contents of and representation in audio data, in addition to questions assessing data provenance and consent. This datasheet is intended to serve both as a reflexive practice for audio dataset creators, documentation standards for audio dataset publishers, and a guide for future audio dataset auditors. We provide the full datasheet in Appendex A.1.

\subsection{Educating and Mobilizing Data Workers}

Documenting existing datasets its not sufficient for improving dataset practices. As Birhane et al. \citet{birhane2024ai} note in their review of audits, ``Most of the academic work we reviewed focused on the process of evaluating AI systems for bias, fairness, or disparate impacts. Conversely, these studies rarely focused on other stages of auditing crucial to accountability in non-academic work, such as discovering harms, communicating audit results, or organizing non-technical interventions and collective action.'' Artists, creators, and other data workers learning about inclusion of their data in AI datasets has been a key catalyst for these communities discussing their wants and needs in regards to inclusion in AI datasets and subsequently organizing and advocating \cite{paris}. To support this collective action, we created \url{https://audio-audit.vercel.app/}, a website that enables anyone to search for their inclusion in prominent audio datasets. Modelled after \url{https://haveibeentrained.com/}, a popular website for assessing inclusion of images in AI datasets, this website enables users to understand how their work is being used in audio AI without needing to download and process these often very large datasets.

\section{Conclusion and Future Work}

In this paper, we conducted a large-scale survey of audio datasets that are used in generative audio models: an audit of broadly inclusive of speech, music, and sound datasets. We found that these datasets exhibit similar patterns and rates of bias and toxicity as text and image datasets, raising the concern that audio models could exhibit similar levels of bias and toxicity as LLMs and image models if these risks are not mitigated.
We found that hundreds of audio datasets are in use, with several open datasets being significantly larger and more widely used than others.
However, we found indications that datasets have recently started becoming more closed and commercial, with past sources of massive datasets, including YouTube and Spotify, taking down datasets or implementing new measures to block crawling of their audio repositories.
The widespread presence of copyrighted material in many of these massive audio data sources, frequent use of proprietary audio datasets in research by corporations, and new commercial perceptions of them raise serious concerns for musicians, voice actors, and other audio workers hosting content on commercial platforms like YouTube or Spotify.

Our audit is just the start of addressing the dataset documentation debt \cite{bender2021dangers} in audio AI. 
We were unable to assess the demographics of the people in audio datasets, and each of music, speech, and sound require specific and unique audits and analyses. Finally, we hope future audits will provide a deeper understanding of acoustic qualities of audio datasets.

We hope dataset developers will take steps to diversify their audio datasets and underlying text content, mitigate biases and toxicity, and remove copyrighted and other non-consensually sourced material from datasets. We hope to see analyses of the content of specific audio datasets in addition to more contextual investigations of specific types of bias and toxicity in these datasets.

\section{Acknowledgements}
We would like to thank Chris Donahue for detailed feedback. We deeply appreciate feedback and insightful discussions from Tim Friedlander , Carin Gilfry, and Matthew Parham from the National Associations of Voice Actors. This work was supported, in part, by NSF CAREER award \#2316287.

\bibliographystyle{ACM-Reference-Format}
\bibliography{main} 


\begin{thebibliography}{174}


\ifx \showCODEN    \undefined \def \showCODEN     #1{\unskip}     \fi
\ifx \showDOI      \undefined \def \showDOI       #1{#1}\fi
\ifx \showISBNx    \undefined \def \showISBNx     #1{\unskip}     \fi
\ifx \showISBNxiii \undefined \def \showISBNxiii  #1{\unskip}     \fi
\ifx \showISSN     \undefined \def \showISSN      #1{\unskip}     \fi
\ifx \showLCCN     \undefined \def \showLCCN      #1{\unskip}     \fi
\ifx \shownote     \undefined \def \shownote      #1{#1}          \fi
\ifx \showarticletitle \undefined \def \showarticletitle #1{#1}   \fi
\ifx \showURL      \undefined \def \showURL       {\relax}        \fi
\providecommand\bibfield[2]{#2}
\providecommand\bibinfo[2]{#2}
\providecommand\natexlab[1]{#1}
\providecommand\showeprint[2][]{arXiv:#2}

\bibitem[no_(2024)]%
        {no_fakes}
 \bibinfo{year}{2024}\natexlab{}.
\newblock \bibinfo{title}{NO FAKES Act of 2024}.
\newblock
\newblock
\urldef\tempurl%
\url{https://files.constantcontact.com/1849eea4801/e18c1685-5a11-4861-9d43-ec84a7588ad7.pdf}
\showURL{%
\tempurl}


\bibitem[Agostinelli et~al\mbox{.}(2023)]%
        {Agostinelli2023musiclm}
\bibfield{author}{\bibinfo{person}{Andrea Agostinelli}, \bibinfo{person}{Timo~I. Denk}, \bibinfo{person}{Zalán Borsos}, \bibinfo{person}{Jesse Engel}, \bibinfo{person}{Mauro Verzetti}, \bibinfo{person}{Antoine Caillon}, \bibinfo{person}{Qingqing Huang}, \bibinfo{person}{Aren Jansen}, \bibinfo{person}{Adam Roberts}, \bibinfo{person}{Marco Tagliasacchi}, \bibinfo{person}{Matt Sharifi}, \bibinfo{person}{Neil Zeghidour}, {and} \bibinfo{person}{Christian Frank}.} \bibinfo{year}{2023}\natexlab{}.
\newblock \bibinfo{title}{MusicLM: Generating Music From Text}.
\newblock
\newblock
\showeprint[arxiv]{2301.11325}~[cs.SD]
\urldef\tempurl%
\url{https://arxiv.org/abs/2301.11325}
\showURL{%
\tempurl}


\bibitem[AI(2022)]%
        {profanity}
\bibfield{author}{\bibinfo{person}{Surge AI}.} \bibinfo{year}{2022}\natexlab{}.
\newblock \bibinfo{title}{The obscenity list}.
\newblock
\newblock
\urldef\tempurl%
\url{https://github.com/surge-ai/profanity}
\showURL{%
\tempurl}


\bibitem[Ardila et~al\mbox{.}(2020)]%
        {ardila2019common}
\bibfield{author}{\bibinfo{person}{Rosana Ardila}, \bibinfo{person}{Megan Branson}, \bibinfo{person}{Kelly Davis}, \bibinfo{person}{Michael Kohler}, \bibinfo{person}{Josh Meyer}, \bibinfo{person}{Michael Henretty}, \bibinfo{person}{Reuben Morais}, \bibinfo{person}{Lindsay Saunders}, \bibinfo{person}{Francis Tyers}, {and} \bibinfo{person}{Gregor Weber}.} \bibinfo{year}{2020}\natexlab{}.
\newblock \showarticletitle{Common Voice: A Massively-Multilingual Speech Corpus}. In \bibinfo{booktitle}{\emph{The Language Resources and Evaluation Conference (LREC)}}. \bibinfo{publisher}{European Language Resources Association}, \bibinfo{address}{Marseille, France}, \bibinfo{pages}{4218--4222}.
\newblock


\bibitem[arXiv(2023)]%
        {arxiv_page}
\bibfield{author}{\bibinfo{person}{arXiv}.} \bibinfo{year}{2023}\natexlab{}.
\newblock \bibinfo{title}{About arXiv}.
\newblock
\newblock
\urldef\tempurl%
\url{https://info.arxiv.org/about/index.html}
\showURL{%
\tempurl}


\bibitem[Bakhturina et~al\mbox{.}(2021)]%
        {bakhturina2021hi}
\bibfield{author}{\bibinfo{person}{Evelina Bakhturina}, \bibinfo{person}{Vitaly Lavrukhin}, \bibinfo{person}{Boris Ginsburg}, {and} \bibinfo{person}{Yang Zhang}.} \bibinfo{year}{2021}\natexlab{}.
\newblock \showarticletitle{{Hi-Fi Multi-Speaker English TTS Dataset}}. In \bibinfo{booktitle}{\emph{INTERSPEECH}}. \bibinfo{publisher}{ISCA}, \bibinfo{address}{Brno, Czechia}, \bibinfo{pages}{2776--2780}.
\newblock


\bibitem[Barnett(2023)]%
        {barnett2023ethical}
\bibfield{author}{\bibinfo{person}{Julia Barnett}.} \bibinfo{year}{2023}\natexlab{}.
\newblock \showarticletitle{The ethical implications of generative audio models: A systematic literature review}. In \bibinfo{booktitle}{\emph{Proceedings of the 2023 AAAI/ACM Conference on AI, Ethics, and Society}}. \bibinfo{publisher}{Association for Computing Machinery (ACM)}, \bibinfo{address}{Montreal, Canada}, \bibinfo{pages}{146--161}.
\newblock


\bibitem[Barnett et~al\mbox{.}(2024)]%
        {barnett2024exploring}
\bibfield{author}{\bibinfo{person}{Julia Barnett}, \bibinfo{person}{Hugo~Flores Garcia}, {and} \bibinfo{person}{Bryan Pardo}.} \bibinfo{year}{2024}\natexlab{}.
\newblock \bibinfo{title}{Exploring Musical Roots: Applying Audio Embeddings to Empower Influence Attribution for a Generative Music Model}.
\newblock
\newblock
\showeprint[arxiv]{2401.14542}~[cs.SD]
\urldef\tempurl%
\url{https://arxiv.org/abs/2401.14542}
\showURL{%
\tempurl}


\bibitem[Batlle-Roca et~al\mbox{.}(2023)]%
        {batlle2023transparency}
\bibfield{author}{\bibinfo{person}{Roser Batlle-Roca}, \bibinfo{person}{Emila G{\'o}mez}, \bibinfo{person}{WeiHsiang Liao}, \bibinfo{person}{Xavier Serra}, {and} \bibinfo{person}{Yuki Mitsufuji}.} \bibinfo{year}{2023}\natexlab{}.
\newblock \bibinfo{title}{Transparency in music-generative {AI}: A systematic literature review}.
\newblock
\newblock
\urldef\tempurl%
\url{https://doi.org/10.21203/rs.3.rs-3708077/v1}
\showURL{%
\tempurl}


\bibitem[Bender et~al\mbox{.}(2021)]%
        {bender2021dangers}
\bibfield{author}{\bibinfo{person}{Emily~M Bender}, \bibinfo{person}{Timnit Gebru}, \bibinfo{person}{Angelina McMillan-Major}, {and} \bibinfo{person}{Shmargaret Shmitchell}.} \bibinfo{year}{2021}\natexlab{}.
\newblock \showarticletitle{On the dangers of stochastic parrots: Can language models be too big?}. In \bibinfo{booktitle}{\emph{Proceedings of the 2021 ACM conference on fairness, accountability, and transparency}}. \bibinfo{publisher}{Association for Computing Machinery (ACM)}, \bibinfo{address}{Virtual Event, Canada}, \bibinfo{pages}{610--623}.
\newblock


\bibitem[Benjamin(2019)]%
        {benjamin2019race}
\bibfield{author}{\bibinfo{person}{Ruha Benjamin}.} \bibinfo{year}{2019}\natexlab{}.
\newblock \bibinfo{booktitle}{\emph{Race after technology: Abolitionist tools for the new Jim code}}.
\newblock \bibinfo{publisher}{John Wiley \& Sons}.
\newblock


\bibitem[Bertin-Mahieux et~al\mbox{.}(2011)]%
        {Bertin-Mahieux2011}
\bibfield{author}{\bibinfo{person}{Thierry Bertin-Mahieux}, \bibinfo{person}{Daniel~P.W. Ellis}, \bibinfo{person}{Brian Whitman}, {and} \bibinfo{person}{Paul Lamere}.} \bibinfo{year}{2011}\natexlab{}.
\newblock \showarticletitle{The Million Song Dataset}. In \bibinfo{booktitle}{\emph{{Proceedings of the 12th International Conference on Music Information Retrieval ({ISMIR} 2011)}}}. \bibinfo{publisher}{International Society for Music Information Retrieval}, \bibinfo{address}{Miami, US}, \bibinfo{pages}{591--596}.
\newblock


\bibitem[Bianchi et~al\mbox{.}(2023)]%
        {bianchi2023easily}
\bibfield{author}{\bibinfo{person}{Federico Bianchi}, \bibinfo{person}{Pratyusha Kalluri}, \bibinfo{person}{Esin Durmus}, \bibinfo{person}{Faisal Ladhak}, \bibinfo{person}{Myra Cheng}, \bibinfo{person}{Debora Nozza}, \bibinfo{person}{Tatsunori Hashimoto}, \bibinfo{person}{Dan Jurafsky}, \bibinfo{person}{James Zou}, {and} \bibinfo{person}{Aylin Caliskan}.} \bibinfo{year}{2023}\natexlab{}.
\newblock \showarticletitle{Easily accessible text-to-image generation amplifies demographic stereotypes at large scale}. In \bibinfo{booktitle}{\emph{Proceedings of the 2023 ACM Conference on Fairness, Accountability, and Transparency}}. \bibinfo{pages}{1493--1504}.
\newblock


\bibitem[Birhane et~al\mbox{.}(2024a)]%
        {birhane2024dark}
\bibfield{author}{\bibinfo{person}{Abeba Birhane}, \bibinfo{person}{Sepehr Dehdashtian}, \bibinfo{person}{Vinay Prabhu}, {and} \bibinfo{person}{Vishnu Boddeti}.} \bibinfo{year}{2024}\natexlab{a}.
\newblock \showarticletitle{The Dark Side of Dataset Scaling: Evaluating Racial Classification in Multimodal Models}. In \bibinfo{booktitle}{\emph{The 2024 ACM Conference on Fairness, Accountability, and Transparency}}. \bibinfo{pages}{1229--1244}.
\newblock


\bibitem[Birhane et~al\mbox{.}(2022)]%
        {birhane2022values}
\bibfield{author}{\bibinfo{person}{Abeba Birhane}, \bibinfo{person}{Pratyusha Kalluri}, \bibinfo{person}{Dallas Card}, \bibinfo{person}{William Agnew}, \bibinfo{person}{Ravit Dotan}, {and} \bibinfo{person}{Michelle Bao}.} \bibinfo{year}{2022}\natexlab{}.
\newblock \showarticletitle{The values encoded in machine learning research}. In \bibinfo{booktitle}{\emph{Proceedings of the 2022 ACM Conference on Fairness, Accountability, and Transparency}}. \bibinfo{pages}{173--184}.
\newblock


\bibitem[Birhane et~al\mbox{.}(2023)]%
        {birhane2024into}
\bibfield{author}{\bibinfo{person}{Abeba Birhane}, \bibinfo{person}{Vinay Prabhu}, \bibinfo{person}{Sang Han}, \bibinfo{person}{Vishnu~Naresh Boddeti}, {and} \bibinfo{person}{Alexandra~Sasha Luccioni}.} \bibinfo{year}{2023}\natexlab{}.
\newblock \showarticletitle{Into the LAION's den: investigating hate in multimodal datasets}. In \bibinfo{booktitle}{\emph{Proceedings of the 37th International Conference on Neural Information Processing Systems}}. \bibinfo{publisher}{Curran Associates, Inc.}, \bibinfo{address}{New Orleans, USA}, \bibinfo{pages}{21268--21284}.
\newblock


\bibitem[Birhane et~al\mbox{.}(2021a)]%
        {birhane2021multimodal}
\bibfield{author}{\bibinfo{person}{Abeba Birhane}, \bibinfo{person}{Vinay~Uday Prabhu}, {and} \bibinfo{person}{Emmanuel Kahembwe}.} \bibinfo{year}{2021}\natexlab{a}.
\newblock \showarticletitle{Multimodal datasets: misogyny, pornography, and malignant stereotypes}.
\newblock \bibinfo{journal}{\emph{arXiv preprint arXiv:2110.01963}} (\bibinfo{year}{2021}).
\newblock


\bibitem[Birhane et~al\mbox{.}(2021b)]%
        {Birhane2021laion}
\bibfield{author}{\bibinfo{person}{Abeba Birhane}, \bibinfo{person}{Vinay~Uday Prabhu}, {and} \bibinfo{person}{Emmanuel Kahembwe}.} \bibinfo{year}{2021}\natexlab{b}.
\newblock \bibinfo{title}{Multimodal datasets: misogyny, pornography, and malignant stereotypes}.
\newblock
\newblock
\showeprint[arxiv]{2110.01963}~[cs.CY]
\urldef\tempurl%
\url{https://arxiv.org/abs/2110.01963}
\showURL{%
\tempurl}


\bibitem[Birhane et~al\mbox{.}(2024b)]%
        {birhane2024ai}
\bibfield{author}{\bibinfo{person}{Abeba Birhane}, \bibinfo{person}{Ryan Steed}, \bibinfo{person}{Victor Ojewale}, \bibinfo{person}{Briana Vecchione}, {and} \bibinfo{person}{Inioluwa~Deborah Raji}.} \bibinfo{year}{2024}\natexlab{b}.
\newblock \showarticletitle{AI auditing: The broken bus on the road to AI accountability}. In \bibinfo{booktitle}{\emph{2024 IEEE Conference on Secure and Trustworthy Machine Learning (SaTML)}}. IEEE, \bibinfo{publisher}{IEEE}, \bibinfo{address}{Toronto, Canada}, \bibinfo{pages}{612--643}.
\newblock


\bibitem[Bloom(2017)]%
        {bloom2017reading}
\bibfield{author}{\bibinfo{person}{James~D Bloom}.} \bibinfo{year}{2017}\natexlab{}.
\newblock \bibinfo{booktitle}{\emph{Reading the male gaze in literature and culture: Studies in erotic epistemology}}.
\newblock \bibinfo{publisher}{Springer}, \bibinfo{address}{Cham, Switzerland}.
\newblock


\bibitem[Bogdanov et~al\mbox{.}(2019)]%
        {bogdanov2019mtg}
\bibfield{author}{\bibinfo{person}{Dmitry Bogdanov}, \bibinfo{person}{Minz Won}, \bibinfo{person}{Philip Tovstogan}, \bibinfo{person}{Alastair Porter}, {and} \bibinfo{person}{Xavier Serra}.} \bibinfo{year}{2019}\natexlab{}.
\newblock \showarticletitle{The {MTG-Jamendo} Dataset for Automatic Music Tagging}. In \bibinfo{booktitle}{\emph{Machine Learning for Music Discovery Workshop, International Conference on Machine Learning (ICML 2019)}}. \bibinfo{publisher}{PMLR}, \bibinfo{address}{Long Beach, CA, United States}, \bibinfo{numpages}{3}~pages.
\newblock
\urldef\tempurl%
\url{http://hdl.handle.net/10230/42015}
\showURL{%
\tempurl}


\bibitem[Bommasani et~al\mbox{.}(2022)]%
        {bommasani2022picking}
\bibfield{author}{\bibinfo{person}{Rishi Bommasani}, \bibinfo{person}{Kathleen~A. Creel}, \bibinfo{person}{Ananya Kumar}, \bibinfo{person}{Dan Jurafsky}, {and} \bibinfo{person}{Percy~S. Liang}.} \bibinfo{year}{2022}\natexlab{}.
\newblock \showarticletitle{Picking on the Same Person: Does Algorithmic Monoculture lead to Outcome Homogenization?}
\newblock \bibinfo{journal}{\emph{Advances in Neural Information Processing Systems}}  \bibinfo{volume}{35} (\bibinfo{date}{Dec.} \bibinfo{year}{2022}), \bibinfo{pages}{3663–3678}.
\newblock


\bibitem[Borsos et~al\mbox{.}(2023a)]%
        {borsos2023audiolm}
\bibfield{author}{\bibinfo{person}{Zal{\'a}n Borsos}, \bibinfo{person}{Rapha{\"e}l Marinier}, \bibinfo{person}{Damien Vincent}, \bibinfo{person}{Eugene Kharitonov}, \bibinfo{person}{Olivier Pietquin}, \bibinfo{person}{Matt Sharifi}, \bibinfo{person}{Dominik Roblek}, \bibinfo{person}{Olivier Teboul}, \bibinfo{person}{David Grangier}, \bibinfo{person}{Marco Tagliasacchi}, {et~al\mbox{.}}} \bibinfo{year}{2023}\natexlab{a}.
\newblock \showarticletitle{Audiolm: a language modeling approach to audio generation}.
\newblock \bibinfo{journal}{\emph{IEEE/ACM transactions on audio, speech, and language processing}}  \bibinfo{volume}{31} (\bibinfo{year}{2023}), \bibinfo{pages}{2523--2533}.
\newblock


\bibitem[Borsos et~al\mbox{.}(2023b)]%
        {borsos2023soundstorm}
\bibfield{author}{\bibinfo{person}{Zal{\'a}n Borsos}, \bibinfo{person}{Matt Sharifi}, \bibinfo{person}{Damien Vincent}, \bibinfo{person}{Eugene Kharitonov}, \bibinfo{person}{Neil Zeghidour}, {and} \bibinfo{person}{Marco Tagliasacchi}.} \bibinfo{year}{2023}\natexlab{b}.
\newblock \showarticletitle{Soundstorm: Efficient parallel audio generation}.
\newblock \bibinfo{journal}{\emph{arXiv preprint arXiv:2305.09636}} (\bibinfo{year}{2023}).
\newblock


\bibitem[Bralios et~al\mbox{.}(2024)]%
        {bralios2024generation}
\bibfield{author}{\bibinfo{person}{Dimitrios Bralios}, \bibinfo{person}{Gordon Wichern}, \bibinfo{person}{Fran{\c{c}}ois~G Germain}, \bibinfo{person}{Zexu Pan}, \bibinfo{person}{Sameer Khurana}, \bibinfo{person}{Chiori Hori}, {and} \bibinfo{person}{Jonathan Le~Roux}.} \bibinfo{year}{2024}\natexlab{}.
\newblock \showarticletitle{Generation or Replication: Auscultating Audio Latent Diffusion Models}. In \bibinfo{booktitle}{\emph{ICASSP 2024-2024 IEEE International Conference on Acoustics, Speech and Signal Processing (ICASSP)}}. IEEE, \bibinfo{publisher}{IEEE}, \bibinfo{address}{Seoul, South Korea}, \bibinfo{pages}{1156--1160}.
\newblock


\bibitem[Carlini et~al\mbox{.}(2022)]%
        {carlini2022quantifying}
\bibfield{author}{\bibinfo{person}{Nicholas Carlini}, \bibinfo{person}{Daphne Ippolito}, \bibinfo{person}{Matthew Jagielski}, \bibinfo{person}{Katherine Lee}, \bibinfo{person}{Florian Tramer}, {and} \bibinfo{person}{Chiyuan Zhang}.} \bibinfo{year}{2022}\natexlab{}.
\newblock \showarticletitle{Quantifying memorization across neural language models}.
\newblock \bibinfo{journal}{\emph{arXiv preprint arXiv:2202.07646}} (\bibinfo{year}{2022}).
\newblock


\bibitem[Champion(2024)]%
        {champion2023anonymizing}
\bibfield{author}{\bibinfo{person}{Pierre Champion}.} \bibinfo{year}{2024}\natexlab{}.
\newblock \bibinfo{title}{Anonymizing Speech: Evaluating and Designing Speaker Anonymization Techniques}.
\newblock
\newblock
\showeprint[arxiv]{2308.04455}~[cs.CR]
\urldef\tempurl%
\url{https://arxiv.org/abs/2308.04455}
\showURL{%
\tempurl}


\bibitem[Cheng et~al\mbox{.}(2024)]%
        {cheng2024dated}
\bibfield{author}{\bibinfo{person}{Jeffrey Cheng}, \bibinfo{person}{Marc Marone}, \bibinfo{person}{Orion Weller}, \bibinfo{person}{Dawn Lawrie}, \bibinfo{person}{Daniel Khashabi}, {and} \bibinfo{person}{Benjamin Van~Durme}.} \bibinfo{year}{2024}\natexlab{}.
\newblock \showarticletitle{Dated Data: Tracing Knowledge Cutoffs in Large Language Models}.
\newblock \bibinfo{journal}{\emph{arXiv preprint arXiv:2403.12958}} (\bibinfo{year}{2024}).
\newblock


\bibitem[Chu et~al\mbox{.}(2023)]%
        {chu2023qwenaudio}
\bibfield{author}{\bibinfo{person}{Yunfei Chu}, \bibinfo{person}{Jin Xu}, \bibinfo{person}{Xiaohuan Zhou}, \bibinfo{person}{Qian Yang}, \bibinfo{person}{Shiliang Zhang}, \bibinfo{person}{Zhijie Yan}, \bibinfo{person}{Chang Zhou}, {and} \bibinfo{person}{Jingren Zhou}.} \bibinfo{year}{2023}\natexlab{}.
\newblock \bibinfo{title}{Qwen-Audio: Advancing Universal Audio Understanding via Unified Large-Scale Audio-Language Models}.
\newblock
\newblock
\showeprint[arxiv]{2311.07919}~[eess.AS]
\urldef\tempurl%
\url{https://arxiv.org/abs/2311.07919}
\showURL{%
\tempurl}


\bibitem[Civit et~al\mbox{.}(2022)]%
        {civit2022systematic}
\bibfield{author}{\bibinfo{person}{Miguel Civit}, \bibinfo{person}{Javier Civit-Masot}, \bibinfo{person}{Francisco Cuadrado}, {and} \bibinfo{person}{Maria~J Escalona}.} \bibinfo{year}{2022}\natexlab{}.
\newblock \showarticletitle{A systematic review of artificial intelligence-based music generation: Scope, applications, and future trends}.
\newblock \bibinfo{journal}{\emph{Expert Systems with Applications}}  \bibinfo{volume}{209} (\bibinfo{year}{2022}), \bibinfo{pages}{118190}.
\newblock


\bibitem[Clifton et~al\mbox{.}(2020)]%
        {clifton2020spotify}
\bibfield{author}{\bibinfo{person}{Ann Clifton}, \bibinfo{person}{Aasish Pappu}, \bibinfo{person}{Sravana Reddy}, \bibinfo{person}{Yongze Yu}, \bibinfo{person}{Jussi Karlgren}, \bibinfo{person}{Ben Carterette}, {and} \bibinfo{person}{Rosie Jones}.} \bibinfo{year}{2020}\natexlab{}.
\newblock \bibinfo{title}{The Spotify Podcast Dataset}.
\newblock
\newblock
\showeprint[arxiv]{2004.04270}~[cs.CL]
\urldef\tempurl%
\url{https://arxiv.org/abs/2004.04270}
\showURL{%
\tempurl}


\bibitem[Cole(2023)]%
        {404laion}
\bibfield{author}{\bibinfo{person}{Samantha Cole}.} \bibinfo{year}{2023}\natexlab{}.
\newblock \bibinfo{title}{Largest Dataset Powering AI Images Removed After Discovery of Child Sexual Abuse Material}.
\newblock
\newblock
\urldef\tempurl%
\url{https://www.404media.co/laion-datasets-removed-stanford-csam-child-abuse/}
\showURL{%
\tempurl}


\bibitem[Commons(2024a)]%
        {cc4}
\bibfield{author}{\bibinfo{person}{Creative Commons}.} \bibinfo{year}{2024}\natexlab{a}.
\newblock \bibinfo{title}{{CC BY 4.0}}.
\newblock
\newblock
\urldef\tempurl%
\url{https://creativecommons.org/licenses/by/4.0/}
\showURL{%
\tempurl}


\bibitem[Commons(2024b)]%
        {cc0}
\bibfield{author}{\bibinfo{person}{Creative Commons}.} \bibinfo{year}{2024}\natexlab{b}.
\newblock \bibinfo{title}{{CC0}}.
\newblock
\newblock
\urldef\tempurl%
\url{https://creativecommons.org/public-domain/cc0/}
\showURL{%
\tempurl}


\bibitem[Commons(2024c)]%
        {odc}
\bibfield{author}{\bibinfo{person}{Open~Data Commons}.} \bibinfo{year}{2024}\natexlab{c}.
\newblock \bibinfo{title}{Open Data Commons Attribution License (ODC-By) v1.0}.
\newblock
\newblock
\urldef\tempurl%
\url{https://opendatacommons.org/licenses/by/summary/}
\showURL{%
\tempurl}


\bibitem[Coscarelli(2023)]%
        {Coscarelli_2023}
\bibfield{author}{\bibinfo{person}{Joe Coscarelli}.} \bibinfo{year}{2023}\natexlab{}.
\newblock \showarticletitle{An {A.I.} Hit of Fake {‘Drake’} and {‘The Weeknd’} Rattles the Music World}.
\newblock \bibinfo{journal}{\emph{The New York Times}} (\bibinfo{date}{Apr} \bibinfo{year}{2023}).
\newblock
\showISSN{0362-4331}
\urldef\tempurl%
\url{https://www.nytimes.com/2023/04/19/arts/music/ai-drake-the-weeknd-fake.html}
\showURL{%
\tempurl}


\bibitem[Cosentino et~al\mbox{.}(2020)]%
        {cosentino2020librimix}
\bibfield{author}{\bibinfo{person}{Joris Cosentino}, \bibinfo{person}{Manuel Pariente}, \bibinfo{person}{Samuele Cornell}, \bibinfo{person}{Antoine Deleforge}, {and} \bibinfo{person}{Emmanuel Vincent}.} \bibinfo{year}{2020}\natexlab{}.
\newblock \bibinfo{title}{LibriMix: An Open-Source Dataset for Generalizable Speech Separation}.
\newblock
\newblock
\showeprint[arxiv]{2005.11262}~[eess.AS]


\bibitem[Criddle and Bryan(2024)]%
        {ftwater}
\bibfield{author}{\bibinfo{person}{Cristina Criddle} {and} \bibinfo{person}{Kenza Bryan}.} \bibinfo{year}{2024}\natexlab{}.
\newblock \showarticletitle{{AI} boom sparks concern over Big Tech’s water consumption}.
\newblock \bibinfo{journal}{\emph{Financial Times}} (\bibinfo{year}{2024}).
\newblock
\urldef\tempurl%
\url{https://www.ft.com/content/6544119e-a511-4cfa-9243-13b8cf855c13}
\showURL{%
\tempurl}


\bibitem[Crocco et~al\mbox{.}(2016)]%
        {crocco2016audio}
\bibfield{author}{\bibinfo{person}{Marco Crocco}, \bibinfo{person}{Marco Cristani}, \bibinfo{person}{Andrea Trucco}, {and} \bibinfo{person}{Vittorio Murino}.} \bibinfo{year}{2016}\natexlab{}.
\newblock \showarticletitle{Audio surveillance: A systematic review}.
\newblock \bibinfo{journal}{\emph{ACM Computing Surveys (CSUR)}} \bibinfo{volume}{48}, \bibinfo{number}{4} (\bibinfo{year}{2016}), \bibinfo{pages}{1--46}.
\newblock


\bibitem[Dalugdug(2024)]%
        {MBW2023}
\bibfield{author}{\bibinfo{person}{Mandy Dalugdug}.} \bibinfo{year}{2024}\natexlab{}.
\newblock \showarticletitle{Major music companies fight back against unlicensed {AI} in new ICMP-led initiative}.
\newblock \bibinfo{journal}{\emph{Music Business Worldwide}} (\bibinfo{date}{apr} \bibinfo{year}{2024}).
\newblock
\newblock
\shownote{\url{https://www.musicbusinessworldwide.com/music-publishers-rightsandai-portal-ai/}}.


\bibitem[Danielescu et~al\mbox{.}(2023)]%
        {DanielescuNB23}
\bibfield{author}{\bibinfo{person}{Andreea Danielescu}, \bibinfo{person}{Sharone~A Horowit-Hendler}, \bibinfo{person}{Alexandria Pabst}, \bibinfo{person}{Kenneth~Michael Stewart}, \bibinfo{person}{Eric~M Gallo}, {and} \bibinfo{person}{Matthew~Peter Aylett}.} \bibinfo{year}{2023}\natexlab{}.
\newblock \showarticletitle{Creating Inclusive Voices for the 21st Century: A Non-Binary Text-to-Speech for Conversational Assistants}. In \bibinfo{booktitle}{\emph{Proceedings of the 2023 CHI Conference on Human Factors in Computing Systems}} (Hamburg, Germany) \emph{(\bibinfo{series}{CHI '23})}. \bibinfo{publisher}{Association for Computing Machinery}, \bibinfo{address}{New York, NY, USA}, Article \bibinfo{articleno}{390}, \bibinfo{numpages}{17}~pages.
\newblock
\showISBNx{9781450394215}
\urldef\tempurl%
\url{https://doi.org/10.1145/3544548.3581281}
\showDOI{\tempurl}


\bibitem[Davis(2024)]%
        {verge_openai_yt}
\bibfield{author}{\bibinfo{person}{Wes Davis}.} \bibinfo{year}{2024}\natexlab{}.
\newblock \bibinfo{title}{OpenAI transcribed over a million hours of YouTube videos to train GPT-4}.
\newblock
\newblock
\urldef\tempurl%
\url{https://www.theverge.com/2024/4/6/24122915/openai-youtube-transcripts-gpt-4-training-data-google}
\showURL{%
\tempurl}


\bibitem[Defferrard et~al\mbox{.}(2017)]%
        {defferrard2016fma}
\bibfield{author}{\bibinfo{person}{Micha{\"e}l Defferrard}, \bibinfo{person}{Kirell Benzi}, \bibinfo{person}{Pierre Vandergheynst}, {and} \bibinfo{person}{Xavier Bresson}.} \bibinfo{year}{2017}\natexlab{}.
\newblock \showarticletitle{FMA: A Dataset For Music Analysis}. In \bibinfo{booktitle}{\emph{18th International Society for Music Information Retrieval Conference}}. \bibinfo{publisher}{International Society for Music Information Retrieval}, \bibinfo{address}{Suzhou, China}, \bibinfo{pages}{316--323}.
\newblock


\bibitem[Deshmukh et~al\mbox{.}(2024)]%
        {deshmukh2024audioentailment}
\bibfield{author}{\bibinfo{person}{Soham Deshmukh}, \bibinfo{person}{Shuo Han}, \bibinfo{person}{Hazim Bukhari}, \bibinfo{person}{Benjamin Elizalde}, \bibinfo{person}{Hannes Gamper}, \bibinfo{person}{Rita Singh}, {and} \bibinfo{person}{Bhiksha Raj}.} \bibinfo{year}{2024}\natexlab{}.
\newblock \bibinfo{title}{Audio Entailment: Assessing Deductive Reasoning for Audio Understanding}.
\newblock
\newblock
\showeprint[arxiv]{2407.18062}~[cs.SD]
\urldef\tempurl%
\url{https://arxiv.org/abs/2407.18062}
\showURL{%
\tempurl}


\bibitem[Dodge et~al\mbox{.}(2021)]%
        {dodge2021documenting}
\bibfield{author}{\bibinfo{person}{Jesse Dodge}, \bibinfo{person}{Maarten Sap}, \bibinfo{person}{Ana Marasovi{\'c}}, \bibinfo{person}{William Agnew}, \bibinfo{person}{Gabriel Ilharco}, \bibinfo{person}{Dirk Groeneveld}, \bibinfo{person}{Margaret Mitchell}, {and} \bibinfo{person}{Matt Gardner}.} \bibinfo{year}{2021}\natexlab{}.
\newblock \showarticletitle{Documenting Large Webtext Corpora: A Case Study on the Colossal Clean Crawled Corpus}. In \bibinfo{booktitle}{\emph{Proceedings of the 2021 Conference on Empirical Methods in Natural Language Processing}}. \bibinfo{publisher}{Association for Computational Linguistics}, \bibinfo{address}{Online and Punta Cana, Dominican Republic}, \bibinfo{pages}{1286--1305}.
\newblock


\bibitem[Donahue et~al\mbox{.}(2023)]%
        {Donahue2023singsong}
\bibfield{author}{\bibinfo{person}{Chris Donahue}, \bibinfo{person}{Antoine Caillon}, \bibinfo{person}{Adam Roberts}, \bibinfo{person}{Ethan Manilow}, \bibinfo{person}{Philippe Esling}, \bibinfo{person}{Andrea Agostinelli}, \bibinfo{person}{Mauro Verzetti}, \bibinfo{person}{Ian Simon}, \bibinfo{person}{Olivier Pietquin}, \bibinfo{person}{Neil Zeghidour}, {and} \bibinfo{person}{Jesse Engel}.} \bibinfo{year}{2023}\natexlab{}.
\newblock \bibinfo{title}{SingSong: Generating musical accompaniments from singing}.
\newblock
\newblock
\showeprint[arxiv]{2301.12662}~[cs.SD]
\urldef\tempurl%
\url{https://arxiv.org/abs/2301.12662}
\showURL{%
\tempurl}


\bibitem[Donahue et~al\mbox{.}(2017)]%
        {donahue2017dance}
\bibfield{author}{\bibinfo{person}{Chris Donahue}, \bibinfo{person}{Zachary~C Lipton}, {and} \bibinfo{person}{Julian McAuley}.} \bibinfo{year}{2017}\natexlab{}.
\newblock \showarticletitle{Dance dance convolution}. In \bibinfo{booktitle}{\emph{International conference on machine learning}}. PMLR, \bibinfo{publisher}{PMLR}, \bibinfo{address}{Sydney, Australia}, \bibinfo{pages}{1039--1048}.
\newblock


\bibitem[Douwes et~al\mbox{.}(2021)]%
        {douwes2021energy}
\bibfield{author}{\bibinfo{person}{Constance Douwes}, \bibinfo{person}{Philippe Esling}, {and} \bibinfo{person}{Jean-Pierre Briot}.} \bibinfo{year}{2021}\natexlab{}.
\newblock \bibinfo{title}{Energy Consumption of Deep Generative Audio Models}.
\newblock
\newblock
\showeprint[arxiv]{2107.02621}~[cs.LG]
\urldef\tempurl%
\url{https://arxiv.org/abs/2107.02621}
\showURL{%
\tempurl}


\bibitem[Dubey et~al\mbox{.}(2023)]%
        {dubey2023icassp}
\bibfield{author}{\bibinfo{person}{Harishchandra Dubey}, \bibinfo{person}{Ashkan Aazami}, \bibinfo{person}{Vishak Gopal}, \bibinfo{person}{Babak Naderi}, \bibinfo{person}{Sebastian Braun}, \bibinfo{person}{Ross Cutler}, \bibinfo{person}{Hannes Gamper}, \bibinfo{person}{Mehrsa Golestaneh}, {and} \bibinfo{person}{Robert Aichner}.} \bibinfo{year}{2023}\natexlab{}.
\newblock \showarticletitle{{ICASSP} 2023 Deep Noise Suppression Challenge}. In \bibinfo{booktitle}{\emph{ICASSP}}. \bibinfo{publisher}{IEEE}, \bibinfo{address}{Rhodes, Greece}, \bibinfo{pages}{725--737}.
\newblock


\bibitem[Everman et~al\mbox{.}(2023)]%
        {everman2023evaluating}
\bibfield{author}{\bibinfo{person}{Brad Everman}, \bibinfo{person}{Trevor Villwock}, \bibinfo{person}{Dayuan Chen}, \bibinfo{person}{Noe Soto}, \bibinfo{person}{Oliver Zhang}, {and} \bibinfo{person}{Ziliang Zong}.} \bibinfo{year}{2023}\natexlab{}.
\newblock \showarticletitle{Evaluating the carbon impact of large language models at the inference stage}. In \bibinfo{booktitle}{\emph{2023 IEEE international performance, computing, and communications conference (IPCCC)}}. IEEE, \bibinfo{publisher}{IEEE}, \bibinfo{address}{Anaheim, USA}, \bibinfo{pages}{150--157}.
\newblock


\bibitem[Feffer et~al\mbox{.}(2023)]%
        {feffer2023deepdrake}
\bibfield{author}{\bibinfo{person}{Michael Feffer}, \bibinfo{person}{Zachary~C Lipton}, {and} \bibinfo{person}{Chris Donahue}.} \bibinfo{year}{2023}\natexlab{}.
\newblock \showarticletitle{Deepdrake ft. bts-gan and taylorvc: an exploratory analysis of musical deepfakes and hosting platforms}. In \bibinfo{booktitle}{\emph{Proceedings of the 2nd Workshop on Human-Centric Music Information Retrieval}}. \bibinfo{publisher}{CEUR Workshop Proceedings}, \bibinfo{address}{Milan, Italy}, \bibinfo{numpages}{7}~pages.
\newblock


\bibitem[Forsgren and Martiros(2022)]%
        {forsgren2022riffusion}
\bibfield{author}{\bibinfo{person}{Seth Forsgren} {and} \bibinfo{person}{Hayk Martiros}.} \bibinfo{year}{2022}\natexlab{}.
\newblock \bibinfo{title}{Riffusion: Stable diffusion for real-time music generation}.
\newblock
\newblock
\urldef\tempurl%
\url{https://riffusion.com/about}
\showURL{%
\tempurl}


\bibitem[Fox et~al\mbox{.}(2023)]%
        {fox2023patchwork}
\bibfield{author}{\bibinfo{person}{Sarah~E Fox}, \bibinfo{person}{Samantha Shorey}, \bibinfo{person}{Esther~Y Kang}, \bibinfo{person}{Dominique Montiel~Valle}, {and} \bibinfo{person}{Estefania Rodriguez}.} \bibinfo{year}{2023}\natexlab{}.
\newblock \showarticletitle{Patchwork: the hidden, human labor of AI integration within essential work}.
\newblock \bibinfo{journal}{\emph{Proceedings of the ACM on Human-Computer Interaction}} \bibinfo{volume}{7}, \bibinfo{number}{CSCW1} (\bibinfo{year}{2023}), \bibinfo{pages}{1--20}.
\newblock


\bibitem[Fuckner et~al\mbox{.}(2023)]%
        {fuckner2023uncovering}
\bibfield{author}{\bibinfo{person}{Marcio Fuckner}, \bibinfo{person}{Sophie Horsman}, \bibinfo{person}{Pascal Wiggers}, {and} \bibinfo{person}{Iskaj Janssen}.} \bibinfo{year}{2023}\natexlab{}.
\newblock \showarticletitle{Uncovering Bias in ASR Systems: Evaluating Wav2vec2 and Whisper for Dutch speakers}. In \bibinfo{booktitle}{\emph{2023 International Conference on Speech Technology and Human-Computer Dialogue (SpeD)}}. IEEE, \bibinfo{pages}{146--151}.
\newblock


\bibitem[Gadre et~al\mbox{.}(2024)]%
        {gadre2024datacomp}
\bibfield{author}{\bibinfo{person}{Samir~Yitzhak Gadre}, \bibinfo{person}{Gabriel Ilharco}, \bibinfo{person}{Alex Fang}, \bibinfo{person}{Jonathan Hayase}, \bibinfo{person}{Georgios Smyrnis}, \bibinfo{person}{Thao Nguyen}, \bibinfo{person}{Ryan Marten}, \bibinfo{person}{Mitchell Wortsman}, \bibinfo{person}{Dhruba Ghosh}, \bibinfo{person}{Jieyu Zhang}, {et~al\mbox{.}}} \bibinfo{year}{2024}\natexlab{}.
\newblock \showarticletitle{Datacomp: In search of the next generation of multimodal datasets}.
\newblock \bibinfo{journal}{\emph{Advances in Neural Information Processing Systems}}  \bibinfo{volume}{36} (\bibinfo{year}{2024}).
\newblock


\bibitem[Garcia et~al\mbox{.}(2023)]%
        {garcia2023vampnet}
\bibfield{author}{\bibinfo{person}{Hugo F~Flores Garcia}, \bibinfo{person}{Prem Seetharaman}, \bibinfo{person}{Rithesh Kumar}, {and} \bibinfo{person}{Bryan Pardo}.} \bibinfo{year}{2023}\natexlab{}.
\newblock \showarticletitle{VampNet: Music Generation via Masked Acoustic Token Modeling}. In \bibinfo{booktitle}{\emph{Ismir 2023 Hybrid Conference}}. \bibinfo{publisher}{International Society for Music Information Retrieval}, \bibinfo{address}{Milan, Italy}, \bibinfo{pages}{359--366}.
\newblock


\bibitem[Garofolo et~al\mbox{.}(1993)]%
        {garofolo1993csr}
\bibfield{author}{\bibinfo{person}{John Garofolo}, \bibinfo{person}{David Graff}, \bibinfo{person}{Doug Paul}, {and} \bibinfo{person}{David Pallett}.} \bibinfo{year}{1993}\natexlab{}.
\newblock \showarticletitle{{CSR-I (WSJ0)} complete ldc93s6a}.
\newblock \bibinfo{journal}{\emph{Web Download. Philadelphia: Linguistic Data Consortium}}  \bibinfo{volume}{83} (\bibinfo{year}{1993}).
\newblock


\bibitem[Gebru et~al\mbox{.}(2021)]%
        {gebru2021datasheets}
\bibfield{author}{\bibinfo{person}{Timnit Gebru}, \bibinfo{person}{Jamie Morgenstern}, \bibinfo{person}{Briana Vecchione}, \bibinfo{person}{Jennifer~Wortman Vaughan}, \bibinfo{person}{Hanna Wallach}, \bibinfo{person}{Hal~Daum{\'e} Iii}, {and} \bibinfo{person}{Kate Crawford}.} \bibinfo{year}{2021}\natexlab{}.
\newblock \showarticletitle{Datasheets for datasets}.
\newblock \bibinfo{journal}{\emph{Commun. ACM}} \bibinfo{volume}{64}, \bibinfo{number}{12} (\bibinfo{year}{2021}), \bibinfo{pages}{86--92}.
\newblock


\bibitem[Gemmeke et~al\mbox{.}(2017)]%
        {audioset}
\bibfield{author}{\bibinfo{person}{Jort~F. Gemmeke}, \bibinfo{person}{Daniel P.~W. Ellis}, \bibinfo{person}{Dylan Freedman}, \bibinfo{person}{Aren Jansen}, \bibinfo{person}{Wade Lawrence}, \bibinfo{person}{R.~Channing Moore}, \bibinfo{person}{Manoj Plakal}, {and} \bibinfo{person}{Marvin Ritter}.} \bibinfo{year}{2017}\natexlab{}.
\newblock \showarticletitle{Audio Set: An ontology and human-labeled dataset for audio events}. In \bibinfo{booktitle}{\emph{Proc. IEEE ICASSP 2017}}. \bibinfo{publisher}{IEEE}, \bibinfo{address}{New Orleans, LA}, \bibinfo{pages}{776--780}.
\newblock


\bibitem[Ghosh et~al\mbox{.}(2024)]%
        {ghosh2024gama}
\bibfield{author}{\bibinfo{person}{Sreyan Ghosh}, \bibinfo{person}{Sonal Kumar}, \bibinfo{person}{Ashish Seth}, \bibinfo{person}{Chandra Kiran~Reddy Evuru}, \bibinfo{person}{Utkarsh Tyagi}, \bibinfo{person}{S Sakshi}, \bibinfo{person}{Oriol Nieto}, \bibinfo{person}{Ramani Duraiswami}, {and} \bibinfo{person}{Dinesh Manocha}.} \bibinfo{year}{2024}\natexlab{}.
\newblock \bibinfo{title}{GAMA: A Large Audio-Language Model with Advanced Audio Understanding and Complex Reasoning Abilities}.
\newblock
\newblock
\showeprint[arxiv]{2406.11768}~[cs.SD]
\urldef\tempurl%
\url{https://arxiv.org/abs/2406.11768}
\showURL{%
\tempurl}


\bibitem[Gong et~al\mbox{.}(2021)]%
        {gong2021ast}
\bibfield{author}{\bibinfo{person}{Yuan Gong}, \bibinfo{person}{Yu-An Chung}, {and} \bibinfo{person}{James Glass}.} \bibinfo{year}{2021}\natexlab{}.
\newblock \bibinfo{title}{AST: Audio Spectrogram Transformer}.
\newblock
\newblock
\showeprint[arxiv]{2104.01778}~[cs.SD]
\urldef\tempurl%
\url{https://arxiv.org/abs/2104.01778}
\showURL{%
\tempurl}


\bibitem[Gong et~al\mbox{.}(2023a)]%
        {gong_ltuas}
\bibfield{author}{\bibinfo{person}{Yuan Gong}, \bibinfo{person}{Alexander~H Liu}, \bibinfo{person}{Hongyin Luo}, \bibinfo{person}{Leonid Karlinsky}, {and} \bibinfo{person}{James Glass}.} \bibinfo{year}{2023}\natexlab{a}.
\newblock \showarticletitle{Joint Audio and Speech Understanding}. In \bibinfo{booktitle}{\emph{2023 IEEE Automatic Speech Recognition and Understanding Workshop (ASRU)}}.
\newblock


\bibitem[Gong et~al\mbox{.}(2023b)]%
        {gong2023listen}
\bibfield{author}{\bibinfo{person}{Yuan Gong}, \bibinfo{person}{Hongyin Luo}, \bibinfo{person}{Alexander~H Liu}, \bibinfo{person}{Leonid Karlinsky}, {and} \bibinfo{person}{James Glass}.} \bibinfo{year}{2023}\natexlab{b}.
\newblock \showarticletitle{Listen, Think, and Understand}.
\newblock \bibinfo{journal}{\emph{arXiv preprint arXiv:2305.10790}} (\bibinfo{year}{2023}).
\newblock


\bibitem[Gong et~al\mbox{.}(2023c)]%
        {gong2023figstep}
\bibfield{author}{\bibinfo{person}{Yichen Gong}, \bibinfo{person}{Delong Ran}, \bibinfo{person}{Jinyuan Liu}, \bibinfo{person}{Conglei Wang}, \bibinfo{person}{Tianshuo Cong}, \bibinfo{person}{Anyu Wang}, \bibinfo{person}{Sisi Duan}, {and} \bibinfo{person}{Xiaoyun Wang}.} \bibinfo{year}{2023}\natexlab{c}.
\newblock \bibinfo{title}{FigStep: Jailbreaking Large Vision-language Models via Typographic Visual Prompts}.
\newblock
\newblock
\showeprint[arxiv]{2311.05608}~[cs.CR]
\urldef\tempurl%
\url{https://arxiv.org/abs/2311.05608}
\showURL{%
\tempurl}


\bibitem[{Google Doodle Team}(2019)]%
        {google_bach}
\bibfield{author}{\bibinfo{person}{{Google Doodle Team}}.} \bibinfo{year}{2019}\natexlab{}.
\newblock \bibinfo{title}{Celebrating {Johann Sebastian Bach} Doodle - {Google} Doodles}.
\newblock
\newblock
\urldef\tempurl%
\url{https://doodles.google/doodle/celebrating-johann-sebastian-bach/}
\showURL{%
\tempurl}


\bibitem[Habib et~al\mbox{.}(2019)]%
        {habib2019semi}
\bibfield{author}{\bibinfo{person}{Raza Habib}, \bibinfo{person}{Soroosh Mariooryad}, \bibinfo{person}{Matt Shannon}, \bibinfo{person}{Eric Battenberg}, \bibinfo{person}{RJ Skerry-Ryan}, \bibinfo{person}{Daisy Stanton}, \bibinfo{person}{David Kao}, {and} \bibinfo{person}{Tom Bagby}.} \bibinfo{year}{2019}\natexlab{}.
\newblock \showarticletitle{Semi-supervised generative modeling for controllable speech synthesis}.
\newblock \bibinfo{journal}{\emph{arXiv preprint arXiv:1910.01709}} (\bibinfo{year}{2019}).
\newblock


\bibitem[Hao et~al\mbox{.}(2024)]%
        {Hao2024SyntheticDI}
\bibfield{author}{\bibinfo{person}{Shuang Hao}, \bibinfo{person}{Wenfeng Han}, \bibinfo{person}{Tao Jiang}, \bibinfo{person}{Yiping Li}, \bibinfo{person}{Haonan Wu}, \bibinfo{person}{Chunlin Zhong}, \bibinfo{person}{Zhangjun Zhou}, {and} \bibinfo{person}{He Tang}.} \bibinfo{year}{2024}\natexlab{}.
\newblock \showarticletitle{Synthetic Data in AI: Challenges, Applications, and Ethical Implications}.
\newblock \bibinfo{journal}{\emph{ArXiv}}  \bibinfo{volume}{abs/2401.01629} (\bibinfo{year}{2024}).
\newblock
\urldef\tempurl%
\url{https://api.semanticscholar.org/CorpusID:266741494}
\showURL{%
\tempurl}


\bibitem[Henderson et~al\mbox{.}(2023)]%
        {henderson2023s}
\bibfield{author}{\bibinfo{person}{Peter Henderson}, \bibinfo{person}{Tatsunori Hashimoto}, {and} \bibinfo{person}{Mark Lemley}.} \bibinfo{year}{2023}\natexlab{}.
\newblock \showarticletitle{Where's the Liability in harmful AI Speech?}
\newblock \bibinfo{journal}{\emph{J. Free Speech L.}}  \bibinfo{volume}{3} (\bibinfo{year}{2023}), \bibinfo{pages}{589}.
\newblock


\bibitem[Hong et~al\mbox{.}(2024)]%
        {hong2024s}
\bibfield{author}{\bibinfo{person}{Rachel Hong}, \bibinfo{person}{William Agnew}, \bibinfo{person}{Tadayoshi Kohno}, {and} \bibinfo{person}{Jamie Morgenstern}.} \bibinfo{year}{2024}\natexlab{}.
\newblock \bibinfo{title}{Who's in and who's out? A case study of multimodal CLIP-filtering in DataComp}.
\newblock
\newblock
\showeprint[arxiv]{2405.08209}~[cs.CY]
\urldef\tempurl%
\url{https://arxiv.org/abs/2405.08209}
\showURL{%
\tempurl}


\bibitem[Hook(2023)]%
        {Hook_2023}
\bibfield{author}{\bibinfo{person}{Steve Hook}.} \bibinfo{year}{2023}\natexlab{}.
\newblock \bibinfo{title}{{AI} hub discord server hit with subpoena by {RIAA}}.
\newblock
\newblock
\urldef\tempurl%
\url{https://www.pcguide.com/ai/ai-hub-discord-server-hit-with-subpoena-by-riaa/}
\showURL{%
\tempurl}


\bibitem[Hoover(2023)]%
        {Hoover}
\bibfield{author}{\bibinfo{person}{Amanda Hoover}.} \bibinfo{year}{2023}\natexlab{}.
\newblock \showarticletitle{Spotify Has an {AI} Music Problem—but Bots Love It}.
\newblock \bibinfo{journal}{\emph{Wired}} (\bibinfo{year}{2023}).
\newblock
\showISSN{1059-1028}
\urldef\tempurl%
\url{https://www.wired.com/story/spotify-ai-music-robot-listeners/}
\showURL{%
\tempurl}


\bibitem[Hope et~al\mbox{.}(2023)]%
        {hopenonbinary}
\bibfield{author}{\bibinfo{person}{Maxwell Hope}, \bibinfo{person}{Charlotte Ward}, {and} \bibinfo{person}{Jason Lilley}.} \bibinfo{year}{2023}\natexlab{}.
\newblock \showarticletitle{Nonbinary American English speakers encode gender in vowel acoustics}. In \bibinfo{booktitle}{\emph{Proceedings of Interspeech}}. \bibinfo{publisher}{ISCA}, \bibinfo{address}{Dublin, Ireland}, \bibinfo{pages}{4713--4717}.
\newblock


\bibitem[Hoskins(2024)]%
        {umgtiktok}
\bibfield{author}{\bibinfo{person}{Peter Hoskins}.} \bibinfo{year}{2024}\natexlab{}.
\newblock \showarticletitle{Universal Music to pull songs from TikTok}.
\newblock \bibinfo{journal}{\emph{BBC}} (\bibinfo{year}{2024}).
\newblock
\urldef\tempurl%
\url{https://www.bbc.com/news/business-68150311}
\showURL{%
\tempurl}


\bibitem[Huang et~al\mbox{.}(2017)]%
        {huang2017counterpoint}
\bibfield{author}{\bibinfo{person}{Cheng-Zhi~Anna Huang}, \bibinfo{person}{Tim Cooijmans}, \bibinfo{person}{Adam Roberts}, \bibinfo{person}{Aaron Courville}, {and} \bibinfo{person}{Douglas Eck}.} \bibinfo{year}{2017}\natexlab{}.
\newblock \showarticletitle{Counterpoint by Convolution}. In \bibinfo{booktitle}{\emph{International Society for Music Information Retrieval (ISMIR)}}. \bibinfo{publisher}{International Society for Music Information Retrieval}, \bibinfo{address}{Suzhou, China}, \bibinfo{pages}{211--218}.
\newblock


\bibitem[Huang et~al\mbox{.}(2018)]%
        {huang2018music}
\bibfield{author}{\bibinfo{person}{Cheng-Zhi~Anna Huang}, \bibinfo{person}{Ashish Vaswani}, \bibinfo{person}{Jakob Uszkoreit}, \bibinfo{person}{Noam Shazeer}, \bibinfo{person}{Ian Simon}, \bibinfo{person}{Curtis Hawthorne}, \bibinfo{person}{Andrew~M Dai}, \bibinfo{person}{Matthew~D Hoffman}, \bibinfo{person}{Monica Dinculescu}, {and} \bibinfo{person}{Douglas Eck}.} \bibinfo{year}{2018}\natexlab{}.
\newblock \showarticletitle{Music transformer}.
\newblock \bibinfo{journal}{\emph{arXiv preprint arXiv:1809.04281}} (\bibinfo{year}{2018}).
\newblock


\bibitem[Hutiri et~al\mbox{.}(2024)]%
        {HutiriHarms24}
\bibfield{author}{\bibinfo{person}{Wiebke Hutiri}, \bibinfo{person}{Orestis Papakyriakopoulos}, {and} \bibinfo{person}{Alice Xiang}.} \bibinfo{year}{2024}\natexlab{}.
\newblock \showarticletitle{Not My Voice! A Taxonomy of Ethical and Safety Harms of Speech Generators}. In \bibinfo{booktitle}{\emph{Proceedings of the 2024 ACM Conference on Fairness, Accountability, and Transparency}} (Rio de Janeiro, Brazil) \emph{(\bibinfo{series}{FAccT '24})}. \bibinfo{publisher}{Association for Computing Machinery}, \bibinfo{address}{New York, NY, USA}, \bibinfo{pages}{359–376}.
\newblock
\showISBNx{9798400704505}
\urldef\tempurl%
\url{https://doi.org/10.1145/3630106.3658911}
\showDOI{\tempurl}


\bibitem[Ito and Johnson(2017)]%
        {ljspeech17}
\bibfield{author}{\bibinfo{person}{Keith Ito} {and} \bibinfo{person}{Linda Johnson}.} \bibinfo{year}{2017}\natexlab{}.
\newblock \bibinfo{title}{The LJ Speech Dataset}.
\newblock \bibinfo{howpublished}{\url{https://keithito.com/LJ-Speech-Dataset/}}.
\newblock


\bibitem[Jacques et~al\mbox{.}(2019)]%
        {RichardCA19}
\bibfield{author}{\bibinfo{person}{Richard Jacques}, \bibinfo{person}{Asbj\o{}rn F\o{}lstad}, \bibinfo{person}{Elizabeth Gerber}, \bibinfo{person}{Jonathan Grudin}, \bibinfo{person}{Ewa Luger}, \bibinfo{person}{Andr\'{e}s Monroy-Hern\'{a}ndez}, {and} \bibinfo{person}{Dakuo Wang}.} \bibinfo{year}{2019}\natexlab{}.
\newblock \showarticletitle{Conversational Agents: Acting on the Wave of Research and Development}. In \bibinfo{booktitle}{\emph{Extended Abstracts of the 2019 CHI Conference on Human Factors in Computing Systems}} (Glasgow, Scotland Uk) \emph{(\bibinfo{series}{CHI EA '19})}. \bibinfo{publisher}{Association for Computing Machinery}, \bibinfo{address}{New York, NY, USA}, \bibinfo{pages}{1–8}.
\newblock
\showISBNx{9781450359719}
\urldef\tempurl%
\url{https://doi.org/10.1145/3290607.3299034}
\showDOI{\tempurl}


\bibitem[Johnson(2023)]%
        {spotify_ai_removal}
\bibfield{author}{\bibinfo{person}{Arianna Johnson}.} \bibinfo{year}{2023}\natexlab{}.
\newblock \showarticletitle{Spotify Removes ‘Tens Of Thousands’ Of AI-Generated Songs: Here’s Why}.
\newblock \bibinfo{journal}{\emph{Forbes}} (\bibinfo{year}{2023}).
\newblock
\urldef\tempurl%
\url{https://www.forbes.com/sites/ariannajohnson/2023/05/09/spotify-removes-tens-of-thousands-of-ai-generated-songs-heres-why/?sh=601d69624f4a}
\showURL{%
\tempurl}


\bibitem[Johnson(2020)]%
        {johnson2020MIT}
\bibfield{author}{\bibinfo{person}{Khari Johnson}.} \bibinfo{year}{2020}\natexlab{}.
\newblock \showarticletitle{{MIT} takes down 80 Million Tiny Images data set due to racist and offensive content}.
\newblock \bibinfo{journal}{\emph{VentureBeat}} (\bibinfo{year}{2020}).
\newblock
\urldef\tempurl%
\url{https://venturebeat.com/ai/mit-takes-down-80-million-tiny-images-data-set-due-to-racist-and-offensive-content/}
\showURL{%
\tempurl}


\bibitem[{Kahn} et~al\mbox{.}(2020)]%
        {librilight}
\bibfield{author}{\bibinfo{person}{J. {Kahn}}, \bibinfo{person}{M. {Rivière}}, \bibinfo{person}{W. {Zheng}}, \bibinfo{person}{E. {Kharitonov}}, \bibinfo{person}{Q. {Xu}}, \bibinfo{person}{P.~E. {Mazaré}}, \bibinfo{person}{J. {Karadayi}}, \bibinfo{person}{V. {Liptchinsky}}, \bibinfo{person}{R. {Collobert}}, \bibinfo{person}{C. {Fuegen}}, \bibinfo{person}{T. {Likhomanenko}}, \bibinfo{person}{G. {Synnaeve}}, \bibinfo{person}{A. {Joulin}}, \bibinfo{person}{A. {Mohamed}}, {and} \bibinfo{person}{E. {Dupoux}}.} \bibinfo{year}{2020}\natexlab{}.
\newblock \showarticletitle{Libri-Light: A Benchmark for ASR with Limited or No Supervision}. In \bibinfo{booktitle}{\emph{ICASSP 2020 - 2020 IEEE International Conference on Acoustics, Speech and Signal Processing (ICASSP)}}. \bibinfo{publisher}{IEEE}, \bibinfo{address}{Barcelona, Spain}, \bibinfo{pages}{7669--7673}.
\newblock
\newblock
\shownote{\url{https://github.com/facebookresearch/libri-light}}.


\bibitem[Kelley and Dickerson(2020)]%
        {kelley2020review}
\bibfield{author}{\bibinfo{person}{Troy Kelley} {and} \bibinfo{person}{Kelly Dickerson}.} \bibinfo{year}{2020}\natexlab{}.
\newblock \showarticletitle{A review of artificial intelligence ({AI}) algorithms for sound classification: Implications for human-robot interaction (hri)}.
\newblock \bibinfo{journal}{\emph{Defense Technical Information Center (DTIC)}} (\bibinfo{year}{2020}).
\newblock


\bibitem[Kendall and Farrington(2023)]%
        {kendall2023corpus}
\bibfield{author}{\bibinfo{person}{Tyler Kendall} {and} \bibinfo{person}{Charlie Farrington}.} \bibinfo{year}{2023}\natexlab{}.
\newblock \bibinfo{title}{The Corpus of Regional {African American} Language. Version 2023.06. Eugene, Ore.: The Online Resources for African American Language Project}.
\newblock
\newblock


\bibitem[Khosrowi et~al\mbox{.}(2023)]%
        {khosrowi2023diffusing}
\bibfield{author}{\bibinfo{person}{Donal Khosrowi}, \bibinfo{person}{Finola Finn}, {and} \bibinfo{person}{Elinor Clark}.} \bibinfo{year}{2023}\natexlab{}.
\newblock \showarticletitle{Diffusing the Creator: Attributing Credit for Generative AI Outputs}. In \bibinfo{booktitle}{\emph{Proceedings of the 2023 AAAI/ACM Conference on AI, Ethics, and Society}}. \bibinfo{publisher}{Association for Computing Machinery (ACM)}, \bibinfo{address}{Montreal, Canada}, \bibinfo{pages}{890--900}.
\newblock


\bibitem[Kim et~al\mbox{.}(2020)]%
        {kim2020glow}
\bibfield{author}{\bibinfo{person}{Jaehyeon Kim}, \bibinfo{person}{Sungwon Kim}, \bibinfo{person}{Jungil Kong}, {and} \bibinfo{person}{Sungroh Yoon}.} \bibinfo{year}{2020}\natexlab{}.
\newblock \showarticletitle{Glow-tts: A generative flow for text-to-speech via monotonic alignment search}.
\newblock \bibinfo{journal}{\emph{Advances in Neural Information Processing Systems}}  \bibinfo{volume}{33} (\bibinfo{year}{2020}), \bibinfo{pages}{8067--8077}.
\newblock


\bibitem[Kim et~al\mbox{.}(2018)]%
        {kim2018crepe}
\bibfield{author}{\bibinfo{person}{Jong~Wook Kim}, \bibinfo{person}{Justin Salamon}, \bibinfo{person}{Peter Li}, {and} \bibinfo{person}{Juan~Pablo Bello}.} \bibinfo{year}{2018}\natexlab{}.
\newblock \showarticletitle{Crepe: A convolutional representation for pitch estimation}. In \bibinfo{booktitle}{\emph{2018 IEEE International Conference on Acoustics, Speech and Signal Processing (ICASSP)}}. IEEE, \bibinfo{publisher}{IEEE}, \bibinfo{address}{Calgary, Canada}, \bibinfo{pages}{161--165}.
\newblock


\bibitem[Kim et~al\mbox{.}(2022)]%
        {kim2022guided}
\bibfield{author}{\bibinfo{person}{Sungwon Kim}, \bibinfo{person}{Heeseung Kim}, {and} \bibinfo{person}{Sungroh Yoon}.} \bibinfo{year}{2022}\natexlab{}.
\newblock \showarticletitle{Guided-tts 2: A diffusion model for high-quality adaptive text-to-speech with untranscribed data}.
\newblock \bibinfo{journal}{\emph{arXiv preprint arXiv:2205.15370}} (\bibinfo{year}{2022}).
\newblock


\bibitem[Kleinberg and Raghavan(2021)]%
        {kleinberg2021algorithmic}
\bibfield{author}{\bibinfo{person}{Jon Kleinberg} {and} \bibinfo{person}{Manish Raghavan}.} \bibinfo{year}{2021}\natexlab{}.
\newblock \showarticletitle{Algorithmic monoculture and social welfare}.
\newblock \bibinfo{journal}{\emph{Proceedings of the National Academy of Sciences}} \bibinfo{volume}{118}, \bibinfo{number}{22} (\bibinfo{date}{June} \bibinfo{year}{2021}), \bibinfo{pages}{e2018340118}.
\newblock
\urldef\tempurl%
\url{https://doi.org/10.1073/pnas.2018340118}
\showDOI{\tempurl}


\bibitem[Koenecke et~al\mbox{.}(2024)]%
        {koenecke2024careless}
\bibfield{author}{\bibinfo{person}{Allison Koenecke}, \bibinfo{person}{Anna Seo~Gyeong Choi}, \bibinfo{person}{Katelyn~X Mei}, \bibinfo{person}{Hilke Schellmann}, {and} \bibinfo{person}{Mona Sloane}.} \bibinfo{year}{2024}\natexlab{}.
\newblock \showarticletitle{Careless Whisper: Speech-to-Text Hallucination Harms}. In \bibinfo{booktitle}{\emph{The 2024 ACM Conference on Fairness, Accountability, and Transparency}}. \bibinfo{pages}{1672--1681}.
\newblock


\bibitem[Law et~al\mbox{.}(2009)]%
        {law2009evaluation}
\bibfield{author}{\bibinfo{person}{Edith Law}, \bibinfo{person}{Kris West}, \bibinfo{person}{Michael~I Mandel}, \bibinfo{person}{Mert Bay}, {and} \bibinfo{person}{J~Stephen Downie}.} \bibinfo{year}{2009}\natexlab{}.
\newblock \showarticletitle{Evaluation of algorithms using games: The case of music tagging.}. In \bibinfo{booktitle}{\emph{ISMIR}}. Citeseer, \bibinfo{publisher}{Internaional Society for Music Information Retrieval}, \bibinfo{address}{Kobe, Japan}, \bibinfo{pages}{387--392}.
\newblock


\bibitem[Lee et~al\mbox{.}(2024)]%
        {lee2024deepfakes}
\bibfield{author}{\bibinfo{person}{H.~P. Lee}, \bibinfo{person}{Y.~J. Yang}, \bibinfo{person}{T.~S. Von~Davier}, \bibinfo{person}{J. Forlizzi}, {and} \bibinfo{person}{S. Das}.} \bibinfo{year}{2024}\natexlab{}.
\newblock \showarticletitle{Deepfakes, Phrenology, Surveillance, and More! A Taxonomy of AI Privacy Risks}. In \bibinfo{booktitle}{\emph{Proceedings of the CHI Conference on Human Factors in Computing Systems}}. \bibinfo{publisher}{Association for Computing Machinery (ACM)}, \bibinfo{address}{Honolulu, USA}, \bibinfo{pages}{1--19}.
\newblock


\bibitem[Lee et~al\mbox{.}(2022)]%
        {lee2022ethics}
\bibfield{author}{\bibinfo{person}{Kyungyun Lee}, \bibinfo{person}{Gladys Hitt}, \bibinfo{person}{Emily Terada}, {and} \bibinfo{person}{Jin~Ha Lee}.} \bibinfo{year}{2022}\natexlab{}.
\newblock \showarticletitle{Ethics of Singing Voice Synthesis: Perceptions of Users and Developers.}. In \bibinfo{booktitle}{\emph{Proceedings of the 23rd International Society for Music Information Retrieval Conference (ISMIR 2022)}}. \bibinfo{publisher}{International Society for Music Information Retrieval}, \bibinfo{address}{Bengaluru, India}, \bibinfo{pages}{733--740}.
\newblock


\bibitem[Lemley(2024)]%
        {lemley2024generative}
\bibfield{author}{\bibinfo{person}{Mark Lemley}.} \bibinfo{year}{2024}\natexlab{}.
\newblock \showarticletitle{How Generative AI Turns Copyright Law Upside Down}.
\newblock \bibinfo{journal}{\emph{Science and Technology Law Review}} \bibinfo{volume}{25}, \bibinfo{number}{2} (\bibinfo{year}{2024}).
\newblock


\bibitem[Li et~al\mbox{.}(2021)]%
        {li2021cleanml}
\bibfield{author}{\bibinfo{person}{Peng Li}, \bibinfo{person}{Xi Rao}, \bibinfo{person}{Jennifer Blase}, \bibinfo{person}{Yue Zhang}, \bibinfo{person}{Xu Chu}, {and} \bibinfo{person}{Ce Zhang}.} \bibinfo{year}{2021}\natexlab{}.
\newblock \showarticletitle{Cleanml: A study for evaluating the impact of data cleaning on ml classification tasks}. In \bibinfo{booktitle}{\emph{2021 IEEE 37th International Conference on Data Engineering (ICDE)}}. IEEE, \bibinfo{pages}{13--24}.
\newblock


\bibitem[LibriVox({[n.\,d.]})]%
        {librivox}
\bibfield{author}{\bibinfo{person}{LibriVox}.} \bibinfo{year}{[n.\,d.]}\natexlab{}.
\newblock
\newblock
\urldef\tempurl%
\url{https://librivox.org/pages/about-librivox/}
\showURL{%
\tempurl}


\bibitem[Liu et~al\mbox{.}(2023a)]%
        {liu2023audioldm}
\bibfield{author}{\bibinfo{person}{Haohe Liu}, \bibinfo{person}{Zehua Chen}, \bibinfo{person}{Yi Yuan}, \bibinfo{person}{Xinhao Mei}, \bibinfo{person}{Xubo Liu}, \bibinfo{person}{Danilo Mandic}, \bibinfo{person}{Wenwu Wang}, {and} \bibinfo{person}{Mark~D. Plumbley}.} \bibinfo{year}{2023}\natexlab{a}.
\newblock \bibinfo{title}{AudioLDM: Text-to-Audio Generation with Latent Diffusion Models}.
\newblock
\newblock
\showeprint[arxiv]{2301.12503}~[cs.SD]
\urldef\tempurl%
\url{https://arxiv.org/abs/2301.12503}
\showURL{%
\tempurl}


\bibitem[Liu et~al\mbox{.}(2023b)]%
        {liu2023fast}
\bibfield{author}{\bibinfo{person}{Yisi Liu}, \bibinfo{person}{Peter Wu}, \bibinfo{person}{Alan~W Black}, {and} \bibinfo{person}{Gopala~K Anumanchipalli}.} \bibinfo{year}{2023}\natexlab{b}.
\newblock \showarticletitle{A Fast and Accurate Pitch Estimation Algorithm Based on the Pseudo Wigner-Ville Distribution}. In \bibinfo{booktitle}{\emph{ICASSP 2023-2023 IEEE International Conference on Acoustics, Speech and Signal Processing (ICASSP)}}. IEEE, \bibinfo{publisher}{IEEE}, \bibinfo{address}{Rhodes Island, Greece}, \bibinfo{pages}{1--5}.
\newblock


\bibitem[Lo(2024)]%
        {Lo_2024}
\bibfield{author}{\bibinfo{person}{Clifford Lo}.} \bibinfo{year}{2024}\natexlab{}.
\newblock \bibinfo{title}{Hong Kong worker transfers HK\$4 million after call from deepfake {UK} firm {‘CFO’}}.
\newblock
\newblock
\urldef\tempurl%
\url{https://www.scmp.com/news/hong-kong/law-and-crime/article/3264684/hong-kong-employee-tricked-paying-out-hk4-million-after-video-call-deepfake-chief-financial-officer}
\showURL{%
\tempurl}


\bibitem[Longpre et~al\mbox{.}(2024)]%
        {longpre2024consent}
\bibfield{author}{\bibinfo{person}{Shayne Longpre}, \bibinfo{person}{Robert Mahari}, \bibinfo{person}{Ariel Lee}, \bibinfo{person}{Campbell Lund}, \bibinfo{person}{Hamidah Oderinwale}, \bibinfo{person}{William Brannon}, \bibinfo{person}{Nayan Saxena}, \bibinfo{person}{Naana Obeng-Marnu}, \bibinfo{person}{Tobin South}, \bibinfo{person}{Cole Hunter}, {et~al\mbox{.}}} \bibinfo{year}{2024}\natexlab{}.
\newblock \showarticletitle{Consent in Crisis: The Rapid Decline of the AI Data Commons}.
\newblock \bibinfo{journal}{\emph{arXiv preprint arXiv:2407.14933}} (\bibinfo{year}{2024}).
\newblock


\bibitem[Lu et~al\mbox{.}(2020)]%
        {lu2020gender}
\bibfield{author}{\bibinfo{person}{Kaiji Lu}, \bibinfo{person}{Piotr Mardziel}, \bibinfo{person}{Fangjing Wu}, \bibinfo{person}{Preetam Amancharla}, {and} \bibinfo{person}{Anupam Datta}.} \bibinfo{year}{2020}\natexlab{}.
\newblock \bibinfo{booktitle}{\emph{Gender Bias in Neural Natural Language Processing}}.
\newblock \bibinfo{publisher}{Springer International Publishing}, \bibinfo{address}{Cham}, \bibinfo{pages}{189--202}.
\newblock
\showISBNx{978-3-030-62077-6}
\urldef\tempurl%
\url{https://doi.org/10.1007/978-3-030-62077-6_14}
\showDOI{\tempurl}


\bibitem[Luccioni et~al\mbox{.}(2023)]%
        {luccioni2023estimating}
\bibfield{author}{\bibinfo{person}{Alexandra~Sasha Luccioni}, \bibinfo{person}{Sylvain Viguier}, {and} \bibinfo{person}{Anne-Laure Ligozat}.} \bibinfo{year}{2023}\natexlab{}.
\newblock \showarticletitle{Estimating the carbon footprint of bloom, a 176b parameter language model}.
\newblock \bibinfo{journal}{\emph{Journal of Machine Learning Research}} \bibinfo{volume}{24}, \bibinfo{number}{253} (\bibinfo{year}{2023}), \bibinfo{pages}{1--15}.
\newblock


\bibitem[Luccioni et~al\mbox{.}(2024)]%
        {luccioni2023power}
\bibfield{author}{\bibinfo{person}{Sasha Luccioni}, \bibinfo{person}{Yacine Jernite}, {and} \bibinfo{person}{Emma Strubell}.} \bibinfo{year}{2024}\natexlab{}.
\newblock \showarticletitle{Power hungry processing: Watts driving the cost of AI deployment?}. In \bibinfo{booktitle}{\emph{The 2024 ACM Conference on Fairness, Accountability, and Transparency}}. \bibinfo{publisher}{Association for Computing Machinery (ACM)}, \bibinfo{address}{Rio de Janeiro, Brazil}, \bibinfo{pages}{85--99}.
\newblock


\bibitem[Lucy et~al\mbox{.}(2024)]%
        {lucy2024aboutme}
\bibfield{author}{\bibinfo{person}{Li Lucy}, \bibinfo{person}{Suchin Gururangan}, \bibinfo{person}{Luca Soldaini}, \bibinfo{person}{Emma Strubell}, \bibinfo{person}{David Bamman}, \bibinfo{person}{Lauren~F. Klein}, {and} \bibinfo{person}{Jesse Dodge}.} \bibinfo{year}{2024}\natexlab{}.
\newblock \bibinfo{title}{AboutMe: Using Self-Descriptions in Webpages to Document the Effects of English Pretraining Data Filters}.
\newblock
\newblock
\showeprint[arxiv]{2401.06408}~[cs.CL]
\urldef\tempurl%
\url{https://arxiv.org/abs/2401.06408}
\showURL{%
\tempurl}


\bibitem[Mahelona et~al\mbox{.}(2023)]%
        {papareo}
\bibfield{author}{\bibinfo{person}{Keoni Mahelona}, \bibinfo{person}{Gianna Leoni}, \bibinfo{person}{Suzanne Duncan}, {and} \bibinfo{person}{Miles Thompson}.} \bibinfo{year}{2023}\natexlab{}.
\newblock \bibinfo{title}{OpenAI's Whisper is another case study in Colonisation}.
\newblock
\newblock
\urldef\tempurl%
\url{https://blog.papareo.nz/whisper-is-another-case-study-in-colonisation/}
\showURL{%
\tempurl}


\bibitem[Manco et~al\mbox{.}(2023)]%
        {manco2023thesong}
\bibfield{author}{\bibinfo{person}{Ilaria Manco}, \bibinfo{person}{Benno Weck}, \bibinfo{person}{Seungheon Doh}, \bibinfo{person}{Minz Won}, \bibinfo{person}{Yixiao Zhang}, \bibinfo{person}{Dmitry Bogdanov}, \bibinfo{person}{Yusong Wu}, \bibinfo{person}{Ke Chen}, \bibinfo{person}{Philip Tovstogan}, \bibinfo{person}{Emmanouil Benetos}, \bibinfo{person}{Elio Quinton}, \bibinfo{person}{György Fazekas}, {and} \bibinfo{person}{Juhan Nam}.} \bibinfo{year}{2023}\natexlab{}.
\newblock \showarticletitle{The Song Describer Dataset: a Corpus of Audio Captions for Music-and-Language Evaluation}. In \bibinfo{booktitle}{\emph{Machine Learning for Audio Workshop at NeurIPS 2023}}.
\newblock


\bibitem[Marx(2024)]%
        {paris}
\bibfield{author}{\bibinfo{person}{Paris Marx}.} \bibinfo{year}{2024}\natexlab{}.
\newblock \showarticletitle{How artists are fighting generative AI}.
\newblock  (\bibinfo{year}{2024}).
\newblock
\urldef\tempurl%
\url{https://disconnect.blog/how-artists-are-fighting-generative-ai/}
\showURL{%
\tempurl}


\bibitem[Mehrish et~al\mbox{.}(2023)]%
        {mehrish2023review}
\bibfield{author}{\bibinfo{person}{Ambuj Mehrish}, \bibinfo{person}{Navonil Majumder}, \bibinfo{person}{Rishabh Bharadwaj}, \bibinfo{person}{Rada Mihalcea}, {and} \bibinfo{person}{Soujanya Poria}.} \bibinfo{year}{2023}\natexlab{}.
\newblock \showarticletitle{A review of deep learning techniques for speech processing}.
\newblock \bibinfo{journal}{\emph{Information Fusion}}  \bibinfo{volume}{99} (\bibinfo{year}{2023}), \bibinfo{pages}{101869}.
\newblock


\bibitem[Milmo(2024)]%
        {Milmo_2024}
\bibfield{author}{\bibinfo{person}{Dan Milmo}.} \bibinfo{year}{2024}\natexlab{}.
\newblock \showarticletitle{UK engineering firm Arup falls victim to £20m deepfake scam}.
\newblock \bibinfo{journal}{\emph{The Guardian}} (\bibinfo{date}{May} \bibinfo{year}{2024}).
\newblock
\showISSN{0261-3077}
\urldef\tempurl%
\url{https://www.theguardian.com/technology/article/2024/may/17/uk-engineering-arup-deepfake-scam-hong-kong-ai-video}
\showURL{%
\tempurl}


\bibitem[Moher et~al\mbox{.}(2015)]%
        {moher2015preferred}
\bibfield{author}{\bibinfo{person}{David Moher}, \bibinfo{person}{Larissa Shamseer}, \bibinfo{person}{Mike Clarke}, \bibinfo{person}{Davina Ghersi}, \bibinfo{person}{Alessandro Liberati}, \bibinfo{person}{Mark Petticrew}, \bibinfo{person}{Paul Shekelle}, \bibinfo{person}{Lesley~A Stewart}, {and} \bibinfo{person}{Prisma-P Group}.} \bibinfo{year}{2015}\natexlab{}.
\newblock \showarticletitle{Preferred reporting items for systematic review and meta-analysis protocols (PRISMA-P) 2015 statement}.
\newblock \bibinfo{journal}{\emph{Systematic reviews}}  \bibinfo{volume}{4} (\bibinfo{year}{2015}), \bibinfo{pages}{1--9}.
\newblock


\bibitem[Morreale et~al\mbox{.}(2023)]%
        {morreale2023data}
\bibfield{author}{\bibinfo{person}{Fabio Morreale}, \bibinfo{person}{Megha Sharma}, {and} \bibinfo{person}{I-Chieh Wei}.} \bibinfo{year}{2023}\natexlab{}.
\newblock \showarticletitle{Data Collection in Music Generation Training Sets: A Critical Analysis}. In \bibinfo{booktitle}{\emph{Ismir 2023 Hybrid Conference}}. \bibinfo{publisher}{International Society for Music Information Retrieval}, \bibinfo{address}{Milan, Italy}, \bibinfo{pages}{37--46}.
\newblock


\bibitem[M{\"u}hlbach and Arora(2020)]%
        {muhlbach2020behind}
\bibfield{author}{\bibinfo{person}{Saskia M{\"u}hlbach} {and} \bibinfo{person}{Payal Arora}.} \bibinfo{year}{2020}\natexlab{}.
\newblock \showarticletitle{Behind the music: How labor changed for musicians through the subscription economy}.
\newblock \bibinfo{journal}{\emph{First Monday}} \bibinfo{volume}{25}, \bibinfo{number}{4} (\bibinfo{year}{2020}).
\newblock


\bibitem[Mulvey(2013)]%
        {mulvey2013visual}
\bibfield{author}{\bibinfo{person}{Laura Mulvey}.} \bibinfo{year}{2013}\natexlab{}.
\newblock \showarticletitle{Visual pleasure and narrative cinema}.
\newblock In \bibinfo{booktitle}{\emph{Feminism and film theory}}. \bibinfo{publisher}{Routledge}, \bibinfo{address}{London, UK}, \bibinfo{pages}{57--68}.
\newblock


\bibitem[Murphy(2024)]%
        {Murphy_2024}
\bibfield{author}{\bibinfo{person}{Margi Murphy}.} \bibinfo{year}{2024}\natexlab{}.
\newblock \showarticletitle{Deepfake Audio of {Biden} Alarms Experts in Lead-Up to US Elections}.
\newblock \bibinfo{journal}{\emph{Bloomberg.com}} (\bibinfo{date}{Jan.} \bibinfo{year}{2024}).
\newblock
\urldef\tempurl%
\url{https://www.bloomberg.com/news/articles/2024-01-23/fake-biden-robocall-message-in-new-hampshire-alarms-election-experts}
\showURL{%
\tempurl}


\bibitem[Nacimiento-Garc{\'\i}a et~al\mbox{.}(2024)]%
        {nacimiento2024gender}
\bibfield{author}{\bibinfo{person}{Eduardo Nacimiento-Garc{\'\i}a}, \bibinfo{person}{Holi~Sunya D{\'\i}az-Kaas-Nielsen}, {and} \bibinfo{person}{Carina~S Gonz{\'a}lez-Gonz{\'a}lez}.} \bibinfo{year}{2024}\natexlab{}.
\newblock \showarticletitle{Gender and Accent Biases in AI-Based Tools for Spanish: A Comparative Study between Alexa and Whisper}.
\newblock \bibinfo{journal}{\emph{Applied Sciences}} \bibinfo{volume}{14}, \bibinfo{number}{11} (\bibinfo{year}{2024}), \bibinfo{pages}{4734}.
\newblock


\bibitem[Nest(2015)]%
        {echno_nest_license}
\bibfield{author}{\bibinfo{person}{The~Echo Nest}.} \bibinfo{year}{2015}\natexlab{}.
\newblock \bibinfo{title}{Terms of Service}.
\newblock
\newblock
\urldef\tempurl%
\url{https://web.archive.org/web/20160307235643/http://developer.echonest.com/terms}
\showURL{%
\tempurl}


\bibitem[News(2024)]%
        {fox59_ai2024}
\bibfield{author}{\bibinfo{person}{Fox~59 News}.} \bibinfo{year}{2024}\natexlab{}.
\newblock \bibinfo{title}{A.I. Bringing Back the 'Grandparent Scam'}.
\newblock
\newblock
\urldef\tempurl%
\url{https://fox59.com/news/a-i-bringing-back-the-grandparent-scam/}
\showURL{%
\tempurl}


\bibitem[Newton-Rex(2024a)]%
        {Newton-Rex_2024_Suno}
\bibfield{author}{\bibinfo{person}{Ed Newton-Rex}.} \bibinfo{year}{2024}\natexlab{a}.
\newblock \bibinfo{title}{Suno is a music {AI} company aiming to generate \$120 billion per year. But is it trained on copyrighted recordings?}
\newblock
\newblock
\urldef\tempurl%
\url{https://www.musicbusinessworldwide.com/suno-is-a-music-ai-company-aiming-to-generate-120-billion-per-year-newton-rex/}
\showURL{%
\tempurl}


\bibitem[Newton-Rex(2024b)]%
        {Newton-Rex_2024_Udio}
\bibfield{author}{\bibinfo{person}{Ed Newton-Rex}.} \bibinfo{year}{2024}\natexlab{b}.
\newblock \bibinfo{title}{Yes… Udio’s output resembles copyrighted music, too.}
\newblock
\newblock
\urldef\tempurl%
\url{https://www.musicbusinessworldwide.com/yes-udios-output-resembles-copyrighted-music-too/}
\showURL{%
\tempurl}


\bibitem[Nicolas and Skinner(2012)]%
        {nicolas2012s}
\bibfield{author}{\bibinfo{person}{Gandalf Nicolas} {and} \bibinfo{person}{Allison~Louise Skinner}.} \bibinfo{year}{2012}\natexlab{}.
\newblock \showarticletitle{“That's So Gay!” Priming the General Negative Usage of the Word Gay Increases Implicit Anti-Gay Bias}.
\newblock \bibinfo{journal}{\emph{The Journal of social psychology}} \bibinfo{volume}{152}, \bibinfo{number}{5} (\bibinfo{year}{2012}), \bibinfo{pages}{654--658}.
\newblock


\bibitem[Nogueira et~al\mbox{.}(2022)]%
        {nogueira2022sound}
\bibfield{author}{\bibinfo{person}{Ana Filipa~Rodrigues Nogueira}, \bibinfo{person}{Hugo~S Oliveira}, \bibinfo{person}{Jose~JM Machado}, {and} \bibinfo{person}{Joao Manuel~RS Tavares}.} \bibinfo{year}{2022}\natexlab{}.
\newblock \showarticletitle{Sound classification and processing of urban environments: A systematic literature review}.
\newblock \bibinfo{journal}{\emph{Sensors}} \bibinfo{volume}{22}, \bibinfo{number}{22} (\bibinfo{year}{2022}), \bibinfo{pages}{8608}.
\newblock


\bibitem[of~California(2024)]%
        {public_domain}
\bibfield{author}{\bibinfo{person}{The~University of California}.} \bibinfo{year}{2024}\natexlab{}.
\newblock \bibinfo{title}{The public domain}.
\newblock
\newblock
\urldef\tempurl%
\url{https://copyright.universityofcalifornia.edu/use/public-domain.html#:~:text=After%20March%201%2C%201989%2C%20all,as%20of%20January%201%2C%202014.}
\showURL{%
\tempurl}


\bibitem[Ojewale et~al\mbox{.}(2024)]%
        {ojewale2024towards}
\bibfield{author}{\bibinfo{person}{Victor Ojewale}, \bibinfo{person}{Ryan Steed}, \bibinfo{person}{Briana Vecchione}, \bibinfo{person}{Abeba Birhane}, {and} \bibinfo{person}{Inioluwa~Deborah Raji}.} \bibinfo{year}{2024}\natexlab{}.
\newblock \showarticletitle{Towards AI Accountability Infrastructure: Gaps and Opportunities in AI Audit Tooling}.
\newblock \bibinfo{journal}{\emph{arXiv preprint arXiv:2402.17861}} (\bibinfo{year}{2024}).
\newblock


\bibitem[Oliveira et~al\mbox{.}(2023)]%
        {Cmltts2023}
\bibfield{author}{\bibinfo{person}{Frederico~S. Oliveira}, \bibinfo{person}{Edresson Casanova}, \bibinfo{person}{Arnaldo~Candido Junior}, \bibinfo{person}{Anderson~S. Soares}, {and} \bibinfo{person}{Arlindo~R. Galv{\~a}o~Filho}.} \bibinfo{year}{2023}\natexlab{}.
\newblock \showarticletitle{CML-TTS: A Multilingual Dataset for Speech Synthesis in Low-Resource Languages}. In \bibinfo{booktitle}{\emph{Text, Speech, and Dialogue}}, \bibfield{editor}{\bibinfo{person}{Kamil Ek{\v{s}}tein}, \bibinfo{person}{Franti{\v{s}}ek P{\'a}rtl}, {and} \bibinfo{person}{Miloslav Konop{\'i}k}} (Eds.). \bibinfo{publisher}{Springer Nature Switzerland}, \bibinfo{address}{Cham}, \bibinfo{pages}{188--199}.
\newblock
\showISBNx{978-3-031-40498-6}


\bibitem[O’Reilly et~al\mbox{.}(2024)]%
        {o2024maskmark}
\bibfield{author}{\bibinfo{person}{Patrick O’Reilly}, \bibinfo{person}{Zeyu Jin}, \bibinfo{person}{Jiaqi Su}, {and} \bibinfo{person}{Bryan Pardo}.} \bibinfo{year}{2024}\natexlab{}.
\newblock \showarticletitle{Maskmark: Robust Neural watermarking for Real and Synthetic Speech}. In \bibinfo{booktitle}{\emph{ICASSP 2024-2024 IEEE International Conference on Acoustics, Speech and Signal Processing (ICASSP)}}. IEEE, \bibinfo{publisher}{IEEE}, \bibinfo{address}{Seoul, South Korea}, \bibinfo{pages}{4650--4654}.
\newblock


\bibitem[Palaniappan et~al\mbox{.}(2014)]%
        {palaniappan2014artificial}
\bibfield{author}{\bibinfo{person}{Rajkumar Palaniappan}, \bibinfo{person}{Kenneth Sundaraj}, {and} \bibinfo{person}{Sebastian Sundaraj}.} \bibinfo{year}{2014}\natexlab{}.
\newblock \showarticletitle{Artificial intelligence techniques used in respiratory sound analysis--a systematic review}.
\newblock \bibinfo{journal}{\emph{Biomedizinische Technik/Biomedical Engineering}} \bibinfo{volume}{59}, \bibinfo{number}{1} (\bibinfo{year}{2014}), \bibinfo{pages}{7--18}.
\newblock


\bibitem[Panayotov et~al\mbox{.}(2015)]%
        {panayotov2015librispeech}
\bibfield{author}{\bibinfo{person}{Vassil Panayotov}, \bibinfo{person}{Guoguo Chen}, \bibinfo{person}{Daniel Povey}, {and} \bibinfo{person}{Sanjeev Khudanpur}.} \bibinfo{year}{2015}\natexlab{}.
\newblock \showarticletitle{Librispeech: an {ASR} corpus based on public domain audio books}. In \bibinfo{booktitle}{\emph{2015 IEEE international conference on acoustics, speech and signal processing (ICASSP)}}. IEEE, \bibinfo{publisher}{IEEE}, \bibinfo{address}{Brisbane, Australia}, \bibinfo{pages}{5206--5210}.
\newblock


\bibitem[Patel(2023)]%
        {Patel_2023}
\bibfield{author}{\bibinfo{person}{Nilay Patel}.} \bibinfo{year}{2023}\natexlab{}.
\newblock \bibinfo{title}{Google and {YouTube} are trying to have it both ways with {AI} and copyright - The Verge}.
\newblock
\newblock
\urldef\tempurl%
\url{https://www.theverge.com/2023/8/22/23841822/google-youtube-ai-copyright-umg-scraping-universal}
\showURL{%
\tempurl}


\bibitem[Paullada et~al\mbox{.}(2021)]%
        {paullada2021data}
\bibfield{author}{\bibinfo{person}{Amandalynne Paullada}, \bibinfo{person}{Inioluwa~Deborah Raji}, \bibinfo{person}{Emily~M Bender}, \bibinfo{person}{Emily Denton}, {and} \bibinfo{person}{Alex Hanna}.} \bibinfo{year}{2021}\natexlab{}.
\newblock \showarticletitle{Data and its (dis) contents: A survey of dataset development and use in machine learning research}.
\newblock \bibinfo{journal}{\emph{Patterns}} \bibinfo{volume}{2}, \bibinfo{number}{11} (\bibinfo{year}{2021}), \bibinfo{numpages}{14}~pages.
\newblock


\bibitem[Peng et~al\mbox{.}(2024)]%
        {peng2024voicecraft}
\bibfield{author}{\bibinfo{person}{Puyuan Peng}, \bibinfo{person}{Po-Yao Huang}, \bibinfo{person}{Shang-Wen Li}, \bibinfo{person}{Abdelrahman Mohamed}, {and} \bibinfo{person}{David Harwath}.} \bibinfo{year}{2024}\natexlab{}.
\newblock \bibinfo{title}{VoiceCraft: Zero-Shot Speech Editing and Text-to-Speech in the Wild}.
\newblock
\newblock
\showeprint[arxiv]{2403.16973}~[eess.AS]
\urldef\tempurl%
\url{https://arxiv.org/abs/2403.16973}
\showURL{%
\tempurl}


\bibitem[Peracha(2022)]%
        {peracha2021js}
\bibfield{author}{\bibinfo{person}{Omar Peracha}.} \bibinfo{year}{2022}\natexlab{}.
\newblock \bibinfo{title}{JS Fake Chorales: a Synthetic Dataset of Polyphonic Music with Human Annotation}.
\newblock
\newblock
\showeprint[arxiv]{2107.10388}~[cs.SD]
\urldef\tempurl%
\url{https://arxiv.org/abs/2107.10388}
\showURL{%
\tempurl}


\bibitem[Prabhu and Birhane(2021)]%
        {prabhu2021large}
\bibfield{author}{\bibinfo{person}{Vinay~Uday Prabhu} {and} \bibinfo{person}{Abeba Birhane}.} \bibinfo{year}{2021}\natexlab{}.
\newblock \showarticletitle{Large datasets: A pyrrhic win for computer vision}. In \bibinfo{booktitle}{\emph{Institute of Electrical and Electronics Engineers/Computer Vision Foundation Conference on Applications of Computer Vision}}. \bibinfo{publisher}{IEEE}, \bibinfo{address}{Virtual}.
\newblock


\bibitem[Pratap et~al\mbox{.}(2020)]%
        {Pratap2020MLSAL}
\bibfield{author}{\bibinfo{person}{Vineel Pratap}, \bibinfo{person}{Qiantong Xu}, \bibinfo{person}{Anuroop Sriram}, \bibinfo{person}{Gabriel Synnaeve}, {and} \bibinfo{person}{Ronan Collobert}.} \bibinfo{year}{2020}\natexlab{}.
\newblock \showarticletitle{MLS: A Large-Scale Multilingual Dataset for Speech Research}. In \bibinfo{booktitle}{\emph{Interspeech 2020}}. \bibinfo{publisher}{ISCA}, \bibinfo{address}{Shanghai, China}, \bibinfo{pages}{2757--2761}.
\newblock


\bibitem[Qi et~al\mbox{.}(2024)]%
        {qi2024visual}
\bibfield{author}{\bibinfo{person}{Xiangyu Qi}, \bibinfo{person}{Kaixuan Huang}, \bibinfo{person}{Ashwinee Panda}, \bibinfo{person}{Peter Henderson}, \bibinfo{person}{Mengdi Wang}, {and} \bibinfo{person}{Prateek Mittal}.} \bibinfo{year}{2024}\natexlab{}.
\newblock \showarticletitle{Visual adversarial examples jailbreak aligned large language models}. In \bibinfo{booktitle}{\emph{Proceedings of the AAAI Conference on Artificial Intelligence}}. \bibinfo{publisher}{AAAI}, \bibinfo{address}{Vancouver, Canada}, \bibinfo{pages}{21527--21536}.
\newblock


\bibitem[Queerinai et~al\mbox{.}(2023)]%
        {queerinai2023queer}
\bibfield{author}{\bibinfo{person}{Organizers~Of Queerinai}, \bibinfo{person}{Anaelia Ovalle}, \bibinfo{person}{Arjun Subramonian}, \bibinfo{person}{Ashwin Singh}, \bibinfo{person}{Claas Voelcker}, \bibinfo{person}{Danica~J Sutherland}, \bibinfo{person}{Davide Locatelli}, \bibinfo{person}{Eva Breznik}, \bibinfo{person}{Filip Klubicka}, \bibinfo{person}{Hang Yuan}, {et~al\mbox{.}}} \bibinfo{year}{2023}\natexlab{}.
\newblock \showarticletitle{Queer in {AI}: a case study in community-led participatory {AI}}. In \bibinfo{booktitle}{\emph{Proceedings of the 2023 ACM Conference on Fairness, Accountability, and Transparency}}. \bibinfo{publisher}{Association for Computing Machinery (ACM)}, \bibinfo{address}{Chicago, USA}, \bibinfo{pages}{1882--1895}.
\newblock


\bibitem[Radford et~al\mbox{.}(2023)]%
        {radford2023robust}
\bibfield{author}{\bibinfo{person}{Alec Radford}, \bibinfo{person}{Jong~Wook Kim}, \bibinfo{person}{Tao Xu}, \bibinfo{person}{Greg Brockman}, \bibinfo{person}{Christine McLeavey}, {and} \bibinfo{person}{Ilya Sutskever}.} \bibinfo{year}{2023}\natexlab{}.
\newblock \showarticletitle{Robust speech recognition via large-scale weak supervision}. In \bibinfo{booktitle}{\emph{International Conference on Machine Learning}}. PMLR, \bibinfo{publisher}{PMLR}, \bibinfo{address}{Honolulu, USA}, \bibinfo{pages}{28492--28518}.
\newblock


\bibitem[Raffel(2016)]%
        {raffel2016learning}
\bibfield{author}{\bibinfo{person}{Colin Raffel}.} \bibinfo{year}{2016}\natexlab{}.
\newblock \bibinfo{booktitle}{\emph{Learning-based methods for comparing sequences, with applications to audio-to-midi alignment and matching}}.
\newblock \bibinfo{publisher}{Columbia University}.
\newblock


\bibitem[Raffel et~al\mbox{.}(2020)]%
        {raffel2020exploring}
\bibfield{author}{\bibinfo{person}{Colin Raffel}, \bibinfo{person}{Noam Shazeer}, \bibinfo{person}{Adam Roberts}, \bibinfo{person}{Katherine Lee}, \bibinfo{person}{Sharan Narang}, \bibinfo{person}{Michael Matena}, \bibinfo{person}{Yanqi Zhou}, \bibinfo{person}{Wei Li}, {and} \bibinfo{person}{Peter~J Liu}.} \bibinfo{year}{2020}\natexlab{}.
\newblock \showarticletitle{Exploring the limits of transfer learning with a unified text-to-text transformer}.
\newblock \bibinfo{journal}{\emph{Journal of machine learning research}} \bibinfo{volume}{21}, \bibinfo{number}{140} (\bibinfo{year}{2020}), \bibinfo{pages}{1--67}.
\newblock


\bibitem[Rudinger et~al\mbox{.}(2017)]%
        {rudinger2017social}
\bibfield{author}{\bibinfo{person}{Rachel Rudinger}, \bibinfo{person}{Chandler May}, {and} \bibinfo{person}{Benjamin Van~Durme}.} \bibinfo{year}{2017}\natexlab{}.
\newblock \showarticletitle{Social bias in elicited natural language inferences}. In \bibinfo{booktitle}{\emph{Proceedings of the First ACL Workshop on Ethics in Natural Language Processing}}. \bibinfo{publisher}{Assocation for Computational Linguistics (ACL)}, \bibinfo{address}{Valencia, Spain}, \bibinfo{pages}{74--79}.
\newblock


\bibitem[Samuelson(2023)]%
        {samuelson2023generative}
\bibfield{author}{\bibinfo{person}{Pamela Samuelson}.} \bibinfo{year}{2023}\natexlab{}.
\newblock \showarticletitle{Generative AI meets copyright}.
\newblock \bibinfo{journal}{\emph{Science}} \bibinfo{volume}{381}, \bibinfo{number}{6654} (\bibinfo{year}{2023}), \bibinfo{pages}{158--161}.
\newblock


\bibitem[Sap et~al\mbox{.}(2019)]%
        {sap2019risk}
\bibfield{author}{\bibinfo{person}{Maarten Sap}, \bibinfo{person}{Dallas Card}, \bibinfo{person}{Saadia Gabriel}, \bibinfo{person}{Yejin Choi}, {and} \bibinfo{person}{Noah~A Smith}.} \bibinfo{year}{2019}\natexlab{}.
\newblock \showarticletitle{The risk of racial bias in hate speech detection}. In \bibinfo{booktitle}{\emph{Proceedings of the 57th annual meeting of the association for computational linguistics}}. \bibinfo{publisher}{Association for Computational Linguistics (ACL)}, \bibinfo{address}{Florence, Italy}, \bibinfo{pages}{1668--1678}.
\newblock


\bibitem[Schuhmann et~al\mbox{.}(2022)]%
        {schuhmann2022laion}
\bibfield{author}{\bibinfo{person}{Christoph Schuhmann}, \bibinfo{person}{Romain Beaumont}, \bibinfo{person}{Richard Vencu}, \bibinfo{person}{Cade Gordon}, \bibinfo{person}{Ross Wightman}, \bibinfo{person}{Mehdi Cherti}, \bibinfo{person}{Theo Coombes}, \bibinfo{person}{Aarush Katta}, \bibinfo{person}{Clayton Mullis}, \bibinfo{person}{Mitchell Wortsman}, {et~al\mbox{.}}} \bibinfo{year}{2022}\natexlab{}.
\newblock \showarticletitle{Laion-5b: An open large-scale dataset for training next generation image-text models}.
\newblock \bibinfo{journal}{\emph{Advances in Neural Information Processing Systems}}  \bibinfo{volume}{35} (\bibinfo{year}{2022}), \bibinfo{pages}{25278--25294}.
\newblock


\bibitem[Shelby et~al\mbox{.}(2023)]%
        {shelby2023sociotechnical}
\bibfield{author}{\bibinfo{person}{Renee Shelby}, \bibinfo{person}{Shalaleh Rismani}, \bibinfo{person}{Kathryn Henne}, \bibinfo{person}{AJung Moon}, \bibinfo{person}{Negar Rostamzadeh}, \bibinfo{person}{Paul Nicholas}, \bibinfo{person}{N'Mah Yilla-Akbari}, \bibinfo{person}{Jess Gallegos}, \bibinfo{person}{Andrew Smart}, \bibinfo{person}{Emilio Garcia}, {et~al\mbox{.}}} \bibinfo{year}{2023}\natexlab{}.
\newblock \showarticletitle{Sociotechnical harms of algorithmic systems: Scoping a taxonomy for harm reduction}. In \bibinfo{booktitle}{\emph{Proceedings of the 2023 AAAI/ACM Conference on AI, Ethics, and Society}}. \bibinfo{pages}{723--741}.
\newblock


\bibitem[Shuyo(2014)]%
        {langdetect}
\bibfield{author}{\bibinfo{person}{Nakatani Shuyo}.} \bibinfo{year}{2014}\natexlab{}.
\newblock \bibinfo{title}{langdetect}.
\newblock
\newblock
\urldef\tempurl%
\url{https://github.com/Mimino666/langdetect}
\showURL{%
\tempurl}


\bibitem[Sigurgeirsson and Ungless(2024)]%
        {sigurgeirsson2024just}
\bibfield{author}{\bibinfo{person}{Atli Sigurgeirsson} {and} \bibinfo{person}{Eddie~L Ungless}.} \bibinfo{year}{2024}\natexlab{}.
\newblock \showarticletitle{Just Because We Camp, Doesn't Mean We Should: The Ethics of Modelling Queer Voices}.
\newblock \bibinfo{journal}{\emph{arXiv preprint arXiv:2406.07504}} (\bibinfo{year}{2024}).
\newblock


\bibitem[Sisario(2024)]%
        {Sisario_2024}
\bibfield{author}{\bibinfo{person}{Ben Sisario}.} \bibinfo{year}{2024}\natexlab{}.
\newblock \showarticletitle{Universal Music Group Pulls Songs From TikTok}.
\newblock \bibinfo{journal}{\emph{The New York Times}} (\bibinfo{date}{Feb.} \bibinfo{year}{2024}).
\newblock
\showISSN{0362-4331}
\urldef\tempurl%
\url{https://www.nytimes.com/2024/02/01/arts/music/universal-group-tiktok-music.html}
\showURL{%
\tempurl}


\bibitem[Sisman et~al\mbox{.}(2020)]%
        {sisman2020overview}
\bibfield{author}{\bibinfo{person}{Berrak Sisman}, \bibinfo{person}{Junichi Yamagishi}, \bibinfo{person}{Simon King}, {and} \bibinfo{person}{Haizhou Li}.} \bibinfo{year}{2020}\natexlab{}.
\newblock \showarticletitle{An overview of voice conversion and its challenges: From statistical modeling to deep learning}.
\newblock \bibinfo{journal}{\emph{IEEE/ACM Transactions on Audio, Speech, and Language Processing}}  \bibinfo{volume}{29} (\bibinfo{year}{2020}), \bibinfo{pages}{132--157}.
\newblock


\bibitem[Snyder et~al\mbox{.}(2015)]%
        {musan2015}
\bibfield{author}{\bibinfo{person}{David Snyder}, \bibinfo{person}{Guoguo Chen}, {and} \bibinfo{person}{Daniel Povey}.} \bibinfo{year}{2015}\natexlab{}.
\newblock \bibinfo{title}{{MUSAN}: {A} {M}usic, {S}peech, and {N}oise {C}orpus}.
\newblock
\newblock
\showeprint{1510.08484}
\newblock
\shownote{arXiv:1510.08484v1}.


\bibitem[Soldaini et~al\mbox{.}(2024)]%
        {soldaini2024dolma}
\bibfield{author}{\bibinfo{person}{Luca Soldaini}, \bibinfo{person}{Rodney Kinney}, \bibinfo{person}{Akshita Bhagia}, \bibinfo{person}{Dustin Schwenk}, \bibinfo{person}{David Atkinson}, \bibinfo{person}{Russell Authur}, \bibinfo{person}{Ben Bogin}, \bibinfo{person}{Khyathi Chandu}, \bibinfo{person}{Jennifer Dumas}, \bibinfo{person}{Yanai Elazar}, \bibinfo{person}{Valentin Hofmann}, \bibinfo{person}{Ananya~Harsh Jha}, \bibinfo{person}{Sachin Kumar}, \bibinfo{person}{Li Lucy}, \bibinfo{person}{Xinxi Lyu}, \bibinfo{person}{Nathan Lambert}, \bibinfo{person}{Ian Magnusson}, \bibinfo{person}{Jacob Morrison}, \bibinfo{person}{Niklas Muennighoff}, \bibinfo{person}{Aakanksha Naik}, \bibinfo{person}{Crystal Nam}, \bibinfo{person}{Matthew~E. Peters}, \bibinfo{person}{Abhilasha Ravichander}, \bibinfo{person}{Kyle Richardson}, \bibinfo{person}{Zejiang Shen}, \bibinfo{person}{Emma Strubell}, \bibinfo{person}{Nishant Subramani}, \bibinfo{person}{Oyvind Tafjord}, \bibinfo{person}{Pete Walsh}, \bibinfo{person}{Luke
  Zettlemoyer}, \bibinfo{person}{Noah~A. Smith}, \bibinfo{person}{Hannaneh Hajishirzi}, \bibinfo{person}{Iz Beltagy}, \bibinfo{person}{Dirk Groeneveld}, \bibinfo{person}{Jesse Dodge}, {and} \bibinfo{person}{Kyle Lo}.} \bibinfo{year}{2024}\natexlab{}.
\newblock \bibinfo{title}{Dolma: an Open Corpus of Three Trillion Tokens for Language Model Pretraining Research}.
\newblock
\newblock
\showeprint[arxiv]{2402.00159}~[cs.CL]
\urldef\tempurl%
\url{https://arxiv.org/abs/2402.00159}
\showURL{%
\tempurl}


\bibitem[Sturm et~al\mbox{.}(2019)]%
        {sturm2019machine}
\bibfield{author}{\bibinfo{person}{Bob~L Sturm}, \bibinfo{person}{Oded Ben-Tal}, \bibinfo{person}{{\'U}na Monaghan}, \bibinfo{person}{Nick Collins}, \bibinfo{person}{Dorien Herremans}, \bibinfo{person}{Elaine Chew}, \bibinfo{person}{Ga{\"e}tan Hadjeres}, \bibinfo{person}{Emmanuel Deruty}, {and} \bibinfo{person}{Fran{\c{c}}ois Pachet}.} \bibinfo{year}{2019}\natexlab{}.
\newblock \showarticletitle{Machine learning research that matters for music creation: A case study}.
\newblock \bibinfo{journal}{\emph{Journal of New Music Research}} \bibinfo{volume}{48}, \bibinfo{number}{1} (\bibinfo{year}{2019}), \bibinfo{pages}{36--55}.
\newblock


\bibitem[Tang et~al\mbox{.}(2024)]%
        {tang2024salmonn}
\bibfield{author}{\bibinfo{person}{Changli Tang}, \bibinfo{person}{Wenyi Yu}, \bibinfo{person}{Guangzhi Sun}, \bibinfo{person}{Xianzhao Chen}, \bibinfo{person}{Tian Tan}, \bibinfo{person}{Wei Li}, \bibinfo{person}{Lu Lu}, \bibinfo{person}{Zejun MA}, {and} \bibinfo{person}{Chao Zhang}.} \bibinfo{year}{2024}\natexlab{}.
\newblock \showarticletitle{{SALMONN}: Towards Generic Hearing Abilities for Large Language Models}. In \bibinfo{booktitle}{\emph{The Twelfth International Conference on Learning Representations}}.
\newblock
\urldef\tempurl%
\url{https://openreview.net/forum?id=14rn7HpKVk}
\showURL{%
\tempurl}


\bibitem[Tenbarge(2024)]%
        {Johansson2024}
\bibfield{author}{\bibinfo{person}{Kat Tenbarge}.} \bibinfo{year}{2024}\natexlab{}.
\newblock \showarticletitle{Scarlett Johansson says she is ‘shocked and outraged’ by GPT-4o’s new voice that sounds like her}.
\newblock \bibinfo{journal}{\emph{NBC News}} (\bibinfo{year}{2024}).
\newblock
\newblock
\shownote{\url{https://www.nbcnews.com/tech/tech-news/scarlett-johansson-shocked-angered-openai-voice-rcna153180}}.


\bibitem[Thiel(2023)]%
        {thiel2023identifying}
\bibfield{author}{\bibinfo{person}{David Thiel}.} \bibinfo{year}{2023}\natexlab{}.
\newblock \bibinfo{booktitle}{\emph{Identifying and eliminating csam in generative ml training data and models}}.
\newblock \bibinfo{type}{{T}echnical {R}eport}. \bibinfo{institution}{Technical report, Stanford University, Palo Alto, CA, 2023. URL https://purl~…}.
\newblock


\bibitem[Tomar et~al\mbox{.}(2023)]%
        {TomarSarcasm23}
\bibfield{author}{\bibinfo{person}{Mohit Tomar}, \bibinfo{person}{Abhisek Tiwari}, \bibinfo{person}{Tulika Saha}, {and} \bibinfo{person}{Sriparna Saha}.} \bibinfo{year}{2023}\natexlab{}.
\newblock \showarticletitle{Your tone speaks louder than your face! Modality Order Infused Multi-modal Sarcasm Detection}. In \bibinfo{booktitle}{\emph{Proceedings of the 31st ACM International Conference on Multimedia}} (Ottawa ON, Canada) \emph{(\bibinfo{series}{MM '23})}. \bibinfo{publisher}{Association for Computing Machinery}, \bibinfo{address}{New York, NY, USA}, \bibinfo{pages}{3926–3933}.
\newblock
\showISBNx{9798400701085}
\urldef\tempurl%
\url{https://doi.org/10.1145/3581783.3612528}
\showDOI{\tempurl}


\bibitem[Tong et~al\mbox{.}(2023)]%
        {tong2024mass}
\bibfield{author}{\bibinfo{person}{Shengbang Tong}, \bibinfo{person}{Erik Jones}, {and} \bibinfo{person}{Jacob Steinhardt}.} \bibinfo{year}{2023}\natexlab{}.
\newblock \showarticletitle{Mass-Producing Failures of Multimodal Systems with Language Models}. In \bibinfo{booktitle}{\emph{Advances in Neural Information Processing Systems}}, \bibfield{editor}{\bibinfo{person}{A.~Oh}, \bibinfo{person}{T.~Naumann}, \bibinfo{person}{A.~Globerson}, \bibinfo{person}{K.~Saenko}, \bibinfo{person}{M.~Hardt}, {and} \bibinfo{person}{S.~Levine}} (Eds.), Vol.~\bibinfo{volume}{36}. \bibinfo{publisher}{Curran Associates, Inc.}, \bibinfo{address}{New Orleans, USA}, \bibinfo{pages}{29292--29322}.
\newblock
\urldef\tempurl%
\url{https://proceedings.neurips.cc/paper_files/paper/2023/file/5d570ed1708bbe19cb60f7a7aff60575-Paper-Conference.pdf}
\showURL{%
\tempurl}


\bibitem[Tu et~al\mbox{.}(2023)]%
        {tu2023many}
\bibfield{author}{\bibinfo{person}{Haoqin Tu}, \bibinfo{person}{Chenhang Cui}, \bibinfo{person}{Zijun Wang}, \bibinfo{person}{Yiyang Zhou}, \bibinfo{person}{Bingchen Zhao}, \bibinfo{person}{Junlin Han}, \bibinfo{person}{Wangchunshu Zhou}, \bibinfo{person}{Huaxiu Yao}, {and} \bibinfo{person}{Cihang Xie}.} \bibinfo{year}{2023}\natexlab{}.
\newblock \bibinfo{title}{How Many Unicorns Are in This Image? A Safety Evaluation Benchmark for Vision LLMs}.
\newblock
\newblock
\showeprint[arxiv]{2311.16101}~[cs.CV]
\urldef\tempurl%
\url{https://arxiv.org/abs/2311.16101}
\showURL{%
\tempurl}


\bibitem[UMG Recordings Inc.~v. Uncharted~Labs(2024)]%
        {udio_litigation}
\bibfield{author}{\bibinfo{person}{1:24-cv-04777~(S.D.N.Y.) UMG Recordings Inc.~v. Uncharted~Labs, Inc.}} \bibinfo{year}{2024}\natexlab{}.
\newblock \bibinfo{title}{1:24-cv-04777, (S.D.N.Y.)}.
\newblock
\newblock
\urldef\tempurl%
\url{https://storage.courtlistener.com/recap/gov.uscourts.nysd.623701/gov.uscourts.nysd.623701.1.0.pdf}
\showURL{%
\tempurl}


\bibitem[Upadhyay et~al\mbox{.}(2023)]%
        {UpadhyayOlder24}
\bibfield{author}{\bibinfo{person}{Pooja Upadhyay}, \bibinfo{person}{Sharon Heung}, \bibinfo{person}{Shiri Azenkot}, {and} \bibinfo{person}{Robin~N. Brewer}.} \bibinfo{year}{2023}\natexlab{}.
\newblock \showarticletitle{Studying Exploration \& Long-Term Use of Voice Assistants by Older Adults}. In \bibinfo{booktitle}{\emph{Proceedings of the 2023 CHI Conference on Human Factors in Computing Systems}} (Hamburg, Germany) \emph{(\bibinfo{series}{CHI '23})}. \bibinfo{publisher}{Association for Computing Machinery}, \bibinfo{address}{New York, NY, USA}, Article \bibinfo{articleno}{848}, \bibinfo{numpages}{11}~pages.
\newblock
\showISBNx{9781450394215}
\urldef\tempurl%
\url{https://doi.org/10.1145/3544548.3580925}
\showDOI{\tempurl}


\bibitem[Vaswani(2017)]%
        {vaswani2017attention}
\bibfield{author}{\bibinfo{person}{A Vaswani}.} \bibinfo{year}{2017}\natexlab{}.
\newblock \showarticletitle{Attention is all you need}.
\newblock \bibinfo{journal}{\emph{Advances in Neural Information Processing Systems}} (\bibinfo{year}{2017}).
\newblock


\bibitem[Veaux et~al\mbox{.}(2017)]%
        {Veaux2017CSTRVC}
\bibfield{author}{\bibinfo{person}{Christophe Veaux}, \bibinfo{person}{Junichi Yamagishi}, {and} \bibinfo{person}{Kirsten MacDonald}.} \bibinfo{year}{2017}\natexlab{}.
\newblock \showarticletitle{CSTR VCTK Corpus: English Multi-speaker Corpus for CSTR Voice Cloning Toolkit}.
\newblock


\bibitem[Wang et~al\mbox{.}(2021)]%
        {wang2021voxpopuli}
\bibfield{author}{\bibinfo{person}{Changhan Wang}, \bibinfo{person}{Morgane Riviere}, \bibinfo{person}{Ann Lee}, \bibinfo{person}{Anne Wu}, \bibinfo{person}{Chaitanya Talnikar}, \bibinfo{person}{Daniel Haziza}, \bibinfo{person}{Mary Williamson}, \bibinfo{person}{Juan Pino}, {and} \bibinfo{person}{Emmanuel Dupoux}.} \bibinfo{year}{2021}\natexlab{}.
\newblock \showarticletitle{VoxPopuli: A Large-Scale Multilingual Speech Corpus for Representation Learning, Semi-Supervised Learning and Interpretation}. In \bibinfo{booktitle}{\emph{Proceedings of the 59th Annual Meeting of the Association for Computational Linguistics and the 11th International Joint Conference on Natural Language Processing (Volume 1: Long Papers)}}. \bibinfo{publisher}{Association for Computational Linguistics}, \bibinfo{address}{Online}, \bibinfo{pages}{993--1003}.
\newblock


\bibitem[Wang et~al\mbox{.}(2020)]%
        {wang2020deepsonar}
\bibfield{author}{\bibinfo{person}{Run Wang}, \bibinfo{person}{Felix Juefei-Xu}, \bibinfo{person}{Yihao Huang}, \bibinfo{person}{Qing Guo}, \bibinfo{person}{Xiaofei Xie}, \bibinfo{person}{Lei Ma}, {and} \bibinfo{person}{Yang Liu}.} \bibinfo{year}{2020}\natexlab{}.
\newblock \showarticletitle{Deepsonar: Towards effective and robust detection of ai-synthesized fake voices}. In \bibinfo{booktitle}{\emph{Proceedings of the 28th ACM international conference on multimedia}}. \bibinfo{pages}{1207--1216}.
\newblock


\bibitem[Wang et~al\mbox{.}(2023)]%
        {wang2023audit}
\bibfield{author}{\bibinfo{person}{Yuancheng Wang}, \bibinfo{person}{Zeqian Ju}, \bibinfo{person}{Xu Tan}, \bibinfo{person}{Lei He}, \bibinfo{person}{Zhizheng Wu}, \bibinfo{person}{Jiang Bian}, {and} \bibinfo{person}{Sheng Zhao}.} \bibinfo{year}{2023}\natexlab{}.
\newblock \bibinfo{title}{AUDIT: Audio Editing by Following Instructions with Latent Diffusion Models}.
\newblock
\newblock
\showeprint[arxiv]{2304.00830}~[cs.SD]


\bibitem[Wang et~al\mbox{.}(2022)]%
        {wang2022deep}
\bibfield{author}{\bibinfo{person}{Yaqin Wang}, \bibinfo{person}{Jin Wei-Kocsis}, \bibinfo{person}{John~A Springer}, {and} \bibinfo{person}{Eric~T Matson}.} \bibinfo{year}{2022}\natexlab{}.
\newblock \showarticletitle{Deep learning in audio classification}. In \bibinfo{booktitle}{\emph{International Conference on Information and Software Technologies}}. Springer, \bibinfo{publisher}{Springer}, \bibinfo{address}{Kaunas, Lithuania}, \bibinfo{pages}{64--77}.
\newblock


\bibitem[Weidinger et~al\mbox{.}(2021)]%
        {weidinger2021ethical}
\bibfield{author}{\bibinfo{person}{Laura Weidinger}, \bibinfo{person}{John Mellor}, \bibinfo{person}{Maribeth Rauh}, \bibinfo{person}{Conor Griffin}, \bibinfo{person}{Jonathan Uesato}, \bibinfo{person}{Po-Sen Huang}, \bibinfo{person}{Myra Cheng}, \bibinfo{person}{Mia Glaese}, \bibinfo{person}{Borja Balle}, \bibinfo{person}{Atoosa Kasirzadeh}, \bibinfo{person}{Zac Kenton}, \bibinfo{person}{Sasha Brown}, \bibinfo{person}{Will Hawkins}, \bibinfo{person}{Tom Stepleton}, \bibinfo{person}{Courtney Biles}, \bibinfo{person}{Abeba Birhane}, \bibinfo{person}{Julia Haas}, \bibinfo{person}{Laura Rimell}, \bibinfo{person}{Lisa~Anne Hendricks}, \bibinfo{person}{William Isaac}, \bibinfo{person}{Sean Legassick}, \bibinfo{person}{Geoffrey Irving}, {and} \bibinfo{person}{Iason Gabriel}.} \bibinfo{year}{2021}\natexlab{}.
\newblock \bibinfo{title}{Ethical and social risks of harm from Language Models}.
\newblock
\newblock
\showeprint[arxiv]{2112.04359}~[cs.CL]
\urldef\tempurl%
\url{https://arxiv.org/abs/2112.04359}
\showURL{%
\tempurl}


\bibitem[Wenzel and Kaufman(2024)]%
        {Wenzel2024Values}
\bibfield{author}{\bibinfo{person}{Kimi Wenzel} {and} \bibinfo{person}{Geoff Kaufman}.} \bibinfo{year}{2024}\natexlab{}.
\newblock \showarticletitle{Designing for Harm Reduction: Communication Repair for Multicultural Users' Voice Interactions}. In \bibinfo{booktitle}{\emph{Proceedings of the CHI Conference on Human Factors in Computing Systems}}. \bibinfo{publisher}{Association for Computing Machinery}, \bibinfo{address}{New York, NY, USA}, Article \bibinfo{articleno}{879}, \bibinfo{numpages}{17}~pages.
\newblock
\showISBNx{9798400703300}
\urldef\tempurl%
\url{https://doi.org/10.1145/3613904.3642900}
\showDOI{\tempurl}


\bibitem[Wu et~al\mbox{.}(2023)]%
        {wu2023tunesformer}
\bibfield{author}{\bibinfo{person}{Shangda Wu}, \bibinfo{person}{Xiaobing Li}, \bibinfo{person}{Feng Yu}, {and} \bibinfo{person}{Maosong Sun}.} \bibinfo{year}{2023}\natexlab{}.
\newblock \bibinfo{title}{TunesFormer: Forming Irish Tunes with Control Codes by Bar Patching}.
\newblock
\newblock
\showeprint[arxiv]{2301.02884}~[cs.SD]
\urldef\tempurl%
\url{https://arxiv.org/abs/2301.02884}
\showURL{%
\tempurl}


\bibitem[Wu et~al\mbox{.}(2024)]%
        {WuTTM24}
\bibfield{author}{\bibinfo{person}{Shih-Lun Wu}, \bibinfo{person}{Chris Donahue}, \bibinfo{person}{Shinji Watanabe}, {and} \bibinfo{person}{Nicholas~J. Bryan}.} \bibinfo{year}{2024}\natexlab{}.
\newblock \showarticletitle{Music ControlNet: Multiple Time-Varying Controls for Music Generation}.
\newblock \bibinfo{journal}{\emph{IEEE/ACM Trans. Audio, Speech and Lang. Proc.}}  \bibinfo{volume}{32} (\bibinfo{date}{may} \bibinfo{year}{2024}), \bibinfo{pages}{2692–2703}.
\newblock
\showISSN{2329-9290}
\urldef\tempurl%
\url{https://doi.org/10.1109/TASLP.2024.3399026}
\showDOI{\tempurl}


\bibitem[Yamagishi et~al\mbox{.}(2019)]%
        {yamagishi2019cstr}
\bibfield{author}{\bibinfo{person}{Junichi Yamagishi}, \bibinfo{person}{Christophe Veaux}, \bibinfo{person}{Kirsten MacDonald}, {et~al\mbox{.}}} \bibinfo{year}{2019}\natexlab{}.
\newblock \showarticletitle{Cstr vctk corpus: English multi-speaker corpus for cstr voice cloning toolkit (version 0.92)}.
\newblock \bibinfo{journal}{\emph{University of Edinburgh. The Centre for Speech Technology Research (CSTR)}} (\bibinfo{year}{2019}), \bibinfo{pages}{271--350}.
\newblock


\bibitem[YouTube(2024a)]%
        {yt_cc}
\bibfield{author}{\bibinfo{person}{YouTube}.} \bibinfo{year}{2024}\natexlab{a}.
\newblock \bibinfo{title}{{Creative Commons}}.
\newblock
\newblock
\urldef\tempurl%
\url{https://support.google.com/youtube/answer/2797468?hl=en}
\showURL{%
\tempurl}


\bibitem[YouTube(2024b)]%
        {yt_license}
\bibfield{author}{\bibinfo{person}{YouTube}.} \bibinfo{year}{2024}\natexlab{b}.
\newblock \bibinfo{title}{Terms of Service}.
\newblock
\newblock
\urldef\tempurl%
\url{https://www.youtube.com/t/terms#27dc3bf5d9}
\showURL{%
\tempurl}


\bibitem[Yu et~al\mbox{.}(2023)]%
        {YuInterface23}
\bibfield{author}{\bibinfo{person}{Ja~Eun Yu}, \bibinfo{person}{Natalie Parde}, {and} \bibinfo{person}{Debaleena Chattopadhyay}.} \bibinfo{year}{2023}\natexlab{}.
\newblock \showarticletitle{“Where is history”: Toward Designing a Voice Assistant to help Older Adults locate Interface Features quickly}. In \bibinfo{booktitle}{\emph{Proceedings of the 2023 CHI Conference on Human Factors in Computing Systems}} (Hamburg, Germany) \emph{(\bibinfo{series}{CHI '23})}. \bibinfo{publisher}{Association for Computing Machinery}, \bibinfo{address}{New York, NY, USA}, Article \bibinfo{articleno}{849}, \bibinfo{numpages}{19}~pages.
\newblock
\showISBNx{9781450394215}
\urldef\tempurl%
\url{https://doi.org/10.1145/3544548.3581447}
\showDOI{\tempurl}


\bibitem[Zen et~al\mbox{.}(2019)]%
        {zen2019libritts}
\bibfield{author}{\bibinfo{person}{Heiga Zen}, \bibinfo{person}{Viet Dang}, \bibinfo{person}{Rob Clark}, \bibinfo{person}{Yu Zhang}, \bibinfo{person}{Ron~J Weiss}, \bibinfo{person}{Ye Jia}, \bibinfo{person}{Zhifeng Chen}, {and} \bibinfo{person}{Yonghui Wu}.} \bibinfo{year}{2019}\natexlab{}.
\newblock \showarticletitle{LibriTTS: A Corpus Derived from LibriSpeech for Text-to-Speech}. In \bibinfo{booktitle}{\emph{Interspeech 2019}}. \bibinfo{publisher}{ISCA}, \bibinfo{address}{Graz, Austria}, \bibinfo{pages}{1526--1530}.
\newblock


\bibitem[Zhang et~al\mbox{.}(2022)]%
        {zhang2022bigssl}
\bibfield{author}{\bibinfo{person}{Yu Zhang}, \bibinfo{person}{Daniel~S Park}, \bibinfo{person}{Wei Han}, \bibinfo{person}{James Qin}, \bibinfo{person}{Anmol Gulati}, \bibinfo{person}{Joel Shor}, \bibinfo{person}{Aren Jansen}, \bibinfo{person}{Yuanzhong Xu}, \bibinfo{person}{Yanping Huang}, \bibinfo{person}{Shibo Wang}, {et~al\mbox{.}}} \bibinfo{year}{2022}\natexlab{}.
\newblock \showarticletitle{Bigssl: Exploring the frontier of large-scale semi-supervised learning for automatic speech recognition}.
\newblock \bibinfo{journal}{\emph{IEEE Journal of Selected Topics in Signal Processing}} \bibinfo{volume}{16}, \bibinfo{number}{6} (\bibinfo{year}{2022}), \bibinfo{pages}{1519--1532}.
\newblock


\bibitem[Zhuo et~al\mbox{.}(2024)]%
        {zhuo2023lyricwhiz}
\bibfield{author}{\bibinfo{person}{Le Zhuo}, \bibinfo{person}{Ruibin Yuan}, \bibinfo{person}{Jiahao Pan}, \bibinfo{person}{Yinghao Ma}, \bibinfo{person}{Yizhi LI}, \bibinfo{person}{Ge Zhang}, \bibinfo{person}{Si Liu}, \bibinfo{person}{Roger Dannenberg}, \bibinfo{person}{Jie Fu}, \bibinfo{person}{Chenghua Lin}, \bibinfo{person}{Emmanouil Benetos}, \bibinfo{person}{Wei Xue}, {and} \bibinfo{person}{Yike Guo}.} \bibinfo{year}{2024}\natexlab{}.
\newblock \bibinfo{title}{LyricWhiz: Robust Multilingual Zero-shot Lyrics Transcription by Whispering to ChatGPT}.
\newblock
\newblock
\showeprint[arxiv]{2306.17103}~[cs.CL]
\urldef\tempurl%
\url{https://arxiv.org/abs/2306.17103}
\showURL{%
\tempurl}


\end{thebibliography}

\clearpage

\appendix

\section{Appendix}

\subsection{Datasheets for audio datasets}
Our audit uncovered a lack of documentation of audio datasets. This lack of documentation complicates assessing issues of bias and representation in these datasets, in addition to other considerations. \citet{gebru2021datasheets} propose datasheets for datasets. Below we propose datasheets for audio datasets. We reproduce the documentation guide in \cite{gebru2021datasheets}, and we add several questions and modifications specific to audio datasets (our modifications and additions in italics). While our paper answers some of these questions for some popular audio datasets, in general audio datasets have a high documentation debt \cite{bender2021dangers}. We hope future work will answer more of these questions for more audio datasets.

\renewcommand\labelenumi{(\theenumi)}
\begin{enumerate}
\item Motivation
\begin{enumerate}
\item What is the purpose of this dataset?
\item Who is the funding source of this dataset?
\item Who is the creator of this dataset?
\textit{
\item What type of task was this dataset primarily composed for—e.g., generative modelling? Classification? Audio separation? Style transfer? TTS? Something else?
\item What is the intended domain of the output of this dataset—speech? Music? General audio non-discriminatory sounds? Intentional generalization of both speech and music? Non-music, non-speech sounds such as sound effects?
\item Was this dataset intended to be exclusively in the audio domain, or multi-modal?
\item Was this dataset created for a specific purpose, or is it intended to be general use?
\item Was this dataset recorded with the intent of being “pure” sound as in separated or clear audio streams, or was it intentionally created with some environmental or background noise present?
\item Was this dataset created to fill a vacancy in our existing data (such as a low-resource language), or was it to add to an existing pool of similar datasets?
\item Was this a participatory effort by the community to build this dataset, or was this created by someone who is tangential to or otherwise does not belong to the community of which the dataset comprises?
\item If separate, what efforts were made to connect with the community for whom this dataset was comprised?
}
\item Any other comments?
\end{enumerate}

\item Composition
\begin{enumerate}
\item What do the instances that comprise this dataset represent (e.g., full songs, 10-second song clips, speech excerpts, singular utterances, 1-5 second sound effects)? Are there multiple types of instances?
\item How many instances are there total?
\textit{
\item What are the descriptive statistics of the time length of these instances in seconds/minutes/hours (e.g., mean, median, sum, mode)?
\item What data does each instance consist of—raw audio files? Metadata about the audio? Midi files? Other forms of symbolic representations?
\item If the dataset contains music:
\begin{enumerate}
\item What are the genres present? Is it an equal distribution?
\item Are they pre-existing/pre-released songs, or were they recorded for the purpose of this dataset?
\item Is the data created from computational generations of symbolic music?
\item Is the data composed of original recordings of music?
\end{enumerate}
\item If the dataset contains speech:
\begin{enumerate}
\item What languages are present? Is it an equal distribution?
\item How many speakers are present in the dataset?
\item Do the speakers speak multiple language or code-switch, and how is this identified?
\item What topics are present in the dataset?
\item Are there a variety of emotions present in the dataset? Is this identified in any way?
\end{enumerate}
}
\item Is there a label or target associated with each instance?
\item Is any desired metadata missing from individual instances?
\item Are there recommended data splits present (e.g., training/validation/test)?
\item What sources of noise are present in this dataset, and are they intentional or unintentional?
\item Are there any duplications present in this dataset (e.g., songs repeated multiple times), and if so, why?
\item Is the dataset self-contained, or does it rely on/link to an external source (e.g., links to song recordings)? If so, why?

\item \textit{Does the dataset contain data that may be copyrighted?}
\item Does the dataset contain anything that may be considered offensive?
\textit{
\item Has all of the data been listened to by an author or member of the data-construction team?
}
\textit{
\item If the dataset contains sounds created by humans:}
\begin{enumerate}

\item \textit{What demographic information about the speakers can you provide? Ex. age, gender, sexual orientation, language, locale/country, and accent.
}
\item Does the dataset contain audio files that could feasibly be used to identify human beings?
\item Does the dataset contain any speech that might be considered confidential or sensitive in any way (e.g., race or ethnic origins, sexual orientations, religious beliefs, political opinions, financial data, health data, biometrics data, or private and secure information such as criminal records or SSNs)?

\item \textit{ Does the dataset contain speech from a population that does not belong to the group they are trying to emulate (e.g., a white American native attempting to mimic an accent from a country to which they have no genuine relation)?
}
\end{enumerate}
\item Any other comments?
\end{enumerate}

\item Collection Process
\begin{enumerate}
\item How was the data associated with each instance acquired?
\item Was it recorded explicitly for the purpose of this dataset, or sourced elsewhere?
\item \textit{Was the audio scraped, recorded, computationally created (e.g., from MIDI files), or created in another manner?}
\textit{
\item Was the dataset sourced from audio in the public domain?
\item What technical setting was the speech recorded from, including the recording environment? What tools were used to process the audio?
\item What is the sampling rate of the audio? How many channels?
}
\item What mechanisms or procedures were used to collect the data?
\textit{
\item Was the data created from a computational algorithm?
\item Who was involved in the audio recordings, and how were they compensated?
\item Were there any copyright agreements struck with the contributors to the dataset?
\item What were the contributors to the dataset told their contribution would entail? Were they thoroughly briefed on the potential extent of the use of their audio, or was it left to general terms?
}
\item Over what timeframe was the audio collected?
\textit{
\item Was the audio collected from any multimodal sources (e.g., stripped from video)?
}
\item Were any ethical review processes conduced (e.g., from an IRB)?
\textit{
\item Has an analysis of the potential impact of the dataset been conducted?
}
\item If the dataset contains sounds created by humans:
\begin{enumerate}
\item Were the individuals present in the dataset notified about the data collection?
\item Did the individuals present in the dataset consent to the data collection and use of their data?
\item If consent was obtained, were the individuals provided with a mechanism to revoke their consent at a later date?
\end{enumerate}
\item Any other comments?
\end{enumerate}

\item Preprocessing/Cleaning/Labeling 
\begin{enumerate}
\item What preprocessing of the data was conducted in order to get it at the final stage the audio is in now?
\textit{
\item If there is metadata present, how was the metadata sourced?
\item Did users consent to release of this metadata?
\item If there is demographic information present about the individuals who recorded the audio, how was that obtained? Self-reported? Scraped? Sourced elsewhere?
}
\item Was the original raw data saved in addition to the final version of the data (if different)?
\textit{
\item Was background noise (or otherwise present environmental noise) intentionally cleaned from this data?
}
\item Is the code or other software used to prepare the data available to be published/released/acknowledged in line with the dataset?
\textit{
\item Are there any transcriptions available of the speech/lyrics? How was this processed? If human annotators, how were the annotators trained to create the transcriptions? If computationally annotated, what was this process? What is the accuracy rate?
\item Were there any content tagging such as hate-speech tags or swear word flagging?
\item Were any redactions of sensitive data performed on this dataset? How was this conducted?
}
\item Any other comments?
\end{enumerate}

\item Uses
\begin{enumerate}
\item Has the dataset been used for any tasks already?
\textit{
\item If so, have the original voices/musicians in the dataset been made aware of the extended/full use of the dataset?
\item Is there any repository linking all papers or systems using (or with access to) the dataset?
}
\item What tasks is the dataset designed for? Which are they suitable for? What could it potentially be used for?
\item Are there tasks for which the dataset should explicitly not be used?
\item Is there any element of the composition of the dataset that may impact how future uses of the dataset may be impacted?
\textit{
\item Is there any part of this dataset that is privately held but can be requested for research purposes?
}
\item Any other comments?
\end{enumerate}

\item Distribution
\begin{enumerate}
\item Will the dataset be distributed to third parties outside of the entity (e.g., company, institutions, organization) on behalf of which the dataset was created?
\textit{
\item Are the creators of audio files aware of this?
}
\item How will the dataset be distributed (e.g., open source, published on GitHub, by email request)?
\item When will the dataset be distributed?
\item Will the dataset be distributed under any copyright or other IP protection?
\end{enumerate}

\item Maintenance
\begin{enumerate}
\item Will the dataset be maintained or otherwise updated after the time of initial release? By whom?
\item How will the owner/curator of the dataset be contacted?
\textit{
\item If an artist discovers their work is present in this dataset, can they request it be removed? How?
}
\item Will the dataset be updated (e.g., to correct any potential errors, adding new files)?
\item How long will the data be retained?
\item If others want to contribute to this data source, is there a mechanism for them to do so?
\end{enumerate}
\end{enumerate}


\subsection{AudioSet Contents} 
\label{app:audioset}
\subsubsection{Video titles}

\begin{figure*}
\includegraphics[width=\textwidth]{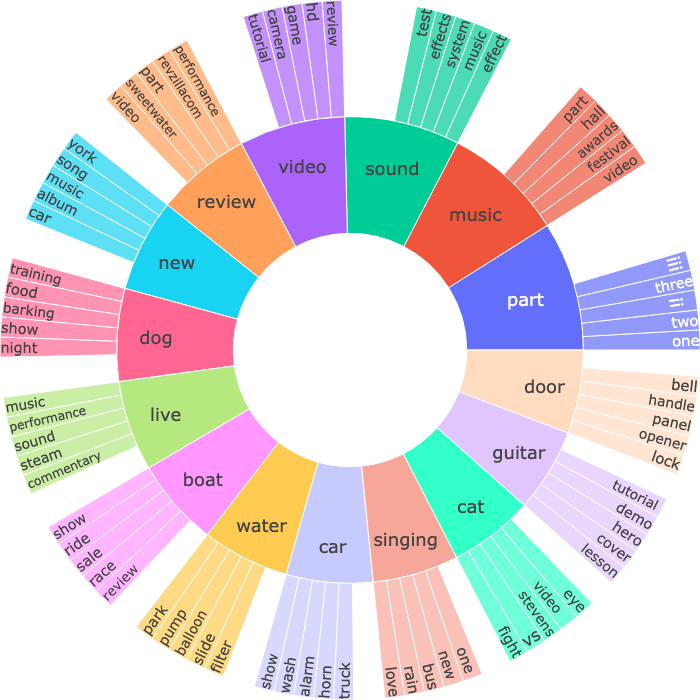}
    \caption{Common bigram breakdown of AudioSet video titles. The inner ring shows the top 25 most common words ignoring stopwords with sizes relatively to scale between words. The outer ring shows the top 5 words that follow each word in the inner ring with constant sizes for readability.}
\label{fig:audioset_title}
\end{figure*}

As AudioSet contains audio from music, speech, and general sounds, we assess its contents by analyzing titles of included texts. We first remove stop words from titles, and then we extract keywords directly from the words that remain as well as common associations with those keywords (Figure \ref{fig:audioset_title}). While YouTube contains a broad range of videos, we find that many sound snippets in AudioSet cluster around specific topics. Video games (``video game'', ``guitar hero''), music (``live music'', ``singing'', ``music festival''), reviews (``review video''), pets (``dog'', ``cat''), and vehicles (``car'', ``boat'') compose a significant fraction of AudioSet.

\subsubsection{Sample transcripts}

In Table \ref{tab:toxic}, we show examples of toxic-detected transcripts and include whether Whisper categorizes the audio as speech or music.

\begin{table*}[ht]
\caption{Ten randomly sampled transcripts in AudioSet that are detected as hateful by \texttt{pysentimento}. Profane words are redacted.}
\label{tab:toxic}
\begin{center}
\begin{tabular}{ll}
\hline
\textbf{Category} & \textbf{Text} \\
\hline
Speech & What? Alicia Fox pulling that hair. She still has. \\
Speech & and then shoot rocks up his *ss. \\
Speech & I don't know how you got here. I'm sorry. You're a real pain in the *ss. You're a \\
& real pain in the *ss. You're a real pain in the *ss. \\
Music & Pour some sh*t, matter of fact, go ahead and drink that Couple more shots, then \\
& we'll get freaky I peeped that, now I need that To the p*ssy like a record, go ahead \\
& and leak it It's real food, a dude up in the plate, party rockin' \\
Music & b*tch \\
Speech & do this without killing us. \\
Music & You're despicable.\\
Speech & Facebook it is. Hey look, I'm putting a video on Facebook. Me too. Look at her. \\
& Get the f*ck out of here, you d**chebag. \\
Speech & that woman had to the order and you you're an idiot no but i didn't come in here \\
& first because i didn't know the order \\
Speech & Stay with your p*ssy! Yes!
\\ \hline
\end{tabular}
\end{center}
\end{table*}

In Table \ref{tab:gay_keyword}, out of the 26 clips in AudioSet that mention \texttt{gay}-related keywords, we show several samples in which this term is used as a derogatory slur and depicts homophobic stereotypes. This problematic usage may propagate to speech generation models trained on AudioSet and amplify social bias \citep{nicolas2012s}.

\begin{table}[ht]
\caption{Sample transcripts in AudioSet that contain either \texttt{gay} or \texttt{gays} keywords. Profane words are redacted.}
\label{tab:gay_keyword}
\begin{center}
\begin{tabular}{l}
\hline
\textbf{Text} \\
\hline
You must be whipped, you must be gay. You must be whipped, whipped, you must be gay. \\ \\
And how was the tea virus? F*ck that intro. It's gay. \\ \\
Why do you suck your thumb? Whoa, whoa, f*ck you motherf*cker. I don't suck my f*cking \\
thumb. He sucks his middle finger. Are you sure? That's gay. You're not gay for that. \\ \\
Oh god, that's gay. Hold on. There we go, spooky! Isn't it f*cking cute? I think it's the cutest \\ thing ever. \\ \\
The zip code must have been like San Francisco as to gay people as to cats are to breeding. \\
There was just ridiculous amounts of cats. And they would not shut up. \\ \\
Why did you want this then? Lol, are you gay? Are you serious? Serious about what? What's \\
f*pping? \\ \\
Ha! Gay!
\\ \hline
\end{tabular}
\end{center}
\end{table}

\clearpage

\subsection{Code and Data Availability}
Code and annotation data are available at \url{https://anonymous.4open.science/r/gen-audio-ethics-F8CF/README.md}.

\end{document}